\documentclass[11pt]{article}
\usepackage[letterpaper,margin=1in]{geometry}

\usepackage{hyperref}

\usepackage{xcolor}
\definecolor{lightblue}{rgb}{0.5,0.5,1.0}
\definecolor{darkred}{rgb}{0.5,0,0}
\definecolor{darkgreen}{rgb}{0,0.5,0}
\definecolor{darkblue}{rgb}{0,0,0.5}
\usepackage{caption}    
\usepackage{subcaption}
\definecolor{lightblue}{rgb}{0.5,0.5,1.0}
\definecolor{darkred}{rgb}{0.5,0,0}
\definecolor{darkgreen}{rgb}{0,0.5,0}
\definecolor{darkblue}{rgb}{0,0,0.5}

\usepackage{pgfplots}
\pgfplotsset{compat=1.13} %I changed this to 1.13 from 1.16 cause my system did not have 1.16. Should change it back.

\usepackage{tabularx}
\usepackage{mathtools}
\usepackage{verbatim}
\usepackage{authblk}
\usepackage{wrapfig}
\usepackage{enumitem}
\usepackage{amssymb}
\usepackage{amsmath}
\usepackage{complexity}
\usepackage{tcolorbox}
\usepackage{nicefrac}
\usepackage{thmtools}
\usepackage{thm-restate}
\usepackage{mathtools}
\usepackage[linesnumbered,ruled,commentsnumbered,longend,algo2e]{algorithm2e}
\usepackage{tikz}
\usetikzlibrary{graphs}
\usetikzlibrary{shapes.misc, positioning}
\usetikzlibrary{shapes,fit}
\usepackage[edges]{forest}
\usetikzlibrary{shapes,arrows,calc,decorations.pathmorphing,arrows.meta}
\usetikzlibrary{shadows.blur}
\usepackage{amsmath}

\usepackage{graphicx}
\usepackage{rotating}

\usepackage[disable]{todonotes} %disable

\newcommand{\Traces}{\textsc{Traces}}
\newcommand{\nauty}{\textsc{nauty}}
\newcommand{\bliss}{\textsc{bliss}}
\newcommand{\saucy}{\textsc{saucy}}
\newcommand{\dejavu}{\textsc{dejavu}}

\newtheorem{lemma}{Lemma}

\SetKwProg{Fn}{function}{}{end}\SetKwFunction{RandomAut}{Automorphisms}
\SetKwProg{Fn}{function}{}{end}\SetKwFunction{RandomIso}{Isomorphism}
\SetKwFunction{FirstLeaf}{FirstLeaf}
\SetKwFunction{RandomWalk}{RandomWalk}
\SetKwFunction{NextNodeDFS}{OneNodeDFS}
\SetKwFunction{NextNodeBFS}{OneNodeBFS}
\SetKwFunction{Some}{Some}
\SetKwFunction{Sift}{Sift}
\SetKwFunction{RandomElement}{RandomElement}
\SetKwFunction{RandomChild}{RandomChild}
\SetKwFunction{NextChild}{NewChild}
\SetKwFunction{NextRandomChild}{NewRandomChild}
\SetKwFunction{Refine}{Refine}
\SetKwFunction{Reduce}{Reduce}
\SetKwFunction{Restore}{Restore}
\SetKwFunction{Subtree}{Subtree}
\SetKwFunction{RRef}{Ref}
\SetKwFunction{SSel}{Sel}
\SetKwFunction{Sift}{Sift}
\SetKwFunction{CertifyAutomorphism}{CertifyAutomorphism}
\SetKwFunction{CertifyIsomorphism}{CertifyIsomorphism}



\DeclareMathOperator{\Sym}{Sym}
\DeclareMathOperator{\supp}{supp}
\DeclareMathOperator{\Aut}{Aut}

\newtheorem{fact}{Fact}
{\bfseries}{\itshape}

\title{Engineering a Preprocessor for Symmetry Detection}
\author[1]{Markus Anders} 
\author[1]{Pascal Schweitzer}
\author[2]{Julian Stie\ss}
\affil[1]{TU Darmstadt}
\affil[2]{RPTU Kaiserslautern-Landau}
\newcommand\blfootnote[1]{%
  \begingroup
  \renewcommand\thefootnote{}\footnote{#1}%
  \addtocounter{footnote}{-1}%
  \endgroup
}
\begin{document}

\maketitle

\begin{abstract}
  State-of-the-art solvers for symmetry detection in combinatorial objects are becoming increasingly sophisticated software libraries.
Most of the solvers were initially designed with inputs from combinatorics in mind (\nauty{}, \bliss{}, \Traces{}, \dejavu{}). They excel at dealing with a complicated core of the input. 
Others focus on practical instances that exhibit sparsity. They excel at dealing with comparatively easy but extremely large substructures of the input (\saucy{}). 
In practice, these differences manifest in significantly diverging performances on different types of graph classes.

We engineer a preprocessor for symmetry detection. 
The result is a tool designed to shrink sparse, large substructures of the input graph.
  On most of the practical instances, the preprocessor improves the overall running time significantly for many of the state-of-the-art solvers.
  At the same time, our benchmarks show that the additional overhead is negligible. 

Overall we obtain single algorithms with competitive performance across all benchmark graphs. 
As such, the preprocessor bridges the disparity between solvers that focus on combinatorial graphs and large practical graphs. 
In fact, on most of the practical instances the combined setup significantly outperforms previous state-of-the-art.
\end{abstract}
\pagestyle{plain}
%\setcounter{page}{0}
%\cfoot{\thepage}

\blfootnote{Supported by the European Research Council (ERC) under the European Union's Horizon 2020 research and innovation programme (EngageS: grant No.~{820148}).}
%-------------------------------------------------------------------------------
%\newpage 
%\null
%\newpage

\section{Introduction}
Exploitation of symmetries is an indispensable instrument in a vast number of algorithmic application areas such as SAT~\cite{DBLP:conf/sat/KatebiSM10, DBLP:conf/sat/Anders22,DBLP:conf/sat/Devriendt0B17}, SMT~\cite{DBLP:conf/cade/DeharbeFMP11}, QBF~\cite{DBLP:conf/sat/KauersS18}, CSP~\cite{DBLP:reference/fai/GentPP06}, ILP~\cite{DBLP:books/daglib/p/Margot10, DBLP:journals/mpc/PfetschR19,DBLP:conf/colognetwente/HojnyP15} and many more.
However, in order to exploit symmetries, we have to compute them first.

Many types of objects can be modelled efficiently as graphs, so that the objects' symmetries correspond to the symmetries of the graph. This includes formulas, equation systems, finite relational structures, and many more (see~\cite{DBLP:conf/stoc/SchweitzerW19}). 
Hence, computing the symmetries of these objects reduces to computing symmetries of graphs.
We refer to the act of computing the symmetries of a graph as \emph{symmetry detection}.

State-of-the-art symmetry detection tools are \nauty{} \cite{DBLP:journals/jsc/McKayP14}, \saucy{} \cite{DBLP:conf/dac/DargaLSM04}, \bliss{} \cite{DBLP:conf/tapas/JunttilaK11}, \Traces{} \cite{DBLP:journals/jsc/McKayP14}, and \dejavu{} \cite{DBLP:conf/esa/AndersS21}.
Given as input a vertex-colored graph, they output all the symmetries of the graph.
All state-of-the-art tools are based around the so-called individualization-refinement (IR) paradigm (Section~\ref{subsec:ir:algo}).
Yet, they substantially differ in the applied  search strategies, pruning invariants, symmetry handling, and various other heuristics (see \cite{DBLP:journals/jsc/McKayP14, DBLP:conf/esa/AndersS21}).
This is also reflected in diverging performances on different graph classes. 

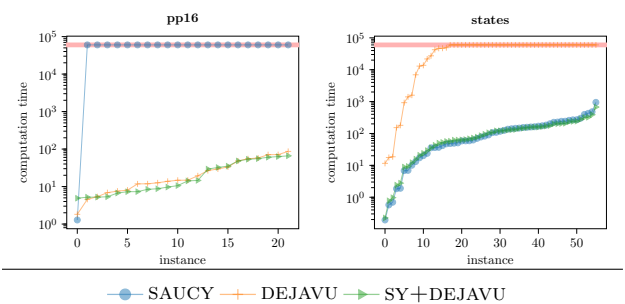
\begin{wrapfigure}{r}{0.5\textwidth}
  \centering
  \scalebox{0.45}{
  % This file was created with tikzplotlib v0.9.17.
\begin{tikzpicture}

\definecolor{color0}{rgb}{0.12156862745098,0.466666666666667,0.705882352941177}
\definecolor{color1}{rgb}{1,0.498039215686275,0.0549019607843137}
\definecolor{color2}{rgb}{0.172549019607843,0.627450980392157,0.172549019607843}

\begin{axis}[
legend cell align={left},
legend style={
  fill opacity=0.8,
  draw opacity=1,
  text opacity=1,
  at={(0.97,0.03)},
  anchor=south east,
  draw=white!80!black
},
log basis y={10},
tick align=outside,
tick pos=left,
title={\textbf{pp16}},
x grid style={white!69.0196078431373!black},
xlabel={instance},
xmin=-1.05, xmax=22.05,
xtick style={color=black},
y grid style={white!69.0196078431373!black},
ylabel={computation time},
ymin=0.750673988650639, ymax=102710.099411587,
ymode=log,
ytick style={color=black}
]
\draw[line width=4pt, red!30] (axis cs:-20,60000) -- (axis cs:600,60000);
\addplot [semithick, color0, opacity=0.5, mark=*, mark size=2.5, mark options={solid}]
table {%
0 1.28503
1 60000
2 60000
3 60000
4 60000
5 60000
6 60000
7 60000
8 60000
9 60000
10 60000
11 60000
12 60000
13 60000
14 60000
15 60000
16 60000
17 60000
18 60000
19 60000
20 60000
21 60000
};
\addlegendentry{\textsc{saucy}}
\addplot [semithick, color1, opacity=0.5, mark=+, mark size=2.5, mark options={solid}]
table {%
0 1.81712
1 4.56074
2 5.37005
3 6.88998
4 7.55267
5 8.04242
6 11.7975
7 11.8585
8 12.6053
9 13.8028
10 14.7303
11 14.8004
12 19.3726
13 26.6251
14 29.5001
15 33.2526
16 48.2377
17 55.1481
18 57.9031
19 70.8897
20 71.0983
21 86.6101
};
\addlegendentry{\textsc{dejavu}}
\addplot [semithick, color2, opacity=0.5, mark=triangle*, mark size=2.5, mark options={solid,rotate=270}]
table {%
0 4.8861
1 5.21386
2 5.23235
3 5.34382
4 6.72088
5 7.2995
6 7.31923
7 8.44222
8 8.75187
9 9.69916
10 10.6007
11 14.172
12 14.575
13 29.276
14 32.3474
15 35.109
16 48.2708
17 53.3948
18 55.5025
19 60.573
20 63.0187
21 66.0008
};
\addlegendentry{\textsc{sy+dejavu}}
\legend{};
\end{axis}

\end{tikzpicture}
  }
  \scalebox{0.45}{
  % This file was created with tikzplotlib v0.9.17.
\begin{tikzpicture}

\definecolor{color0}{rgb}{0.12156862745098,0.466666666666667,0.705882352941177}
\definecolor{color1}{rgb}{1,0.498039215686275,0.0549019607843137}
\definecolor{color2}{rgb}{0.172549019607843,0.627450980392157,0.172549019607843}

\begin{axis}[
legend cell align={left},
legend style={
  fill opacity=0.8,
  draw opacity=1,
  text opacity=1,
  at={(0.97,0.03)},
  anchor=south east,
  draw=white!80!black
},
log basis y={10},
tick align=outside,
tick pos=left,
title={\textbf{states}},
x grid style={white!69.0196078431373!black},
xlabel={instance},
xmin=-2.75, xmax=57.75,
xtick style={color=black},
y grid style={white!69.0196078431373!black},
ylabel={computation time},
ymin=0.102924158716217, ymax=112902.938872106,
ymode=log,
ytick style={color=black}
]
\draw[line width=4pt, red!30] (axis cs:-20,60000) -- (axis cs:600,60000);
\addplot [semithick, color0, opacity=0.5, mark=*, mark size=2.5, mark options={solid}]
table {%
0 0.193674
1 0.575171
2 0.69796
3 1.85285
4 1.91012
5 6.84823
6 6.97194
7 10.0871
8 13.1548
9 17.3229
10 20.1963
11 23.8269
12 35.575
13 36.2086
14 36.8254
15 41.711
16 47.2663
17 48.2689
18 49.5755
19 50.8008
20 58.3577
21 58.7193
22 59.0965
23 62.438
24 70.1958
25 77.5967
26 82.7435
27 90.2511
28 106.513
29 108.828
30 117.149
31 125.007
32 136.182
33 140.789
34 145.398
35 147.541
36 153.534
37 156.587
38 160.063
39 162.769
40 169.67
41 174.311
42 183.014
43 201.37
44 224.427
45 228.175
46 240.828
47 242.314
48 258.449
49 267.713
50 270.157
51 300.536
52 395.799
53 427.971
54 480.101
55 946.239
};
\addlegendentry{\textsc{saucy}}
\addplot [semithick, color1, opacity=0.5, mark=+, mark size=2.5, mark options={solid}]
table {%
0 11.5541
1 17.6688
2 18.621
3 152.816
4 180.254
5 925.98
6 1407.75
7 1586.54
8 6938.86
9 12972.3
10 13747.3
11 21783
12 27333.5
13 42440.9
14 46843.2
15 47186.2
16 49089.1
17 60000
18 60000
19 60000
20 60000
21 60000
22 60000
23 60000
24 60000
25 60000
26 60000
27 60000
28 60000
29 60000
30 60000
31 60000
32 60000
33 60000
34 60000
35 60000
36 60000
37 60000
38 60000
39 60000
40 60000
41 60000
42 60000
43 60000
44 60000
45 60000
46 60000
47 60000
48 60000
49 60000
50 60000
51 60000
52 60000
53 60000
54 60000
55 60000
};
\addlegendentry{\textsc{dejavu}}
\addplot [semithick, color2, opacity=0.5, mark=triangle*, mark size=2.5, mark options={solid,rotate=270}]
table {%
0 0.224378
1 0.776011
2 0.970205
3 2.39819
4 2.77155
5 8.85428
6 9.28767
7 12.1627
8 15.6891
9 20.8127
10 23.8721
11 29.2245
12 35.4288
13 42.7383
14 49.8322
15 52.6286
16 56.5003
17 57.2649
18 62.0057
19 62.2811
20 63.2245
21 64.7339
22 68.4758
23 71.9643
24 76.8057
25 85.7059
26 94.8037
27 105.563
28 108.841
29 121.142
30 121.185
31 123.71
32 127.432
33 130.686
34 132.67
35 143.121
36 146.191
37 152.458
38 157.126
39 158.296
40 158.92
41 163.806
42 170.815
43 176.771
44 200.362
45 202.455
46 205.111
47 208.701
48 234.206
49 239.468
50 243.022
51 279.174
52 319.326
53 322.613
54 387.79
55 674.437
};
\addlegendentry{\textsc{sy+dejavu}}
\legend{};
\end{axis}

\end{tikzpicture}
  }
  %\vspace{-0.2cm}
  \hrule{}
  \vspace{0.1cm}
  \scalebox{0.8}{
    \begin{tikzpicture}
    \begin{axis}[%
        hide axis,
        xmin=10,
        xmax=50,
        ymin=0,
        ymax=0.1,
        legend style={draw=none,legend cell align=left, legend columns=-1}
        ]
        \definecolor{color0}{rgb}{0.12156862745098,0.466666666666667,0.705882352941177}
\definecolor{color1}{rgb}{1,0.498039215686275,0.0549019607843137}
\definecolor{color2}{rgb}{0.172549019607843,0.627450980392157,0.172549019607843}
    \addlegendimage{semithick, color0, opacity=0.5, mark=*, mark size=2.5, mark options={solid}}
    \addlegendentry{\textsc{saucy}}
    \addlegendimage{semithick, color1, opacity=0.5, mark=+, mark size=2.5, mark options={solid}}
    \addlegendentry{\textsc{dejavu}}
    \addlegendimage{semithick, color2, opacity=0.5, mark=triangle*, mark size=2.5, mark options={solid,rotate=270}}
    \addlegendentry{\textsc{sy+dejavu}}
    \end{axis}
\end{tikzpicture}
    }
  \caption{Comparing solvers on a difficult combinatorial set of graphs (\textbf{pp16}, left) and large practical graphs (\textbf{states}, right). Timeout is $60s$ (red bar). \textsc{sy+dejavu} refers to \textsc{dejavu} with the preprocessor described in this paper.} \label{fig:introgarnish}
\end{wrapfigure}

We want to highlight two examples where the diverging performance between the solvers is notable, namely ``practical graphs'' and ``combinatorial graphs''.
For large practical graphs, such as graphs arising in SAT, QBF, MIP, or road networks, the solver \saucy{} outperforms all other solvers significantly (see, e.g., the results in \cite{DBLP:conf/sat/Anders22} or the benchmarks of this paper in Section~\ref{sec:benchmarks}). Indeed, designed with satisfiability-checking in mind, \saucy{} has been delicately engineered specifically for these types of graphs.
Intuitively, graphs arising from practical applications tend to be large in size but comparatively simple in their structure.
On the other hand, on almost all graph classes that are difficult relative to their size (e.g., projective planes, CFI graphs, and other regular combinatorial objects) \Traces{} and \dejavu{} readily outperform other solvers due to their more sophisticated search strategies (see \cite{DBLP:conf/esa/AndersS21} and \cite{DBLP:journals/jsc/McKayP14} for a more nuanced discussion).
In Figure~\ref{fig:introgarnish}, we demonstrate the large disparity between \saucy{} and \dejavu{} on a difficult graph class from combinatorics and a typical class of large practical graphs.

Only having solvers available that are geared towards specific types of graphs is of course an undesirable situation. This for example means we have to choose a solver and thus understand the type of input we are faced with. Also, we will struggle with inputs that are combinations of the different kinds of graphs.
Quite naturally, it is desirable instead to have a single solver performing well on all graphs.

A commonly used paradigm to make solvers for computational problems more widely applicable is to add a preprocessor.
The use of preprocessors has indeed already led to countless success stories, in particular in SAT, QBF or MaxSAT \cite{DBLP:conf/sat/EenB05, DBLP:conf/cade/BiereLS11, DBLP:conf/sat/KorhonenBSJ17}. 
In these applications, it is nowadays standard to apply a preprocessor to all inputs. 
In contrast to this, to date, no preprocessor has been available for symmetry detection.
In fact McKay and Piperno~\cite{DBLP:journals/jsc/McKayP14} explicitly highlight that in their opinion ``graphs of [particular types] ought to be handled by preprocessing'' before using their tools.

Beyond increasing performance, there are various other benefits to having a preprocessor.
Firstly, the problem of initially simplifying the graph can be tackled independently from the design of the main solver. This is especially desirable since implementations of state-of-the-art symmetry detection solvers are complex and detailed descriptions of the inner workings largely unavailable. 
Secondly, in turn, a preprocessor could even reduce the complexity of solver implementations if certain cases are reliably handled before running the solver. 
Lastly, implementing strategies in a common preprocessor makes them available to all the solvers simultaneously. 

Given the lack of an existing preprocessor for symmetry detection, \Traces{}, for example, has complicated subroutines that simplify some low-degree vertices before (and sometimes during) the computation
(see the implementation \cite{nautyTracesweb}).
Overall, the question is whether it is possible to design a common preprocessor that can
simplify inputs and is beneficial to all state-of-the-art solvers.

\textbf{Contribution.} We implement the first preprocessor \textsc{sassy} for symmetry detection. It is compatible by design with all state-of-the-art symmetry detection tools.
Our benchmarks (Section~\ref{sec:benchmarks}) corroborate that solver configurations using the preprocessor significantly outperform state-of-the-art on many practical graph classes. At the same time, the preprocessor introduces only a negligible overhead even on graphs on which the preprocessing techniques do not take effect.

The preprocessor bridges the disparity that exists between solvers that focus on difficult combinatorial graphs (\Traces{}, \dejavu{}, \bliss{}, \nauty{}) and those that focus on large practical graphs (\saucy{}). Through the use of the preprocessor, the former kind of solvers now outperform \saucy{} on most practical graphs. On top of that, on most of the sets, \saucy{} itself is accelerated through the use of the preprocessor as well. 

\textbf{Techniques.}
The preprocessor implements mainly techniques to handle graphs that are sparse, both in the input and output (e.g., practical graphs). In particular, it is made up of the following building blocks which we discuss throughout the paper:
\begin{enumerate}
  \item A framework to capture reduction techniques for input graphs. In particular, it captures the reconstruction of symmetries from the reduced graph back to the input graph, both theoretically and practically (Section~\ref{sec:toolbox}).
  \item A technique to efficiently remove vertices of degree $0$ and $1$ (Section~\ref{sec:deg0} and Section~\ref{sec:deg1}).
  \item Partial removal of degree $2$ vertices avoiding the introduction of colored or directed edges (Section~\ref{sec:deg2}).
  \item An IR-based probing technique for ``sparse automorphisms'' (Section~\ref{sec:probe}).  
  \item Exploiting connected components and homogeneous connections using the concept of quotient graphs (Section~\ref{sec:quotient}). 
\end{enumerate}
While \textsc{sassy} is the first universal preprocessor for symmetry detection, we want to remark that a flavor of (2) is already implemented in \Traces{}.
All the techniques other than (2) are novel contributions, however, we do want to mention that (4) and (5) draw some inspiration from existing techniques of solvers. We explain this in detail in the respective sections.

\section{Philosophy of the Preprocessor}
When designing a preprocessor, one of the main challenges is to map out which techniques and methods fall within the responsibility of the preprocessor and which task should be resolved by the main algorithm. Another delicate matter are the preprocessor/main solver and the user/preprocessor interfaces.

In the design of our preprocessor we were guided by conceptual principles as well as technical requirements.

\textbf{Conceptual principles.} On a conceptual level, our goal is to design efficient preprocessing subroutines that simplify the task of computing symmetries. Naturally, a preprocessor should only apply 
procedures that are comparatively fast in relation to the running time of the main algorithm.

The design of our preprocessor is centered around the so-called \emph{color refinement} algorithm. Color refinement is a powerful heuristic for symmetry detection. It is continuously and repeatedly applied in all state-of-the-art solvers. Thus, procedures that run within or close to color-refinement-time are safe to apply.

The general idea of the preprocessor is to remove or factor out substructures of the graph that are already resolved by an application of color refinement.
The main difficulty lies in detecting and exploiting these substructures as efficiently as possible.
Essentially, any part that can be handled efficiently ought to be carefully handled using precisely the right technique.

While the main solvers themselves are capable and somewhat efficient at handling such substructures (in particular, able to do this in polynomial time), not handling these substructures before handling more complicated parts or not handling them as efficiently as possible slows down the process as a whole.

Overall, we need to balance efficiency, effectiveness, and generality for our subroutines.

\textbf{Technical requirements.} On a technical level, we want our preprocessor to be compatible with all state-of-the-art solvers. Hence, we need to use an interface that is universal for all the existing tools. 
All tools read vertex-colored graphs and output symmetries. Hence, this is the interface that the preprocessor uses as well.

\begin{figure}
  \centering
  \scalebox{0.9}{
  \begin{tikzpicture}[scale=0.7]
    \draw[draw=black!50,rounded corners=0.125cm,fill=white,blur shadow={shadow blur steps=5}] (0,0)     rectangle (4,8);
    \node[align=center] (n1) at (2,7.5) {preprocessor};
    \node[align=center] (n2) at (2,6-0.5) {graph\\reduction};
    \node[align=center] (n3) at (2,2) {symmetry\\lift};
    \draw[draw=black,rounded corners=0.125cm] (0.5,0.5) rectangle (3.5,3.5);
    \draw[draw=black,rounded corners=0.125cm] (0.5,4.5-0.5) rectangle (3.5,7.5-0.5);

    \node[align=center] (n2) at (-5+2.5,2+0.3) {symmetries of};
    \draw [{Stealth[scale=1.25]}-](-5.5,2) -- (0.5,2);
    \node[align=center] (n2) at (-5+2.5,2-0.3) {$G$};

    \node[align=center] (n2) at (-5+2.5,5.5+0.3 + 0.75) {graph};
    \draw [-{Stealth[scale=1.25]}](-5.5,5.5 + 0.75) -- (0.5,5.5 + 0.75);
    \node[align=center] (n2) at (-5+2.5,5.5-0.3 + 0.75) {$G$};

    \node[align=center] (n2) at (-5+2.5,5.5+0.3 - 0.75) {symmetries of};
    \draw [{Stealth[scale=1.25]}-](-5.5,5.5 - 0.75) -- (0.5,5.5 - 0.75);
    \node[align=center] (n2) at (-5+2.5,5.5-0.3 - 0.75) {$G$};

    \draw[draw=black!50,rounded corners=0.125cm,fill=white,blur shadow={shadow blur steps=5}] (6+3.5,0) rectangle (10+3.5,8);
    \node[align=center] (n1) at (8+3.5,4) {main\\solver};

    \node[align=center] (n2) at (3.5+3,2+0.3) {symmetries of};
    \draw [{Stealth[scale=1.25]}-](3.5,2)   -- (6+3.5,2);
    \node[align=center] (n2) at (3.5+3,2-0.3) {$G'$};

    \node[align=center] (n2) at (3.5+3,5.5+0.3) {reduced graph};
    \draw [-{Stealth[scale=1.25]}](3.5,5.5) -- (6+3.5,5.5);
    \node[align=center] (n2) at (3.5+3,5.5-0.3) {$G'$};
  \end{tikzpicture}}
  \caption{Our proposed preprocessor/main solver and user/preprocessor interfaces. The preprocessor may already determine some (or all) symmetries of $G$ during graph reduction. The reduced instance is then passed on to the main solver.} \label{fig:interface}
\end{figure}
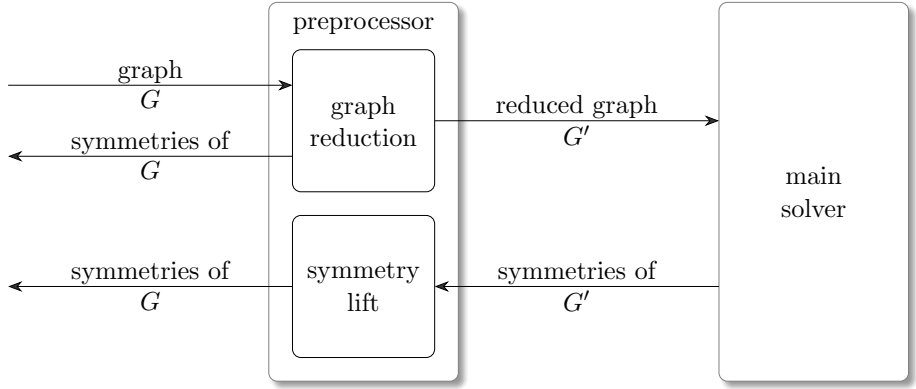

The preprocessor reads a vertex-colored graph and outputs a reduced vertex-colored graph passed to a main solver.
Moreover, the preprocessor may already determine some or all of the symmetries, and immediately outputs these to the user.
There is one more technicality: symmetries of the reduced graph which are computed by the main solver are, by definition, not symmetries of the original graph.
To rectify this, the preprocessor employs a backward-translation (i.e., a form of postprocessing) to lift symmetries that were discovered by the main solver back to being symmetries of the original input graph. 

Our design is illustrated in Figure~\ref{fig:interface}.

\section{Preliminaries}
A graph $G$ is a finite, simple, undirected graph, unless stated otherwise.
The neighborhood of a vertex~$v$ is denoted~$N(v)$, its \emph{degree} is $\deg(v) := |N(v)|$. 
For a set of vertices~$V'\subseteq V(G)$ the \emph{neighborhood} is the set~$N[V']\coloneqq (\bigcup_{v\in V'} N(v))\setminus V'$. 

A \emph{coloring} of a graph~$G$ is a map~$\pi\colon V(G)\rightarrow \mathcal{C}$ from the vertices to some set of colors. A (color) \emph{class}~$C$ is a set~$\pi^{-1}(c)$ of vertices of the same color.
A coloring $\pi$ is referred to as \emph{discrete} whenever $\pi$ is injective. In other words, in a discrete coloring each vertex has its own unique color.
Unless stated otherwise, we work with \emph{colored graphs} $G = (V, E, \pi)$ which consist of vertex set~$V$, edge set~$E$, and a coloring~$\pi$. Slightly abusing notation the pair~$(G,\pi)$ for an uncolored graph~$G=(V,E)$ is identified with~$(V,E,\pi)$. 
For a subset of the vertices~$V'\subseteq V$ the \emph{induced subgraph} of~$G=(V,E,\pi)$ is~$G[V']= (V',E',\pi|_{V'})$ where~$E'= \{e\in E\mid e\subseteq V'\times V'\}$. 

A bijection~$\varphi: V \mapsto V$ is called an \emph{automorphism} (symmetry) whenever $(\varphi(V), \varphi(E)) = (V, E)$ (applying $\varphi$ element-wise to the vertices in the edges of $E$).
If $G$ is colored, $\varphi$ also has to respect colors (i.e., satisfy~$\pi(\varphi(v))= \pi(v)$). 
The number of automorphisms can be exponential in the size of the graph.
Automorphisms form a permutation group under the composition operation. 
The \emph{automorphism group} containing all automorphisms of a (colored) graph~$G$ is~$\Aut(G)$. 
The \emph{support} of an automorphism $\varphi\in \Aut(G)$ is $\supp(\varphi) := \{\varphi(x) \neq x \;|\; x \in V(G)\}$, i.e., vertices not fixed by the automorphism.
A subset of automorphisms $S \subseteq \Aut(G)$ is a \emph{generating set} of $\Aut(G)$, whenever exhaustively composing permutations of $S$  leads to all elements of $\Aut(G)$. We write $\langle S \rangle = \Aut(G)$. 
This enables a concise encoding of $\Aut(G)$. Solvers generally only output a generating set of $\Aut(G)$.

\subsection{Color Refinement}
The color refinement algorithm is a well-studied procedure~\cite{DBLP:conf/icalp/AndersSW21,DBLP:journals/mst/BerkholzBG17,DBLP:journals/jsc/McKayP14}. 
It is a basic indispensable subroutine for isomorphism and symmetry computations.
Intuitively, given a colored graph, it splits apart colors in a specific way to produce a ``refined'' coloring. 
Crucially, this process does \emph{not} change the symmetries of the graph, while the refined coloring already restricts possible solutions.

Formally, a coloring~$\pi$ of a graph is \emph{equitable} if for all pairs of (not necessarily distinct) color classes~$C_1,C_2$, all vertices in~$C_1$ have the same number of neighbors in~$C_2$ (i.e.,~$|N(v)\cap C_2|= |N(v')\cap C_2|$ for all~$v,v'\in C_1$.)
Given a coloring~$\pi$, color refinement computes an equitable refinement~$\pi'$ (i.e., an equitable coloring~$\pi'$ for which~$\pi'(v)=\pi'(v')$ implies~$\pi(v)=\pi(v')$). 
In fact, it computes the coarsest equitable refinement.
Crucially, automorphisms of~$G=(V,E,\pi)$ are also automorphisms of~$G=(V,E,\pi')$ (and vice versa). It is thus beneficial and routine to work with~$\pi'$ instead of~$\pi$.

Color refinement can be implemented in such a way that it admits a worst case running time of~$\Theta((n+m) (\log n))$ (see~\cite{DBLP:journals/mst/BerkholzBG17}). From an implementation perspective it is the most crucial subroutine and therefore highly engineered.

\SetKwProg{Fn}{function}{}{end}
\SetKwFunction{Refine}{ColorRefinement}

\subsection{Quotient Graph}
For an equitable coloring~$\pi$ of an (otherwise uncolored) graph~$G$, the \emph{quotient graph} $Q(G, \pi)$ captures information regarding the number of neighbors that vertices in one color class have in another color class.
A quotient graph is a complete directed graph in which every vertex has a self-loop.
The vertex set of $Q(G, \pi)$ is $V(Q(G, \pi)) := \pi(V(G))$, i.e., the set of colors of vertices under~$\pi$.
The vertices of~$Q(G, \pi)$ are colored with the color they represent in~$G$.
We color the edge $(c_1, c_2)$ with the number of neighbors a vertex color $c_1$ has of color $c_2$ (possibly $c_1 = c_2$). Recall that, since $\pi$ is equitable, all vertices of $c_1$ have the same number of neighbors in $c_2$. 
Two graphs are \emph{indistinguishable by color refinement} if and only if their quotient graphs with respect to the coarsest equitable coloring are equal (see e.g.~\cite{DBLP:conf/icalp/AndersSW21}).

\subsection{Individualization Refinement Algorithms}\label{subsec:ir:algo}
The individualization-refinement framework is a general framework for algorithms computing isomorphisms, automorphisms and canonical labellings~(see~\cite{DBLP:journals/jsc/McKayP14}).
These algorithms generally work on a special tree, the so-called \emph{IR tree}. 
We give a brief description of how IR trees are constructed. Many details of the construction are not required to understand the contents of this paper. Only the techniques described in Section~\ref{sec:probe} rely on parts of the construction.
For a more extensive description see \cite{DBLP:journals/jsc/McKayP14,DBLP:conf/esa/AndersS21}.

\begin{figure}
  \begin{center}
    \begin{tikzpicture}[scale=1]
      \tikzset{decoration={snake,amplitude=.4mm,segment length=2mm,
      post length=0mm,pre length=0mm}, every node/.style={inner sep=0.5,outer sep=0}}

      \node[draw,circle,scale=0.66] at (2,0.5) (n_00) {
        \begin{tikzpicture}[scale=0.33]
        \node[draw,circle,fill=black,scale=0.2,minimum width = 2em] at (0.5,0) (n_0) {};
        \node[draw,circle,fill=black,scale=0.2,minimum width = 2em] at (1,1) (n_1) {};
        \node[draw,circle,fill=black,scale=0.2,minimum width = 2em] at (0,1) (n_2) {};

        \node[draw,circle,fill=black,scale=0.2,minimum width = 2em] at (3,0) (n_0') {};
        \node[draw,circle,fill=black,scale=0.2,minimum width = 2em] at (3,1) (n_1') {};
        \node[draw,circle,fill=black,scale=0.2,minimum width = 2em] at (2,1) (n_2') {};
        \node[draw,circle,fill=black,scale=0.2,minimum width = 2em] at (2,0) (n_3') {};
        
        \node[draw,circle,fill=black,scale=0.2,minimum width = 2em] at (1,2.5) (n_3) {};  
        \draw (n_1) -- (n_0);
        \draw (n_1) -- (n_2);
        \draw (n_2) -- (n_0);
        \draw (n_1') -- (n_0');
        \draw (n_0') -- (n_3');
        \draw (n_2') -- (n_3');
        \draw (n_2') -- (n_1');

        \draw[draw=black!75] (n_0) -- (n_3);
        \draw[draw=black!75] (n_1) -- (n_3);
        \draw[draw=black!75] (n_2) -- (n_3);
        \draw[draw=black!75] (n_0') -- (n_3);
        \draw[draw=black!75] (n_1') -- (n_3);
        \draw[draw=black!75] (n_2') -- (n_3);
        \draw[draw=black!75] (n_3') -- (n_3);
      \end{tikzpicture}
      };  
      \node[draw,circle,scale=0.66,fill=white] at (2,-1.5+0.5) (n_01) {
        \begin{tikzpicture}[scale=0.33]
        \node[draw,circle,fill=black,scale=0.2,minimum width = 2em] at (0.5,0) (n_0) {};
        \node[draw,circle,fill=black,scale=0.2,minimum width = 2em] at (1,1) (n_1) {};
        \node[draw,circle,fill=black,scale=0.2,minimum width = 2em] at (0,1) (n_2) {};

        \node[draw,circle,fill=black,scale=0.2,minimum width = 2em] at (3,0) (n_0') {};
        \node[draw,circle,fill=black,scale=0.2,minimum width = 2em] at (3,1) (n_1') {};
        \node[draw,circle,fill=black,scale=0.2,minimum width = 2em] at (2,1) (n_2') {};
        \node[draw,circle,fill=black,scale=0.2,minimum width = 2em] at (2,0) (n_3') {};
        
        \node[draw,circle,fill=orange,scale=0.2,minimum width = 2em] at (1,2.5) (n_3) {};  
        \draw (n_1) -- (n_0);
        \draw (n_1) -- (n_2);
        \draw (n_2) -- (n_0);
        \draw (n_1') -- (n_0');
        \draw (n_0') -- (n_3');
        \draw (n_2') -- (n_3');
        \draw (n_2') -- (n_1');

        \draw[draw=black!75] (n_0) -- (n_3);
        \draw[draw=black!75] (n_1) -- (n_3);
        \draw[draw=black!75] (n_2) -- (n_3);
        \draw[draw=black!75] (n_0') -- (n_3);
        \draw[draw=black!75] (n_1') -- (n_3);
        \draw[draw=black!75] (n_2') -- (n_3);
        \draw[draw=black!75] (n_3') -- (n_3);
      \end{tikzpicture}
      };
      \draw[decorate] (n_00) -- (n_01);

      \node[draw,circle,scale=0.66,fill=white] at (-4,-3) (n_10) {
        \begin{tikzpicture}[scale=0.33]
        \node[draw,circle,fill=red,scale=0.2,minimum width = 2em] at (0.5,0) (n_0) {};
        \node[draw,circle,fill=black,scale=0.2,minimum width = 2em] at (1,1) (n_1) {};
        \node[draw,circle,fill=black,scale=0.2,minimum width = 2em] at (0,1) (n_2) {};

        \node[draw,circle,fill=black,scale=0.2,minimum width = 2em] at (3,0) (n_0') {};
        \node[draw,circle,fill=black,scale=0.2,minimum width = 2em] at (3,1) (n_1') {};
        \node[draw,circle,fill=black,scale=0.2,minimum width = 2em] at (2,1) (n_2') {};
        \node[draw,circle,fill=black,scale=0.2,minimum width = 2em] at (2,0) (n_3') {};
        
        \node[draw,circle,fill=orange,scale=0.2,minimum width = 2em] at (1,2.5) (n_3) {};  
        \draw (n_1) -- (n_0);
        \draw (n_1) -- (n_2);
        \draw (n_2) -- (n_0);
        \draw (n_1') -- (n_0');
        \draw (n_0') -- (n_3');
        \draw (n_2') -- (n_3');
        \draw (n_2') -- (n_1');

        \draw[draw=black!75] (n_0) -- (n_3);
        \draw[draw=black!75] (n_1) -- (n_3);
        \draw[draw=black!75] (n_2) -- (n_3);
        \draw[draw=black!75] (n_0') -- (n_3);
        \draw[draw=black!75] (n_1') -- (n_3);
        \draw[draw=black!75] (n_2') -- (n_3);
        \draw[draw=black!75] (n_3') -- (n_3);
      \end{tikzpicture}
      };
      \node[draw,circle,scale=0.66,fill=white] at (-4,-4.5) (n_11) {
        \begin{tikzpicture}[scale=0.33]
        \node[draw,circle,fill=red,scale=0.2,minimum width = 2em] at (0.5,0) (n_0) {};
        \node[draw,circle,fill=cyan,scale=0.2,minimum width = 2em] at (1,1) (n_1) {};
        \node[draw,circle,fill=cyan,scale=0.2,minimum width = 2em] at (0,1) (n_2) {};

        \node[draw,circle,fill=black,scale=0.2,minimum width = 2em] at (3,0) (n_0') {};
        \node[draw,circle,fill=black,scale=0.2,minimum width = 2em] at (3,1) (n_1') {};
        \node[draw,circle,fill=black,scale=0.2,minimum width = 2em] at (2,1) (n_2') {};
        \node[draw,circle,fill=black,scale=0.2,minimum width = 2em] at (2,0) (n_3') {};
        
        \node[draw,circle,fill=orange,scale=0.2,minimum width = 2em] at (1,2.5) (n_3) {};  
        \draw (n_1) -- (n_0);
        \draw (n_1) -- (n_2);
        \draw (n_2) -- (n_0);
        \draw (n_1') -- (n_0');
        \draw (n_0') -- (n_3');
        \draw (n_2') -- (n_3');
        \draw (n_2') -- (n_1');

        \draw[draw=black!75] (n_0) -- (n_3);
        \draw[draw=black!75] (n_1) -- (n_3);
        \draw[draw=black!75] (n_2) -- (n_3);
        \draw[draw=black!75] (n_0') -- (n_3);
        \draw[draw=black!75] (n_1') -- (n_3);
        \draw[draw=black!75] (n_2') -- (n_3);
        \draw[draw=black!75] (n_3') -- (n_3);
      \end{tikzpicture}
      };
      \draw[decorate] (n_10) -- (n_11);

      \node[draw,circle,scale=0.66,fill=white] at (-2,-3) (n_20) {
        \begin{tikzpicture}[scale=0.33]
        \node[draw,circle,fill=black,scale=0.2,minimum width = 2em] at (0.5,0) (n_0) {};
        \node[draw,circle,fill=red,scale=0.2,minimum width = 2em] at (1,1) (n_1) {};
        \node[draw,circle,fill=black,scale=0.2,minimum width = 2em] at (0,1) (n_2) {};

        \node[draw,circle,fill=black,scale=0.2,minimum width = 2em] at (3,0) (n_0') {};
        \node[draw,circle,fill=black,scale=0.2,minimum width = 2em] at (3,1) (n_1') {};
        \node[draw,circle,fill=black,scale=0.2,minimum width = 2em] at (2,1) (n_2') {};
        \node[draw,circle,fill=black,scale=0.2,minimum width = 2em] at (2,0) (n_3') {};
        
        \node[draw,circle,fill=orange,scale=0.2,minimum width = 2em] at (1,2.5) (n_3) {};  
        \draw (n_1) -- (n_0);
        \draw (n_1) -- (n_2);
        \draw (n_2) -- (n_0);
        \draw (n_1') -- (n_0');
        \draw (n_0') -- (n_3');
        \draw (n_2') -- (n_3');
        \draw (n_2') -- (n_1');

        \draw[draw=black!75] (n_0) -- (n_3);
        \draw[draw=black!75] (n_1) -- (n_3);
        \draw[draw=black!75] (n_2) -- (n_3);
        \draw[draw=black!75] (n_0') -- (n_3);
        \draw[draw=black!75] (n_1') -- (n_3);
        \draw[draw=black!75] (n_2') -- (n_3);
        \draw[draw=black!75] (n_3') -- (n_3);
      \end{tikzpicture}
      };
      \node[draw,circle,scale=0.66,fill=white] at (-2,-4.5) (n_21) {
        \begin{tikzpicture}[scale=0.33]
        \node[draw,circle,fill=cyan,scale=0.2,minimum width = 2em] at (0.5,0) (n_0) {};
        \node[draw,circle,fill=red,scale=0.2,minimum width = 2em] at (1,1) (n_1) {};
        \node[draw,circle,fill=cyan,scale=0.2,minimum width = 2em] at (0,1) (n_2) {};

        \node[draw,circle,fill=black,scale=0.2,minimum width = 2em] at (3,0) (n_0') {};
        \node[draw,circle,fill=black,scale=0.2,minimum width = 2em] at (3,1) (n_1') {};
        \node[draw,circle,fill=black,scale=0.2,minimum width = 2em] at (2,1) (n_2') {};
        \node[draw,circle,fill=black,scale=0.2,minimum width = 2em] at (2,0) (n_3') {};
        
        \node[draw,circle,fill=orange,scale=0.2,minimum width = 2em] at (1,2.5) (n_3) {};  
        \draw (n_1) -- (n_0);
        \draw (n_1) -- (n_2);
        \draw (n_2) -- (n_0);
        \draw (n_1') -- (n_0');
        \draw (n_0') -- (n_3');
        \draw (n_2') -- (n_3');
        \draw (n_2') -- (n_1');

        \draw[draw=black!75] (n_0) -- (n_3);
        \draw[draw=black!75] (n_1) -- (n_3);
        \draw[draw=black!75] (n_2) -- (n_3);
        \draw[draw=black!75] (n_0') -- (n_3);
        \draw[draw=black!75] (n_1') -- (n_3);
        \draw[draw=black!75] (n_2') -- (n_3);
        \draw[draw=black!75] (n_3') -- (n_3);
      \end{tikzpicture}
      };
      \draw[decorate] (n_20) -- (n_21);

      \node[draw,circle,scale=0.66,fill=white] at (0,-3) (n_30) {
        \begin{tikzpicture}[scale=0.33]
        \node[draw,circle,fill=black,scale=0.2,minimum width = 2em] at (0.5,0) (n_0) {};
        \node[draw,circle,fill=black,scale=0.2,minimum width = 2em] at (1,1) (n_1) {};
        \node[draw,circle,fill=red,scale=0.2,minimum width = 2em] at (0,1) (n_2) {};

        \node[draw,circle,fill=black,scale=0.2,minimum width = 2em] at (3,0) (n_0') {};
        \node[draw,circle,fill=black,scale=0.2,minimum width = 2em] at (3,1) (n_1') {};
        \node[draw,circle,fill=black,scale=0.2,minimum width = 2em] at (2,1) (n_2') {};
        \node[draw,circle,fill=black,scale=0.2,minimum width = 2em] at (2,0) (n_3') {};
        
        \node[draw,circle,fill=orange,scale=0.2,minimum width = 2em] at (1,2.5) (n_3) {};  
        \draw (n_1) -- (n_0);
        \draw (n_1) -- (n_2);
        \draw (n_2) -- (n_0);
        \draw (n_1') -- (n_0');
        \draw (n_0') -- (n_3');
        \draw (n_2') -- (n_3');
        \draw (n_2') -- (n_1');

        \draw[draw=black!75] (n_0) -- (n_3);
        \draw[draw=black!75] (n_1) -- (n_3);
        \draw[draw=black!75] (n_2) -- (n_3);
        \draw[draw=black!75] (n_0') -- (n_3);
        \draw[draw=black!75] (n_1') -- (n_3);
        \draw[draw=black!75] (n_2') -- (n_3);
        \draw[draw=black!75] (n_3') -- (n_3);
      \end{tikzpicture}
      };
      \node[draw,circle,scale=0.66,fill=white] at (0,-4.5) (n_31) {
        \begin{tikzpicture}[scale=0.33]
        \node[draw,circle,fill=cyan,scale=0.2,minimum width = 2em] at (0.5,0) (n_0) {};
        \node[draw,circle,fill=cyan,scale=0.2,minimum width = 2em] at (1,1) (n_1) {};
        \node[draw,circle,fill=red,scale=0.2,minimum width = 2em] at (0,1) (n_2) {};

        \node[draw,circle,fill=black,scale=0.2,minimum width = 2em] at (3,0) (n_0') {};
        \node[draw,circle,fill=black,scale=0.2,minimum width = 2em] at (3,1) (n_1') {};
        \node[draw,circle,fill=black,scale=0.2,minimum width = 2em] at (2,1) (n_2') {};
        \node[draw,circle,fill=black,scale=0.2,minimum width = 2em] at (2,0) (n_3') {};
        
        \node[draw,circle,fill=orange,scale=0.2,minimum width = 2em] at (1,2.5) (n_3) {};  
        \draw (n_1) -- (n_0);
        \draw (n_1) -- (n_2);
        \draw (n_2) -- (n_0);
        \draw (n_1') -- (n_0');
        \draw (n_0') -- (n_3');
        \draw (n_2') -- (n_3');
        \draw (n_2') -- (n_1');

        \draw[draw=black!75] (n_0) -- (n_3);
        \draw[draw=black!75] (n_1) -- (n_3);
        \draw[draw=black!75] (n_2) -- (n_3);
        \draw[draw=black!75] (n_0') -- (n_3);
        \draw[draw=black!75] (n_1') -- (n_3);
        \draw[draw=black!75] (n_2') -- (n_3);
        \draw[draw=black!75] (n_3') -- (n_3);
      \end{tikzpicture}
      };
      \draw[decorate] (n_30) -- (n_31);

      \node[draw,circle,scale=0.66,fill=white] at (2,-3) (n_40) {
        \begin{tikzpicture}[scale=0.33]
        \node[draw,circle,fill=black,scale=0.2,minimum width = 2em] at (0.5,0) (n_0) {};
        \node[draw,circle,fill=black,scale=0.2,minimum width = 2em] at (1,1) (n_1) {};
        \node[draw,circle,fill=black,scale=0.2,minimum width = 2em] at (0,1) (n_2) {};

        \node[draw,circle,fill=red,scale=0.2,minimum width = 2em] at (3,0) (n_0') {};
        \node[draw,circle,fill=black,scale=0.2,minimum width = 2em] at (3,1) (n_1') {};
        \node[draw,circle,fill=black,scale=0.2,minimum width = 2em] at (2,1) (n_2') {};
        \node[draw,circle,fill=black,scale=0.2,minimum width = 2em] at (2,0) (n_3') {};
        
        \node[draw,circle,fill=orange,scale=0.2,minimum width = 2em] at (1,2.5) (n_3) {};  
        \draw (n_1) -- (n_0);
        \draw (n_1) -- (n_2);
        \draw (n_2) -- (n_0);
        \draw (n_1') -- (n_0');
        \draw (n_0') -- (n_3');
        \draw (n_2') -- (n_3');
        \draw (n_2') -- (n_1');

        \draw[draw=black!75] (n_0) -- (n_3);
        \draw[draw=black!75] (n_1) -- (n_3);
        \draw[draw=black!75] (n_2) -- (n_3);
        \draw[draw=black!75] (n_0') -- (n_3);
        \draw[draw=black!75] (n_1') -- (n_3);
        \draw[draw=black!75] (n_2') -- (n_3);
        \draw[draw=black!75] (n_3') -- (n_3);
      \end{tikzpicture}
      };
      \node[draw,circle,scale=0.66,fill=white] at (2,-4.5) (n_41) {
        \begin{tikzpicture}[scale=0.33]
        \node[draw,circle,fill=black,scale=0.2,minimum width = 2em] at (0.5,0) (n_0) {};
        \node[draw,circle,fill=black,scale=0.2,minimum width = 2em] at (1,1) (n_1) {};
        \node[draw,circle,fill=black,scale=0.2,minimum width = 2em] at (0,1) (n_2) {};

        \node[draw,circle,fill=red,scale=0.2,minimum width = 2em] at (3,0) (n_0') {};
        \node[draw,circle,fill=darkgreen,scale=0.2,minimum width = 2em] at (3,1) (n_1') {};
        \node[draw,circle,fill=cyan,scale=0.2,minimum width = 2em] at (2,1) (n_2') {};
        \node[draw,circle,fill=darkgreen,scale=0.2,minimum width = 2em] at (2,0) (n_3') {};
        
        \node[draw,circle,fill=orange,scale=0.2,minimum width = 2em] at (1,2.5) (n_3) {};  
        \draw (n_1) -- (n_0);
        \draw (n_1) -- (n_2);
        \draw (n_2) -- (n_0);
        \draw (n_1') -- (n_0');
        \draw (n_0') -- (n_3');
        \draw (n_2') -- (n_3');
        \draw (n_2') -- (n_1');

        \draw[draw=black!75] (n_0) -- (n_3);
        \draw[draw=black!75] (n_1) -- (n_3);
        \draw[draw=black!75] (n_2) -- (n_3);
        \draw[draw=black!75] (n_0') -- (n_3);
        \draw[draw=black!75] (n_1') -- (n_3);
        \draw[draw=black!75] (n_2') -- (n_3);
        \draw[draw=black!75] (n_3') -- (n_3);
      \end{tikzpicture}
      };
      \draw[decorate] (n_40) -- (n_41);

      \node[draw,circle,scale=0.66,fill=white] at (4,-3) (n_50) {
        \begin{tikzpicture}[scale=0.33]
        \node[draw,circle,fill=black,scale=0.2,minimum width = 2em] at (0.5,0) (n_0) {};
        \node[draw,circle,fill=black,scale=0.2,minimum width = 2em] at (1,1) (n_1) {};
        \node[draw,circle,fill=black,scale=0.2,minimum width = 2em] at (0,1) (n_2) {};

        \node[draw,circle,fill=black,scale=0.2,minimum width = 2em] at (3,0) (n_0') {};
        \node[draw,circle,fill=red,scale=0.2,minimum width = 2em] at (3,1) (n_1') {};
        \node[draw,circle,fill=black,scale=0.2,minimum width = 2em] at (2,1) (n_2') {};
        \node[draw,circle,fill=black,scale=0.2,minimum width = 2em] at (2,0) (n_3') {};
        
        \node[draw,circle,fill=orange,scale=0.2,minimum width = 2em] at (1,2.5) (n_3) {};  
        \draw (n_1) -- (n_0);
        \draw (n_1) -- (n_2);
        \draw (n_2) -- (n_0);
        \draw (n_1') -- (n_0');
        \draw (n_0') -- (n_3');
        \draw (n_2') -- (n_3');
        \draw (n_2') -- (n_1');

        \draw[draw=black!75] (n_0) -- (n_3);
        \draw[draw=black!75] (n_1) -- (n_3);
        \draw[draw=black!75] (n_2) -- (n_3);
        \draw[draw=black!75] (n_0') -- (n_3);
        \draw[draw=black!75] (n_1') -- (n_3);
        \draw[draw=black!75] (n_2') -- (n_3);
        \draw[draw=black!75] (n_3') -- (n_3);
      \end{tikzpicture}
      };
      \node[draw,circle,scale=0.66,fill=white] at (4,-4.5) (n_51) {
        \begin{tikzpicture}[scale=0.33]
        \node[draw,circle,fill=black,scale=0.2,minimum width = 2em] at (0.5,0) (n_0) {};
        \node[draw,circle,fill=black,scale=0.2,minimum width = 2em] at (1,1) (n_1) {};
        \node[draw,circle,fill=black,scale=0.2,minimum width = 2em] at (0,1) (n_2) {};

        \node[draw,circle,fill=darkgreen,scale=0.2,minimum width = 2em] at (3,0) (n_0') {};
        \node[draw,circle,fill=red,scale=0.2,minimum width = 2em] at (3,1) (n_1') {};
        \node[draw,circle,fill=darkgreen,scale=0.2,minimum width = 2em] at (2,1) (n_2') {};
        \node[draw,circle,fill=cyan,scale=0.2,minimum width = 2em] at (2,0) (n_3') {};
        
        \node[draw,circle,fill=orange,scale=0.2,minimum width = 2em] at (1,2.5) (n_3) {};  
        \draw (n_1) -- (n_0);
        \draw (n_1) -- (n_2);
        \draw (n_2) -- (n_0);
        \draw (n_1') -- (n_0');
        \draw (n_0') -- (n_3');
        \draw (n_2') -- (n_3');
        \draw (n_2') -- (n_1');

        \draw[draw=black!75] (n_0) -- (n_3);
        \draw[draw=black!75] (n_1) -- (n_3);
        \draw[draw=black!75] (n_2) -- (n_3);
        \draw[draw=black!75] (n_0') -- (n_3);
        \draw[draw=black!75] (n_1') -- (n_3);
        \draw[draw=black!75] (n_2') -- (n_3);
        \draw[draw=black!75] (n_3') -- (n_3);
      \end{tikzpicture}
      };
      \draw[decorate] (n_50) -- (n_51);

      \node[draw,circle,scale=0.66,fill=white] at (6,-3) (n_60) {
        \begin{tikzpicture}[scale=0.33]
        \node[draw,circle,fill=black,scale=0.2,minimum width = 2em] at (0.5,0) (n_0) {};
        \node[draw,circle,fill=black,scale=0.2,minimum width = 2em] at (1,1) (n_1) {};
        \node[draw,circle,fill=black,scale=0.2,minimum width = 2em] at (0,1) (n_2) {};

        \node[draw,circle,fill=black,scale=0.2,minimum width = 2em] at (3,0) (n_0') {};
        \node[draw,circle,fill=black,scale=0.2,minimum width = 2em] at (3,1) (n_1') {};
        \node[draw,circle,fill=red,scale=0.2,minimum width = 2em] at (2,1) (n_2') {};
        \node[draw,circle,fill=black,scale=0.2,minimum width = 2em] at (2,0) (n_3') {};
        
        \node[draw,circle,fill=orange,scale=0.2,minimum width = 2em] at (1,2.5) (n_3) {};  
        \draw (n_1) -- (n_0);
        \draw (n_1) -- (n_2);
        \draw (n_2) -- (n_0);
        \draw (n_1') -- (n_0');
        \draw (n_0') -- (n_3');
        \draw (n_2') -- (n_3');
        \draw (n_2') -- (n_1');

        \draw[draw=black!75] (n_0) -- (n_3);
        \draw[draw=black!75] (n_1) -- (n_3);
        \draw[draw=black!75] (n_2) -- (n_3);
        \draw[draw=black!75] (n_0') -- (n_3);
        \draw[draw=black!75] (n_1') -- (n_3);
        \draw[draw=black!75] (n_2') -- (n_3);
        \draw[draw=black!75] (n_3') -- (n_3);
      \end{tikzpicture}
      };
      \node[draw,circle,scale=0.66,fill=white] at (6,-4.5) (n_61) {
        \begin{tikzpicture}[scale=0.33]
        \node[draw,circle,fill=black,scale=0.2,minimum width = 2em] at (0.5,0) (n_0) {};
        \node[draw,circle,fill=black,scale=0.2,minimum width = 2em] at (1,1) (n_1) {};
        \node[draw,circle,fill=black,scale=0.2,minimum width = 2em] at (0,1) (n_2) {};

        \node[draw,circle,fill=cyan,scale=0.2,minimum width = 2em] at (3,0) (n_0') {};
        \node[draw,circle,fill=darkgreen,scale=0.2,minimum width = 2em] at (3,1) (n_1') {};
        \node[draw,circle,fill=red,scale=0.2,minimum width = 2em] at (2,1) (n_2') {};
        \node[draw,circle,fill=darkgreen,scale=0.2,minimum width = 2em] at (2,0) (n_3') {};
        
        \node[draw,circle,fill=orange,scale=0.2,minimum width = 2em] at (1,2.5) (n_3) {};  
        \draw (n_1) -- (n_0);
        \draw (n_1) -- (n_2);
        \draw (n_2) -- (n_0);
        \draw (n_1') -- (n_0');
        \draw (n_0') -- (n_3');
        \draw (n_2') -- (n_3');
        \draw (n_2') -- (n_1');

        \draw[draw=black!75] (n_0) -- (n_3);
        \draw[draw=black!75] (n_1) -- (n_3);
        \draw[draw=black!75] (n_2) -- (n_3);
        \draw[draw=black!75] (n_0') -- (n_3);
        \draw[draw=black!75] (n_1') -- (n_3);
        \draw[draw=black!75] (n_2') -- (n_3);
        \draw[draw=black!75] (n_3') -- (n_3);
      \end{tikzpicture}
      };
      \draw[decorate] (n_60) -- (n_61);
      \node[draw,circle,scale=0.66,fill=white] at (8,-3) (n_70) {
        \begin{tikzpicture}[scale=0.33]
        \node[draw,circle,fill=black,scale=0.2,minimum width = 2em] at (0.5,0) (n_0) {};
        \node[draw,circle,fill=black,scale=0.2,minimum width = 2em] at (1,1) (n_1) {};
        \node[draw,circle,fill=black,scale=0.2,minimum width = 2em] at (0,1) (n_2) {};

        \node[draw,circle,fill=black,scale=0.2,minimum width = 2em] at (3,0) (n_0') {};
        \node[draw,circle,fill=black,scale=0.2,minimum width = 2em] at (3,1) (n_1') {};
        \node[draw,circle,fill=black,scale=0.2,minimum width = 2em] at (2,1) (n_2') {};
        \node[draw,circle,fill=red,scale=0.2,minimum width = 2em] at (2,0) (n_3') {};
        
        \node[draw,circle,fill=orange,scale=0.2,minimum width = 2em] at (1,2.5) (n_3) {};  
        \draw (n_1) -- (n_0);
        \draw (n_1) -- (n_2);
        \draw (n_2) -- (n_0);
        \draw (n_1') -- (n_0');
        \draw (n_0') -- (n_3');
        \draw (n_2') -- (n_3');
        \draw (n_2') -- (n_1');

        \draw[draw=black!75] (n_0) -- (n_3);
        \draw[draw=black!75] (n_1) -- (n_3);
        \draw[draw=black!75] (n_2) -- (n_3);
        \draw[draw=black!75] (n_0') -- (n_3);
        \draw[draw=black!75] (n_1') -- (n_3);
        \draw[draw=black!75] (n_2') -- (n_3);
        \draw[draw=black!75] (n_3') -- (n_3);
      \end{tikzpicture}
      };
      \node[draw,circle,scale=0.66,fill=white] at (8,-4.5) (n_71) {
        \begin{tikzpicture}[scale=0.33]
        \node[draw,circle,fill=black,scale=0.2,minimum width = 2em] at (0.5,0) (n_0) {};
        \node[draw,circle,fill=black,scale=0.2,minimum width = 2em] at (1,1) (n_1) {};
        \node[draw,circle,fill=black,scale=0.2,minimum width = 2em] at (0,1) (n_2) {};

        \node[draw,circle,fill=darkgreen,scale=0.2,minimum width = 2em] at (3,0) (n_0') {};
        \node[draw,circle,fill=cyan,scale=0.2,minimum width = 2em] at (3,1) (n_1') {};
        \node[draw,circle,fill=darkgreen,scale=0.2,minimum width = 2em] at (2,1) (n_2') {};
        \node[draw,circle,fill=red,scale=0.2,minimum width = 2em] at (2,0) (n_3') {};
        
        \node[draw,circle,fill=orange,scale=0.2,minimum width = 2em] at (1,2.5) (n_3) {};  
        \draw (n_1) -- (n_0);
        \draw (n_1) -- (n_2);
        \draw (n_2) -- (n_0);
        \draw (n_1') -- (n_0');
        \draw (n_0') -- (n_3');
        \draw (n_2') -- (n_3');
        \draw (n_2') -- (n_1');

        \draw[draw=black!75] (n_0) -- (n_3);
        \draw[draw=black!75] (n_1) -- (n_3);
        \draw[draw=black!75] (n_2) -- (n_3);
        \draw[draw=black!75] (n_0') -- (n_3);
        \draw[draw=black!75] (n_1') -- (n_3);
        \draw[draw=black!75] (n_2') -- (n_3);
        \draw[draw=black!75] (n_3') -- (n_3);
      \end{tikzpicture}
      };
      \draw[decorate] (n_70) -- (n_71);

      \draw[-{Stealth[length=2mm, width=2mm]},black] (n_01) to[in=90, out=270, looseness=0.8] (n_10);
      \draw[-{Stealth[length=2mm, width=2mm]},black] (n_01) to[in=90, out=270, looseness=0.8] (n_20);
      \draw[-{Stealth[length=2mm, width=2mm]},black] (n_01) to[in=90, out=270, looseness=0.8] (n_30);
      \draw[-{Stealth[length=2mm, width=2mm]},black] (n_01) to[in=90, out=270, looseness=0.8] (n_40);
      \draw[-{Stealth[length=2mm, width=2mm]},black] (n_01) to[in=90, out=270, looseness=0.8] (n_50);
      \draw[-{Stealth[length=2mm, width=2mm]},black] (n_01) to[in=90, out=270, looseness=0.8] (n_60);
      \draw[-{Stealth[length=2mm, width=2mm]},black] (n_01) to[in=90, out=270, looseness=0.8] (n_70);

      \foreach \n in {1,...,3}{
      \node[draw=white,fill=white] at (-6-0.5     + 2*\n, -5.5) (n_\n1_1) {};
      \node[draw=white,fill=white] at (-6-0.15 + 2*\n, -5.5) (n_\n1_2) {};
      \node[draw=white,fill=white] at (-6+0.15 + 2*\n, -5.5) (n_\n1_3) {};
      \node[draw=white,fill=white] at (-6+0.5 + 2*\n, -5.5) (n_\n1_4) {};

      \draw[-{Stealth[length=2mm, width=2mm]},black] (n_\n1) -- (n_\n1_1);
      \draw[-{Stealth[length=2mm, width=2mm]},black] (n_\n1) -- (n_\n1_2);
      \draw[-{Stealth[length=2mm, width=2mm]},black] (n_\n1) -- (n_\n1_3);
      \draw[-{Stealth[length=2mm, width=2mm]},black] (n_\n1) -- (n_\n1_4);
        }

        \foreach \n in {4,...,7}{
      \node[draw=white,fill=white] at (-6     + 2*\n, -5.5) (n_\n1_1) {};
      \node[draw=white,fill=white] at (-6-0.5 + 2*\n, -5.5) (n_\n1_2) {};
      \node[draw=white,fill=white] at (-6+0.5 + 2*\n, -5.5) (n_\n1_3) {};
      \draw[-{Stealth[length=2mm, width=2mm]},black] (n_\n1) -- (n_\n1_1);
      \draw[-{Stealth[length=2mm, width=2mm]},black] (n_\n1) -- (n_\n1_2);
      \draw[-{Stealth[length=2mm, width=2mm]},black] (n_\n1) -- (n_\n1_3);
        }
    \end{tikzpicture}
  \end{center}

        \caption{Construction of an IR-tree. Pairs of nodes connected by a squiggly line correspond to one node in the IR tree. The upper node in a pair (other than the root) illustrates the individualization of a vertex, whereas the lower node in the pair shows the final coloring obtained after color refinement has been applied to the upper node.} \label{fig:IR}
\end{figure}
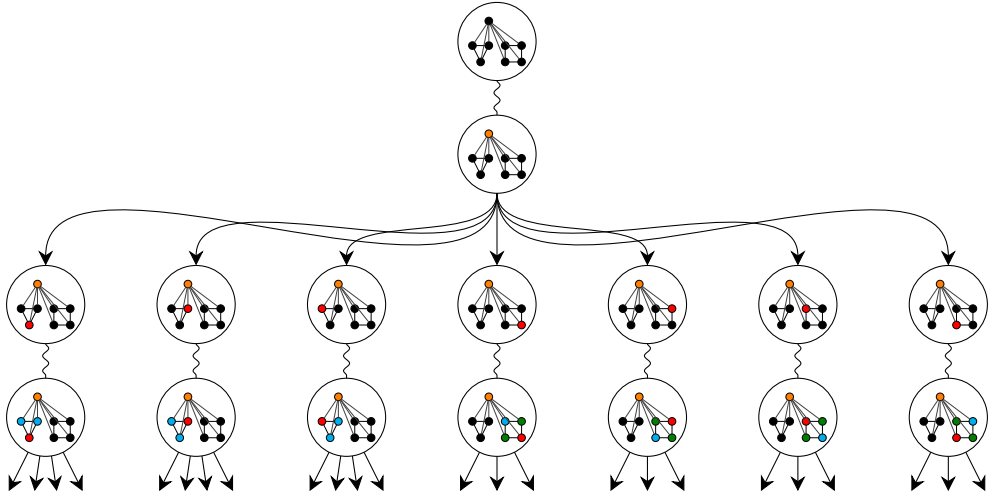

Each node $x$ of the tree has a corresponding equitable coloring $\pi_x$ of the input graph.
The leaves correspond to discrete colorings. 
The most important property of IR trees is that they are isomorphism-invariant, meaning that on $G$ and $\varphi(G)$ (where $\varphi$ is an isomorphism) we obtain isomorphic IR trees.

Let $(G, \pi)$ be the input graph. Let $\pi'$ be the coarsest equitable refinement of $\pi$. 
We let the root of the IR tree correspond to $\pi'$. 

In each node $x$ of an IR tree, a non-trivial color class from the corresponding coloring $\pi_x$ is chosen (i.e., a $C = \pi_x^{-1}(c)$ with $|C| > 1$, $C$ must be chosen isomorphism-invariantly).
If there is no non-trivial color class, then $x$ is a leaf and its corresponding coloring is discrete.
Otherwise, for each $v \in C$, we define $x_v$ as a child of $x$ in the IR tree.
Let $\pi_{x_v}$ denote the coloring corresponding to $x_v$. 
We may obtain $\pi_{x_v}$ from $\pi_x$ as follows.
Starting from $\pi_x$, we first artificially single out $v$ (i.e.,~\emph{individualize} $v$). This means we set $\pi_{x_v}(v) := c'$ where $c' \notin \pi(V(G))$ (again, $c'$ is chosen isomorphism-invariantly).
Then, we refine the coloring using color refinement, obtaining the equitable coloring $\pi_{x_v}$.
Figure~\ref{fig:IR} shows a sketch of obtaining an IR tree from a graph.

We can derive automorphisms from IR trees. If $\pi_1, \pi_2$ are leaves of the tree, i.e., discrete colorings, then $\varphi := \pi_1^{-1}\circ \pi_2$ defines a permutation on $V(G)$. While $\varphi$ is not guaranteed to be an automorphism, we can efficiently test whether it is (by checking whether $\varphi(G) = G$). With this method all of~$\Aut(G)$ can be computed. This follows essentially from the fact that comparing all pairs of leaves in this way will give us all automorphisms of $G$ (or rather a generating set of $\Aut(G)$ when automorphism pruning is applied; see \cite{DBLP:journals/jsc/McKayP14}).

Practical implementations of course apply numerous strategies to perform the search more efficiently, such as automorphism pruning, invariants, or search strategies which may omit parts of the IR tree. For a more comprehensive description of IR trees and practical techniques see \cite{DBLP:journals/jsc/McKayP14,DBLP:conf/esa/AndersS21}. 

\section{A Toolbox for Reducing Graphs} \label{sec:toolbox}
We now embark on our journey of describing techniques that simplify a graph for symmetry detection. 
The goal is always to efficiently reduce the number of vertices and edges of the graph.
However, we need to be wary of some technicalities.
Whenever we alter the graph, we need to make sure that either no symmetries are lost, or that we output the symmetries that would be lost immediately. 
Furthermore, we have to ensure that after preprocessing is done symmetries of the reduced graph can be mapped back to symmetries of the original graph. After all, we are interested in symmetries of the original graph.
In order to ease this process, we first lay out some general techniques that we use throughout the paper. 

The first type of technique we describe modifies an input graph~$G$ on vertex set~$V$ to another graph~$G'$ with vertex set~$V'\subseteq V$ so that
\begin{enumerate}
\item $\Aut(G)|_{V'}\subseteq \Aut(G')$ and \hfill (symmetry preservation)
\item $\Aut(G)|_{V'}\supseteq \Aut(G')$. \hfill (symmetry lifting)
\end{enumerate}

Here by $\Aut(G)|_{V'}$ we mean the set of maps obtained by restricting the domain of each $\varphi \in \Aut(G)$ to $V'$ (and the range to~$\varphi(V')$).
If conditions (1) and (2) hold, $V'$ must also be invariant under $\Aut(G)$.

Under these conditions the restriction to~$V'$ is a natural homomorphism~$p\colon \Aut(G)\rightarrow \Aut(G')$. The orbit-stabilizer theorem (see~\cite[Theorem~2.16]{handbook}) implies then that if~$S'\subseteq \Aut(G)$ is a set of lifts of a generating set~$S$ of~$\Aut(G')$, i.e.~$p(S')= S$, then~$\Aut(G)= \langle S', \ker(p)\rangle$ (where $\langle \Gamma \rangle$ denotes the group \emph{generated} by $\Gamma$, see \cite{seress_2003}). Here~$\ker(p) = \{\varphi\in \Aut(G)\mid p(\varphi) \neq 1 \}$ is the \emph{kernel} of~$p$ and $1$ denotes the identity.

 Overall this enables us to separate the computation of~$\Aut(G)$ into computing automorphisms of the removed parts of the graph and the automorphisms of the reduced graph.

Crucial for the techniques is now that~$G'$ and a generating set of~$\ker(p)$ can be efficiently computed from~$G$, and that the set of lifts~$S'$ can be efficiently computed from a generating set of~$\Aut(G')$.

In particular, we require an efficient postprocessing technique for lifting of automorphisms to parts that were reduced.
For this we introduce two tools that are used throughout the design of the preprocessor.

\paragraph{Canonical Representation Strings.} During preprocessing, the parts we remove from the original graph might be symmetrical to (i.e., in the same orbit as) other parts of the graph. 
So, after symmetries of the reduced graph have been computed, we need to lift symmetries of the reduced graph to symmetries of the original graph.
In particular, the lifted symmetries must map all the removed parts correctly. 
To simplify the lifting of symmetries we introduce \emph{representation strings} associated with the remaining vertices. These encode the nature (i.e., the ``isomorphism type'') of the vertices that were removed. The encoding is stored in the color of a suitable vertex that remains.
If a remaining vertex is then mapped to another vertex, the corresponding subgraphs represented by the strings are then mapped to each other in a canonical way.

We define this process formally through a \emph{representation mapping} $\mathcal{R}(v): V \mapsto V^*$ from the vertices to sequences of vertices as follows.
Assume we have a graph $G := (V, E, \pi)$ which is reduced to $G' := (V', E', \pi')$ with  $V' \subseteq V$ and $E' \subseteq E$. We require the following:
\begin{enumerate}
\item It holds that $\mathcal{R}(v) := vS$ with $S \in V^*$ for all $v \in V'$, i.e., each remaining vertex must represent itself first.
\item It holds that  $\mathcal{R}(v) := \epsilon$ for all $v \in V \smallsetminus V'$, i.e., a removed vertex does not represent any vertex.
\item For each deleted vertex $v \in V \smallsetminus V'$ there is at most one $v' \in V'$ and at most one $i \in \mathbb{N}$ such that $v := \mathcal{R}(v')_i$, i.e., each deleted vertex is represented by at most one remaining vertex, once.
\end{enumerate}
For each automorphism of the remaining graph $\varphi \in \Aut(G')$ we now define its lifted bijection $\varphi_\mathcal{R}(v) \in \Sym(V)$ (the symmetric group on $V$). We define~$\varphi_\mathcal{R}(v) :=$
\[
   \begin{cases} 
    \varphi(v) &\text{if } v \in V'\\
    \mathcal{R}(\varphi(v'))_i          &\text{if } v = \mathcal{R}(v')_i \text{ for } v' \in V', i \in \mathbb{N}\\
     v          &\text{if } v' \neq \mathcal{R}(v')_i \text{ for all }  v' \in V',i \in \mathbb{N}.
  \end{cases}
\]
We call $\mathcal{R}$ a \emph{canonical representation mapping} if $\varphi_\mathcal{R} \in \Aut(G)$ for all $\varphi \in \Aut(G')$.

We note by definition, canonical representation mappings can be chained, i.e., if we reduce a graph $G$ multiple times, we can simply apply the respective canonical representation mappings in reverse until we reach an automorphism of $G$.
We can even rewrite chained canonical representation mappings into a single map by essentially composing the functions. (More accurately we have to interpret strings of strings as simple strings using concatenation.) 

\paragraph{Sparse Automorphisms and Restoration.} 
A concept that we implicitly use throughout the following sections is sparse encodings of automorphisms.  
A conventional way to do this is the cycle notation of permutations, i.e., store only for each non-fixed element its image \cite{handbook}.
The precise encoding used is of no importance, however.
Crucially, automorphisms ought to be encoded using space that is proportional to the size of their support, i.e., in $\mathcal{O}(|\supp(\varphi)|)$. This means, in particular, that sparse automorphisms, i.e., automorphisms that fix almost all points, are succinctly encoded.

Using a canonical representation mapping $\mathcal{R}$ and sparse automorphism encodings, automorphisms of a reduced graph $G'$ can be efficiently lifted to automorphisms of the original graph $G$. Indeed, lifts can be computed in time (and in space) linear in the size of the support of the lift, by replacing vertices by their represented strings.

\begin{fact}
Given $\varphi \in \Aut(G')$, the lift $\varphi_\mathcal{R} \in \Aut(G)$ can be computed in time $\mathcal{O}(|\supp(\varphi_\mathcal{R})|)$.
\end{fact} 

Let us remark that often canonical representations in fact ensure that lifted supports are as small as possible. We say that a representation mapping~$\mathcal{R}$ \emph{respects kernel orbits} if it has the property that~$v_1 \in \mathcal{R}(v) \Leftrightarrow v_2 \in \mathcal{R}(v)$ whenever~$v_1$ and~$v_2$ are in the same orbit of~$\ker(p)$. 
\begin{fact}
If~$\mathcal{R}$ respects kernel orbits then $p(\psi) = \varphi$ implies that~$|\supp(\varphi_\mathcal{R})|\leq |\supp(\psi)|$.
\end{fact} 
All representations we describe subsequently respect kernel orbits. 

We should remark that none of the state-of-the-art solvers except for \saucy{} feature an interface for sparse automorphisms, i.e., an interface that enables access to an automorphism in time $\mathcal{O}(|\supp(\varphi)|)$. 
Instead, access is only possible in $\Omega(|V|)$. 
If a user-application uses the interface for sparse automorphisms correctly, this can yield substantial running time benefits on graphs that contain a large number of sparse automorphisms (which is the case for many practical graphs).
Most solvers internally incur a cost of $\Omega(|V|)$ to handle automorphisms anyway, in turn making the sparse interface unnecessary.
Since this is not true for our preprocessor and to ensure potential running time benefits to user-applications, automorphisms found by the preprocessor are of course accessible in a sparse manner. 
 
\section{Removing low degree vertices} \label{sec:lowdeg}
The first class of efficient reduction techniques we describe removes vertices of low degree. 
We propose strategies for vertices of degree $0, 1$ and $2$ (analogously $n-1, n-2$).

We want to remark that techniques for preprocessing vertices of degree $0$ and~$1$ can be found in the implementation of \Traces{} \cite{nautyTracesweb}, even though the description in \cite{DBLP:journals/jsc/McKayP14} describes differing techniques and does not mention them explicitly. %\pascal{replace ``differs'' with does not mention them explicitly?}.
The \Traces{} implementation for degree $0, 1$ differs from our proposed strategy in that it does \emph{not} compute color refinement before removing degree $0$ and $1$ vertices, which leads to different challenges and techniques. 
\Traces{} also includes code for dealing with degree $2$ vertices \emph{during} its main computations, which we discuss further below.

\subsection{Degree $0$ Vertices} \label{sec:deg0}
Preprocessing vertices of degree $0$ (and analogously $n-1$) is simple. The algorithm detects color classes consisting of vertices of degree $0$. We let~$V'$ be the set of vertices of degree larger than~$0$.
By simply removing vertices of degree $0$ and not representing them in $\mathcal{R}$ at all, $\mathcal{R}$ indeed defines a canonical representation mapping.

The kernel~$\ker(p)$ of the restriction~$p$ onto~$V'$ is computed as follows. For each color class of degree 0 vertices in~$G$ we output generators for the symmetric group on the class.

\subsection{Degree $1$ Vertices} \label{sec:deg1}
Exhaustively removing all vertices of degree $1$ (and analogously $n-2$) essentially removes all tree-like appendages from graphs.
It is well-known that applying color refinement produces the orbit partitioning on these tree-like appendages -- with the notable exception of not determining whether the roots of these appendages are in the same orbit or not.

We can remove degree $1$ vertices recursively.
Let~$G$ be a graph that contains degree $1$ vertices.
We describe $G'$ and $\mathcal{R}$ where we remove a color class of degree $1$ vertices.

Let $C$ denote such a color class of degree $1$ vertices. Since the coloring is equitable, all neighbors of vertices of $C$ are in the same color class~$P$. In case~$P=C$ we have connected components of size 2. This case can be handled similar to the reduction of degree~0 vertices, so we assume~$P\neq C$.
We partition $C$ into classes $C_1, \dots{}, C_m$ where $c \in C_i$ is adjacent to $p_i \in P$.
For the representation mapping, we set $\mathcal{R}(p_i) := p_iC_i$ (where $C_i$ may appear in arbitrary order).
We set $G' := G \smallsetminus \{C\}$. 
The coloring $\pi$ remains unchanged. 
Note that $\pi$ is still an equitable coloring for $G'$. The kernel~$\ker(p)$ is the direct product of the symmetric group $\Sym(C_i)$ for each $i \in \{1, \dots{}, m\}$ (and points outside~$C$ are fixed).
The process can then be repeated until all vertices of degree $1$ are removed.

By construction, the reduction is symmetry preserving and symmetry lifting, thus~$\Aut(G)= \langle S', \ker(p)\rangle$. 
Here, as before,~$S$ is a generating set for~$\Aut(G')$ and~$S'$ a corresponding set of lifts.

\subsection{Degree $2$ Vertices} \label{sec:deg2}
If we allow graphs produced by our preprocessor to contain directed, colored edges, there is a simple reduction that removes all vertices of degree $2$: we may encode the multiset of paths between two vertices~$v_1$ and~$v_2$ with~$\deg(v_i) \geq 3$ whose internal vertices all have degree 2 as one directed, colored edge between~$v_1$ and~$v_2$ (see also \cite[Proof of Lemma 15]{DBLP:journals/jacm/KieferPS19}).

There are, however, drawbacks to this approach: first of all, there are solvers that do not implement directed and colored edges. 
Secondly, even when solvers implement these edges, using directed and colored edges comes at the price of additional overhead \cite{DBLP:conf/wea/Piperno18}.
Intuitively, while removing all degree $2$ vertices can certainly cause a significant size-reduction, some of the complexity of the removed path is only shifted into the color encoding of the edges. In turn, we require refinements to take into account edge colors. This complicates color refinement, the central subroutine.

For these reasons, if possible, we prefer to remove degree $2$ vertices in a way that does not require the introduction of directed or colored edges.
Generally, it is unclear how to efficiently replace paths of colored vertices by uncolored edges.
We want to mention that the technique of \Traces{} does not remove vertices of degree $2$, but handles automorphisms solely permuting degree $2$ vertices using special-purpose code \cite{DBLP:journals/jsc/McKayP14}. 
In particular, this means that degree $2$ vertices still incur cost during color refinement computations during IR search.

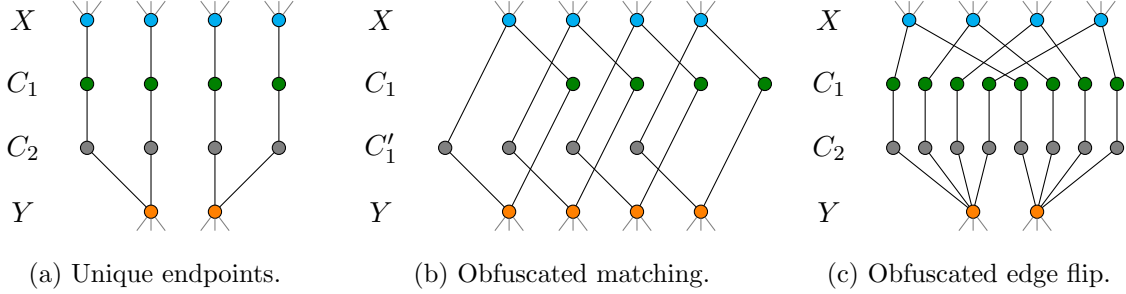
\begin{figure}
  \centering 
  \begin{subfigure}[b]{0.32\textwidth}
    \centering
  \begin{tikzpicture}[every node/.style={inner sep=0.5,outer sep=0, minimum size = 5pt, draw=black}, scale=0.85]
    \node[circle, fill=orange] (x0) at (1,0) {};
    \node[circle, fill=orange] (x1) at (2,0) {};

    \node[circle, fill=gray] (c0) at (0,1) {};
    \node[circle, fill=gray] (c1) at (1,1) {};
    \node[circle, fill=gray] (c2) at (2,1) {};
    \node[circle, fill=gray] (c3) at (3,1) {};

    \node[circle, fill=darkgreen] (d0) at (0,2) {};
    \node[circle, fill=darkgreen] (d1) at (1,2) {};
    \node[circle, fill=darkgreen] (d2) at (2,2) {};
    \node[circle, fill=darkgreen] (d3) at (3,2) {};

    \node[circle, fill=cyan] (y0) at (0,3) {};
    \node[circle, fill=cyan] (y1) at (1,3) {};
    \node[circle, fill=cyan] (y2) at (2,3) {};
    \node[circle, fill=cyan] (y3) at (3,3) {};

    \draw (x0) -- (c0);
    \draw (x0) -- (c1);
    \draw (x1) -- (c2);
    \draw (x1) -- (c3);

    \foreach \i in {0,...,3}{
      \draw (c\i) -- (d\i);
      \draw (y\i) -- (d\i);
    }

    \foreach \i in {0,...,3}{
      \foreach \j in {0,...,2}{
        \node[fill=none,draw=none] (y\i_\j) at (\i -0.3 + \j*0.3, 3 + 0.4) {};
      \draw[draw=black!50] (y\i) -- (y\i_\j);
      }
    }

    \foreach \i in {0,...,1}{
      \foreach \j in {0,...,2}{
        \node[fill=none,draw=none] (x\i_\j) at (1 + \i -0.3 + \j*0.3, 0 - 0.4) {};
      \draw[draw=black!50] (x\i) -- (x\i_\j);
      }
    }

    \node[draw=none,fill=none] (l0) at (-1, 0) {$Y$};
    \node[draw=none,fill=none] (l0) at (-1, 1) {$C_2$};
    \node[draw=none,fill=none] (l0) at (-1, 2) {$C_1$};
    \node[draw=none,fill=none] (l0) at (-1, 3) {$X$};
  \end{tikzpicture}
    \caption{Unique endpoints.} \label{fig:unique_endpoints}
  \end{subfigure}
  \begin{subfigure}[b]{0.32\textwidth}
    \centering
  \begin{tikzpicture}[every node/.style={inner sep=0.5,outer sep=0, minimum size = 5pt, draw=black}, scale=0.85]
    \node[circle, fill=orange] (x0) at (0,0) {};
    \node[circle, fill=orange] (x1) at (1,0) {};
    \node[circle, fill=orange] (x2) at (2,0) {};
    \node[circle, fill=orange] (x3) at (3,0) {};

    \node[circle, fill=gray] (c0) at (0-1,1) {};
    \node[circle, fill=gray] (c1) at (1-1,1) {};
    \node[circle, fill=gray] (c2) at (2-1,1) {};
    \node[circle, fill=gray] (c3) at (3-1,1) {};

    \node[circle, fill=darkgreen] (d0) at (0+1,2) {};
    \node[circle, fill=darkgreen] (d1) at (1+1,2) {};
    \node[circle, fill=darkgreen] (d2) at (2+1,2) {};
    \node[circle, fill=darkgreen] (d3) at (3+1,2) {};

    \node[circle, fill=cyan] (y0) at (0,3) {};
    \node[circle, fill=cyan] (y1) at (1,3) {};
    \node[circle, fill=cyan] (y2) at (2,3) {};
    \node[circle, fill=cyan] (y3) at (3,3) {};

    \foreach \i in {0,...,3}{
      \draw (y\i) -- (c\i);
      \draw (x\i) -- (c\i);
      \draw (y\i) -- (d\i);
      \draw (x\i) -- (d\i);
    }

    \foreach \i in {0,...,3}{
      \foreach \j in {0,...,2}{
        \node[fill=none,draw=none] (y\i_\j) at (\i -0.3 + \j*0.3, 3 + 0.4) {};
      \draw[draw=black!50] (y\i) -- (y\i_\j);
      }
    }

    \foreach \i in {0,...,3}{
      \foreach \j in {0,...,2}{
        \node[fill=none,draw=none] (x\i_\j) at (\i -0.3 + \j*0.3, 0 - 0.4) {};
      \draw[draw=black!50] (x\i) -- (x\i_\j);
      }
    }

    \node[draw=none,fill=none] (l0) at (-2, 0) {$Y$};
    \node[draw=none,fill=none] (l0) at (-2, 1) {$C_1'$};
    \node[draw=none,fill=none] (l0) at (-2, 2) {$C_1$};
    \node[draw=none,fill=none] (l0) at (-2, 3) {$X$};
  \end{tikzpicture}
  \caption{Obfuscated matching.} \label{fig:obfuscated_matching}
  \end{subfigure}
  \begin{subfigure}[b]{0.32\textwidth}
    \centering
  \begin{tikzpicture}[every node/.style={inner sep=0.5,outer sep=0, minimum size = 5pt, draw=black}, scale=0.85]
    \node[circle, fill=orange] (x0) at (1,0) {};
    \node[circle, fill=orange] (x1) at (2,0) {};

    \foreach \i in {0,...,7}{
      \node[circle, fill=gray]      (c\i) at (-0.25 + \i*0.5,1) {};
      \node[circle, fill=darkgreen] (d\i) at (-0.25 + \i*0.5,2) {};
    }

    \node[circle, fill=cyan] (y0) at (0,3) {};
    \node[circle, fill=cyan] (y1) at (1,3) {};
    \node[circle, fill=cyan] (y2) at (2,3) {};
    \node[circle, fill=cyan] (y3) at (3,3) {};

    \foreach \i in {0,...,7}{
      \draw (c\i) -- (d\i);
    }

    \foreach \i in {0,...,3}{
      \draw (x0) -- (c\i);
    }
    \foreach \i in {4,...,7}{
      \draw (x1) -- (c\i);
    }
    
    \foreach \i in {0,...,3}{
      \draw (d\i) -- (y\i);
    }

    \foreach \i in {4,...,7}{
      \pgfmathtruncatemacro{\il}{\i - 4}
      \draw (d\i) -- (y\il);
    }

    \foreach \i in {0,...,3}{
      \foreach \j in {0,...,2}{
        \node[fill=none,draw=none] (y\i_\j) at (\i -0.3 + \j*0.3, 3 + 0.4) {};
      \draw[draw=black!50] (y\i) -- (y\i_\j);
      }
    }

    \foreach \i in {0,...,1}{
      \foreach \j in {0,...,2}{
        \node[fill=none,draw=none] (x\i_\j) at (1 + \i -0.3 + \j*0.3, 0 - 0.4) {};
      \draw[draw=black!50] (x\i) -- (x\i_\j);
      }
    }

    \node[draw=none,fill=none] (l0) at (-1.25, 0) {$Y$};
    \node[draw=none,fill=none] (l0) at (-1.25, 1) {$C_2$};
    \node[draw=none,fill=none] (l0) at (-1.25, 2) {$C_1$};
    \node[draw=none,fill=none] (l0) at (-1.25, 3) {$X$};
  \end{tikzpicture}
  \caption{Obfuscated edge flip.} \label{fig:obfuscated_edgeflip}
  \end{subfigure}
  \caption{Reducible degree 2 patterns.}
\end{figure}

\textbf{Non-branching paths with unique endpoint.} 
We describe a heuristic which we found to be often applicable in practical data sets.
It encodes
paths with internal vertices of degree 2 that run between two color classes by a set of edges connecting the endpoints directly. However, it only does so if the set of paths can be reconstructed unambiguously from the set of edges.
In particular, the inserted edges may not interfere with existing edges.

We detect 
paths of length $t$ between distinct color classes $X$ and $Y$ whose internal vertices have degree 2. In each vertex of~$X$ exactly one such path should start (see Figure~\ref{fig:unique_endpoints}).
More formally, suppose~$X=C_0,C_1,\ldots,C_t,C_{t+1}=Y$ are colors so that
\begin{itemize}
\item vertices in~$X$ do not have neighbors in~$Y$,
\item for~$i\in \{1,\ldots,t\}$ vertices in~$C_i$ have degree 2,
\item for~$i\in \{1,\ldots,t\}$ vertices in~$C_i$ have a neighbor in~$C_{i-1}$ and~$C_{i+1}$, and
\item every node in~$X$ has exactly one neighbor in~$C_1$ 
\end{itemize} 
then we define~$G'=(V',E')$ via~$V'\coloneqq V-(C_1\cup\cdots \cup C_t)$ and~$E'\coloneqq E(G[V']) \cup E''$, where~$E''$ consists of pairs~$(x,y)$ for which there is a path~$(x,c_1,\ldots,c_t,y)$ with~$c_i\in C_i$. The corresponding representation map is~$\mathcal{R}(x)= xc_1c_2\cdots c_t$, where~$(x,c_1,\ldots,c_t,y)$ is the unique path from~$x$ to some vertex~$y\in Y$ with~$c_i\in C_i$. 

Note that the newly introduced edges~$E''$ form a biregular bipartite graph between~$X$ and~$Y$ in which vertices of~$X$ have degree 1. It is not difficult to check that this yields a canonical representation map that respects kernel orbits.

\textbf{Obfuscated Matchings.} 
The preprocessor has special fast code for the particular case in which~$|X|=|Y|$. In this case~$E'$ encodes a perfect matching between~$X$ and~$Y$.

A slight extension of the technique checks for other choices of~$C_i$ whether they also satisfy the required properties and yield exactly the same matching~$E''$. 
In fact, if there is another matching via color classes~$C'_1,\ldots,C'_{t'}$ between $X$ and $Y$ which encodes~$E''$ , we also delete vertices in the~$C'_i$ (see Figure~\ref{fig:obfuscated_matching}).
The special purpose code uses arrays and can efficiently check whether matchings coincide.

We should mention that in the implementation, we only perform the check for paths of length $t=1$. 
It turns out that the special case of~$t=1$ and in fact multiple such paths encoding the same matching is very common in particular on the MIP and SAT benchmarks.

\textbf{Obfuscated Edge Flip.} A case that also can be handled efficiently and is not covered by previous techniques is where $X$ and $Y$ are connected by $|X||Y|$ equally-colored, unique paths. 
In this case, each vertex $x \in X$ is connected to all $y \in Y$ by a path (see Figure~\ref{fig:obfuscated_edgeflip}). 
It is easy to see that deleting all such paths is both symmetry preserving and symmetry lifting (this is related to the edge flip described in Section~\ref{subsec:edge_flip}). 

Formally, suppose~$X=C_0,C_1,\ldots,C_t,C_{t+1}=Y$ are colors so that
\begin{itemize}
\item for~$i\in \{1,\ldots,t\}$ vertices in~$C_i$ have degree 2,
\item for~$i\in \{1,\ldots,t\}$ vertices in~$C_i$ have a neighbor in~$C_{i-1}$ and~$C_{i+1}$, and
\item every node in~$X$ has exactly $|Y|$ neighbors in~$C_1$, where the corresponding paths end in all $y \in Y$.
\end{itemize} 
The technique in turn removes all $C_0,C_1,\ldots,C_t$ from the graph.

Let us now consider computing the lift of this reduction. Unfortunately, canonical representation strings are not sufficient to express the lift: we need to determine how $C_0,C_1,\ldots,C_t$ are mapped, and this depends on \emph{both} the vertices of $X$ and $Y$.
We can not simply attach $C_0,C_1,\ldots,C_t$ to the canonical representation strings of one of the color classes.
However, if we know how both $X$ and $Y$ are mapped, it is trivial to reconstruct the original symmetry: assume a symmetry maps $x \in X$ to $x'$ and $y \in Y$ to $y'$. 
This just means that in the lift, we need to map the path connecting $x$ to $y$ to the path connecting $x'$ to $y'$.
Hence, the lift can still be computed very easily and efficiently.

In the implementation, we do write vertices of $C_0,C_1,\ldots,C_t$ into both the representation strings of $X$ and $Y$, breaking the formal requirement of not having double entries. However, we use an encoding trick to denote the double entries, which triggers special code during the reconstruction of the symmetries.

\section{Probing for Sparse Automorphisms} \label{sec:probe}
We now propose a strategy for probing for sparse automorphisms.
The strategy is inspired by a heuristic introduced by \saucy{}: the heuristic that \saucy{} (and in part \Traces{}) implements is the fast detection and exploitation of ``sparse automorphisms''.
Essentially, \saucy{} is able to not only exploit sparsity in the input (e.g., as in low degree vertices of the graph) but also in the output (as in automorphisms with small support) \cite{DBLP:conf/dac/DargaLSM04}.

While the preprocessor already exploits sparsity using the low-degree vertex techniques defined above, in this section, we discuss a more general technique that is independent of specific substructures of the graph.  
More specifically, we propose a probing strategy for sparse automorphisms.
If successful and automorphisms are discovered, we ``divide them out'', breaking the symmetry by individualization.

The strategy implemented in \saucy{} continuously checks for two colorings $\pi_1, \pi_2$ whether interchanging vertices in color classes of size $1$ (i.e., \emph{singleton vertices}) and fixing all other vertices yields an automorphism of the graph. 
More formally, we define the permutation~$\varphi_{\pi_1, \pi_2}(v) :=$
\[ \begin{cases} 
v &\text{ if } |\pi_1^{-1}(\pi_1(v))| \neq 1 \vee |\pi_2^{-1}(\pi_2(v))| \neq 1\\ 
\pi_2^{-1}(\pi_1(v)) &\text{ otherwise.}
\end{cases}\]
Then, we simply check whether $\varphi_{\pi_1, \pi_2}$ is indeed an automorphism of $G$. 
This check can be done in time 
$\mathcal{O}(\Sigma_{v\in \supp (\varphi_{\pi_1, \pi_2})} 1+ \deg(v))$. For this we only need to check whether $\varphi_{\pi_1, \pi_2}$ induces an automorphism on the support of $\varphi_{\pi_1, \pi_2}$ and on its neighborhood.

\saucy{} performs the check for local automorphisms during the depth-first search of the IR tree. It can then store the information about the automorphism and internally exploit its existence. For preprocessing purposes, however, we want to make the graph simpler or smaller. In particular, we want to be able to immediately divide out discovered automorphisms. This should be independent of the type of subsequent strategy used. 
We therefore propose a new search strategy that is specifically designed to be used for preprocessing.

Our probing strategy only searches for automorphisms which can be used directly to reduce the graph. The idea is as follows. For a color class that we want to reduce, we attempt to collect automorphisms that transitively permute all the vertices in the entire color class. This certifies that the color class is an orbit. We can then individualize an arbitrary vertex of the color class. In contrast, if we only have some automorphisms that together do not act transitively on the color class, it is not clear how to manipulate the graph favorably. In particular, since some of the vertices may not be in the same orbit, we do not know which vertex to individualize.

\subsection{$L$-Bounded IR Probing on a Color Class}

\SetKw{Break}{break}
\SetKwProg{Fn}{function}{}{end}
\SetKwFunction{flatir}{BoundedProbeIR}
\begin{algorithm2e}[h!] 
	\SetAlgoLined
	\SetAlgoNoEnd
	\caption{Bounded IR probing in a color class $C_{probe}$ up to a path of length $L$.}\label{alg:flatir}
	\Fn{\flatir{G, $\pi$, $C_{probe}$, $L$}}{
		\SetKwInOut{Input}{Input}
		\SetKwInOut{Output}{Output}
		\Input{graph $G = (V, E, \pi)$ where $\pi$ is equitable, color class $C_{probe}$ of $\pi$, length bound $L$}
		\Output{(equitable) coloring $\pi'$, set of automorphisms $\Phi$}
    $\Phi := \{\}$ \tcp*{set of automorphisms}
    Pick vertices~$v_1,v_2\in \pi^{-1}(C_{probe})$\;
    \For{$i\in \{1,2\}$}
    {
    $\pi_i$ := $\pi$\;    
    individualize $v_i$ in $\pi_i$\;
    \Refine{G, $\pi_i$}\; 
    }
    $L_C$ := $[C_{probe}]$ \tcp*{list of color classes}
    \While{$|L_C| < L$}{
      \If{$\varphi_{\pi_1, \pi_2}(G, \pi) = (G, \pi)$}{\Break{}\tcp*{automorphism found}}
      $C$ := non-trivial color class of $\pi_1$\;
      $L_C$ += $[C]$ \tcp*{append $C$ to $L_C$}
      individualize some $v \in \pi_1^{-1}(C)$ in $\pi_1$\;
      \Refine{G, $\pi_1$}\;
      individualize some $v \in \pi_2^{-1}(C)$ in $\pi_2$\;
      \Refine{G, $\pi_2$}\; 
    }

    \eIf{$\varphi_{\pi_1, \pi_2}(G, \pi) \neq (G, \pi)$}{\Return{$\pi$, $\emptyset$\tcp*{probing failed}}}{
      $\Phi$ := $\Phi \cup \{\varphi\}$\;
    }

    \For{$w \in C_{probe} \smallsetminus \{v_1, v_2\}$}{
      reset $\pi_2$ to $\pi$ \tcp*{essentially $\pi_2 := \pi$}
      individualize $w$ in $\pi_2$\;
      \Refine{G, $\pi_2$}\; 
      \For{$C \in L_C$}{
      individualize some $v \in \pi_2^{-1}(C)$ in $\pi_2$\;
      \Refine{G, $\pi_2$}\;
    }
    \eIf{$\varphi_{\pi_1, \pi_2}(G, \pi) \neq (G, \pi)$}{\Return{$\pi$, $\emptyset$}\tcp*{probing failed}}{
      $\Phi$ := $\Phi \cup \{\varphi\}$\;
    }
    } 

    individualize $v_1$ in $\pi$ \tcp*{success; individualize $v_1$ in~$(G,\pi)$}
		\Return{\Refine{G, $\pi$}, $\Phi$}\; 
	}
\end{algorithm2e}

We now describe the bounded IR probing algorithm. Philosophically, the algorithm is a blend of random path probing as used by \Traces{} and \dejavu{} with the sparse automorphism detection used by \saucy{}.

\emph{(Description of Algorithm~\ref{alg:flatir}.)} (See Algorithm~\ref{alg:flatir} for the pseudocode.)
The algorithm expects as input a colored graph~$G = (V, E, \pi)$, a color class $C_{probe} = \pi^{-1}(c)$ as well as a length bound $L$.
It outputs a set of automorphisms~$\Phi$ and a coloring~$\pi'$ refining~$\pi$. If the probing was unsuccessful then~$\Phi=\{\}$ and~$\pi'=\pi$. Otherwise~$\langle\Phi \rangle$ acts transitively on~$C_{probe}$ and~$\pi'$  is obtained from~$\pi$ by individualizing a vertex and refining.

We compute arbitrary IR paths (i.e., a rooted path in the IR tree) starting with an individualization of a vertex in $C_{probe}$. 
The path is only computed up to a length of $L$.

Initially, the algorithm examines two of these paths concurrently, starting in two different vertices~$v_1,v_2\in C_{probe}$. It checks after each individualization whether the automorphism~$\varphi_{\pi_1, \pi_2}$ (defined, as above, mapping corresponding singletons) is an automorphism. If this happens to be the case after having performed, say, $L'$ individualizations, we bound all subsequent paths by $L'$.

Afterwards for each vertex~$w\in C_{probe}\setminus \{v_1,v_2\}$ we compute an IR path starting with the individualization of~$w$. We hope to find an automorphism mapping~$v_1$ to~$w$. If we discover an automorphism for each~$w$, we return the set of automorphisms~$\Phi$, individualize~$v_1$ in~$\pi$, refine to obtain~$\pi'$ and return~$\Phi$ and~$\pi'$.

\emph{(Correctness of Algorithm~\ref{alg:flatir}.)} 
Correctness of the algorithm follows simply from the fact that we certify all automorphisms. That is, every map claimed to be an automorphism is indeed an automorphism. 
Since this certification is done for each automorphism, this certifies the fact that $C_{probe}$ is an orbit of $\Aut(G, \pi)$. 
Since we return all automorphisms required to construct the orbit (i.e., we return~$\Phi$), we have $\langle \Phi \cup \Aut(G, \pi') \rangle = \Aut(G, \pi)$ by the orbit-stabilizer theorem (see~\cite{handbook}).

\emph{(Implementation of Algorithm~\ref{alg:flatir}.)}
We want to make some further remarks on the implementation of the algorithm. In fact, even though it can be implemented very efficiently, it generally has to be used sparingly.
Overall we need to decide when and how often to employ the probing strategy and also which depth bound~$L$ to use.
We actually have three restricted probing strategies described next, each of which has dedicated code.

\subsection{High-level Strategies for Probing}
We describe how IR probing is applied in our preprocessor.

\textbf{$1$-IR probing, arbitrary class size.}
The first variant probes with length limit $L=1$ each color class of the graph. 
In order to make this more efficient, the implementation performs an additional orbit calculation: if we already know $v_1$ and $w$ are in the same orbit with respect to automorphisms found so far, we do not have to compute a path for $w$. The orbit calculations are performed using a variant of union-find.

\textbf{$\infty$-IR probing, class size $2$.}
The second variant is probing without a length limit for a \emph{single} color class of size $2$.
But if probing already succeeds in $L=1$, we set $\pi = \pi_1$ and pick a new class of size $2$ to continue.

For this particular case, the algorithm can indeed use fewer auxiliary data structures and there is no need for additional orbit calculations described above.

\textbf{$\infty$-IR probing, class size $B$.}
The last variant probes one color class bounded by some size $B$ with no length limit.
Essentially, we do the same as in the case of class size $2$. However, here more auxiliary structures as well as the orbit calculations are used.
Thus, the resulting procedure is significantly more expensive and thus less preferred in the high-level strategy described further below. 
In the implementation we use $B = 8$.

\section{Exploiting the Quotient Graph} \label{sec:quotient}
We now introduce another set of techniques which make use of the quotient graph~$Q(G, \pi)$.

\subsection{Edge Flip and Removal of Trivial Components} \label{subsec:edge_flip}
First, we describe how to efficiently \emph{flip edges} between color classes.
Let $C_1, C_2$ be two distinct color classes of $\pi$.
Assume they are connected by $m$ edges. 
The maximum number of edges between $C_1$ and $C_2$ is $|C_1||C_2|$.
If $|C_1||C_2| - m < m$, we can flip every edge to a non-edge, and every non-edge to an edge, reducing the total number of edges in the graph. 
Since this operation is isomorphism-invariant and reversible, the automorphism group of the graph does not change.

When applying edge flips repeatedly and exhaustively, vertices in color classes of size $1$ (singleton vertices) will become vertices of degree $0$. In fact, instead of performing edge flips in which singletons are involved, we can remove singletons directly without changing the automorphism group.

We want to remark that in the implementation, we use one canonical representation mapping to keep track of all removed vertices. This also includes removed singletons.
Hence, we use string representations throughout all the techniques described in the paper.
In addition to acting as a global canonical representation mapping, we also allow a renaming of vertices, which enables us to map all remaining vertices into the interval $\{1,2,\ldots,n\}$, whenever $n$ vertices remain.

\subsection{Connected Components}
A strategy more general than removing singletons is to exploit connected components of the quotient graph. 

Consider the quotient graph~$Q=Q(G, \pi)$ of a graph~$G$ with respect to a vertex coloring~$\pi$. The (weakly) connected components of~$Q$ partition the vertex set of $G$ into parts that are homogeneously connected. This allows us to treat components independently:
\begin{lemma}
If~$D_1,\ldots, D_t$ are the connected components of the quotient graph $Q(G, \pi)$, then we have~$\Aut(G,\pi)= \Pi_{i=1}^t \Aut((G,\pi)[D_i])$.
\end{lemma}
By flipping edges between two color classes we can only ever shrink the components of~$Q(G, \pi)$. It is therefore beneficial to first exhaustively flip edges and then consider connected components (see also~\cite{DBLP:journals/tocl/KieferSS22}).

These types of components have previously been employed for isomorphism and automorphism testing~\cite{DBLP:journals/dam/Goldberg83,DBLP:conf/tapas/JunttilaK11}. (In these contexts flips are not employed but rather edges in the quotient graph are characterized by non-homogeneous connections, which is equivalent.)

Regarding the implementation, we compute the connected components of the quotient graph without explicitly computing the quotient graph.
We first perform edge flips for all fully connected color classes, i.e., whenever the number of edges between $C_1, C_2$ equals $|C_1||C_2|$.
Then, we modify a basic algorithm for computing connected components as follows: usually, the algorithm determines for a vertex $v$ its neighborhood $N(v)$ and adds this neighborhood to the connected component of $v$.
Our modification simply also adds $\pi^{-1}\pi(v)$ in addition to $N(v)$ (i.e., it adds entire color classes). 
In turn, the algorithm gives us a partition of the vertices into the components of the quotient graph.

We use this to perform the variants of Algorithm~\ref{alg:flatir} for each component of the quotient graph separately.

We want to mention that after preprocessing is done, we could, theoretically, also use the components of the quotient graph to make independent calls to the main solver on the subgraphs induced by the components. These would, in turn, be smaller, and their handling could be parallelized. However, in our testing, after preprocessing is done, usually only one component is left, or there is one very large component and several smaller ones. We thus, at least so far, did not find it beneficial to use independent solver calls.

\section{Scheduling of Techniques}
We now describe when and how the preprocessor combines the techniques described in the previous sections.

The first step of the preprocessor is to apply color refinement to produce an equitable coloring.
The coloring remains equitable throughout the entire algorithm, by reapplying color refinement whenever necessary (i.e., for the probing techniques).
Beyond this, our implementation allows the user to freely specify a schedule for the various techniques.

The schedule used to produce the benchmarks is as follows.
We first remove singletons.
Then, we remove vertices of degree $0$ and $1$, and apply the heuristics described for vertices of degree $2$.
Next, we flip edges and apply the different variants probing for sparse automorphisms while making use of quotient graph components. 
This is followed again by the removal of singletons.
Lastly, we repeat the schedule as long as the graph still contains vertices of degree $0$ or $1$ and the number of vertices of the graph shrunk by at least $25$\%.
Note that this ensures that the schedule is only repeated at most a logarithmic number of times in the original graph size.

The implementation is called \textsc{sassy}.
It is implemented in C++ and uses the color refinement of \dejavu{} (which is itself an amalgam of color refinement implementations in \Traces{} and \saucy{}).
The implementation is open source and freely available at \cite{sassygithub}.

\section{Benchmarks}\label{sec:benchmarks}
We split the benchmark section into three main parts: first, we check whether applying the preprocessor speeds up state-of-the-art solvers on graph classes where the preprocessing techniques are supposedly effective. At the same time we check whether we introduced excessive overhead on graphs where the techniques are not effective. 
Secondly, we compare the performance of solver configurations using the preprocessor to state-of-the-art \saucy{} and \Traces{} on a wide range of practical data sets.
In the last part of the benchmarks, which can be found in Appendix~\ref{sec:ablation}, we analyze the separate impact of each of the different techniques used in the preprocessor on some of the graphs where the techniques collectively had an impact.

In the following, whenever we apply the preprocessor followed by an execution of a main \textsc{solver}, we write \textsc{sy+solver} for the combined configuration.
All benchmarks were run on the same machine featuring an Intel Core i7 9700K, 64GB of RAM on Ubuntu 20.04. We used \textsc{nauty/Traces} 2.6, \textsc{saucy} 3.0, \textsc{bliss} 0.73 and \textsc{dejavu} 1.2 ($1$\% error bound and $4$ threads for~\textsc{dejavu}). 
We ran all benchmarks $3$ consecutive times. 
We report the average and standard deviation.

Traditionally, the default way to test symmetry detection solvers is to first randomly permute all given benchmark graphs \cite{DBLP:journals/jsc/McKayP14,DBLP:conf/esa/AndersS21}. 
However, we feel that for many of the practical graphs, it is not clear whether this is the right way to test the tools: the initial order is often not arbitrary and may indeed encode information.
Furthermore, on practical graphs, solvers usually run in polynomial-time: aspects such as cache-efficiency and efficient handling of easy cases become more important than, for example, pruning techniques.
Therefore, benchmarks on non-permuted instances might be more useful to practitioners.  

In any case, we ran all benchmarks both ways: firstly, in the traditional manner of randomly permuting the instances. Secondly, using the unaltered initial order of the instances.
We denote sets that were randomly permuted with \textbf{(p)}.

\subsection{Preprocessed versus Unprocessed}
We prepared two collections of instances to test the impact of applying the preprocessor for each solver individually.

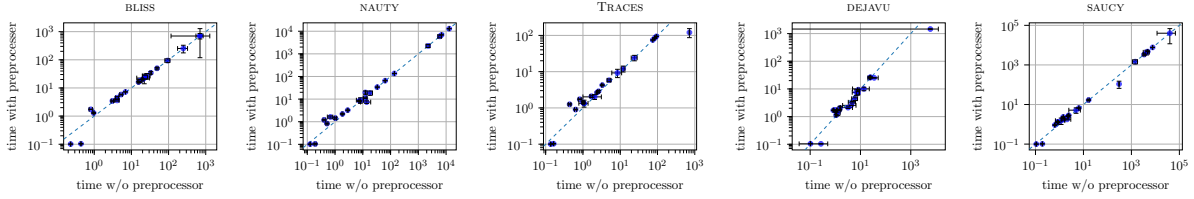
\begin{figure*}[t!]
  \centering
  \scalebox{0.45}{ 
% This file was created with tikzplotlib v0.9.17.
\begin{tikzpicture}

\definecolor{color0}{rgb}{0.12156862745098,0.466666666666667,0.705882352941177}

\begin{axis}[
log basis x={10},
log basis y={10},
tick align=outside,
tick pos=left,
title={\textsc{bliss}},
x grid style={white!69.0196078431373!black},
xlabel={time w/o preprocessor},
xmajorgrids,
xmin=0.150507143291502, xmax=1978.7310243054,
xmode=log,
xtick style={color=black},
y grid style={white!69.0196078431373!black},
ylabel={time with preprocesser},
ymajorgrids,
ymin=0.0628045438171943, ymax=2055.32700796037,
ymode=log,
scale=0.66,
ytick style={color=black}
]
\addplot [draw=blue, fill=blue, mark=*, only marks, opacity=0.7]
table{%
x  y
4.12329333333333 4.50414
6.92227 7.22009333333333
15.5788666666667 16.1007
33.2673666666667 34.5472666666667
0.811611 1.73631666666667
0.965889333333333 1.29618333333333
23.5590333333333 24.5406
94.2520333333333 93.9969333333333
5.28763333333333 5.83772333333333
25.7940333333333 26.5104666666667
701.2 700.105666666667
49.0621666666667 49.9085333333333
22.2417333333333 22.4894666666667
16.9769 17.8759
4.12532666666667 3.61647333333333
0.440568 0.103946
0.232507333333333 0.101429666666667
249.376666666667 252.427333333333
697.992 704.995
3.08457666666667 3.45211
19.9876333333333 20.0791333333333
};
\path [draw=black]
(axis cs:3.71897407982745,4.50414)
--(axis cs:4.52761258683922,4.50414);

\path [draw=black]
(axis cs:6.9152265018161,7.22009333333333)
--(axis cs:6.9293134981839,7.22009333333333);

\path [draw=black]
(axis cs:15.0707360644699,16.1007)
--(axis cs:16.0869972688635,16.1007);

\path [draw=black]
(axis cs:31.8407551467045,34.5472666666667)
--(axis cs:34.6939781866289,34.5472666666667);

\path [draw=black]
(axis cs:0.730512875923874,1.73631666666667)
--(axis cs:0.892709124076126,1.73631666666667);

\path [draw=black]
(axis cs:0.937884508625993,1.29618333333333)
--(axis cs:0.993894158040674,1.29618333333333);

\path [draw=black]
(axis cs:19.928376969962,24.5406)
--(axis cs:27.1896896967047,24.5406);

\path [draw=black]
(axis cs:79.5602855248223,93.9969333333333)
--(axis cs:108.943781141844,93.9969333333333);

\path [draw=black]
(axis cs:4.85930352922336,5.83772333333333)
--(axis cs:5.71596313744331,5.83772333333333);

\path [draw=black]
(axis cs:19.7143315444113,26.5104666666667)
--(axis cs:31.8737351222554,26.5104666666667);

\path [draw=black]
(axis cs:116.618320188986,700.105666666667)
--(axis cs:1285.78167981101,700.105666666667);

\path [draw=black]
(axis cs:46.2635353460228,49.9085333333333)
--(axis cs:51.8607979873105,49.9085333333333);

\path [draw=black]
(axis cs:13.9323899792797,22.4894666666667)
--(axis cs:30.551076687387,22.4894666666667);

\path [draw=black]
(axis cs:14.8471677773955,17.8759)
--(axis cs:19.1066322226045,17.8759);

\path [draw=black]
(axis cs:3.58258896412929,3.61647333333333)
--(axis cs:4.66806436920405,3.61647333333333);

\path [draw=black]
(axis cs:0.435804847822432,0.103946)
--(axis cs:0.445331152177568,0.103946);

\path [draw=black]
(axis cs:0.231620311975705,0.101429666666667)
--(axis cs:0.233394354690962,0.101429666666667);

\path [draw=black]
(axis cs:174.757518121836,252.427333333333)
--(axis cs:323.995815211498,252.427333333333);

\path [draw=black]
(axis cs:545.142808651141,704.995)
--(axis cs:850.841191348859,704.995);

\path [draw=black]
(axis cs:2.86104803012192,3.45211)
--(axis cs:3.30810530321142,3.45211);

\path [draw=black]
(axis cs:19.1747527667336,20.0791333333333)
--(axis cs:20.8005138999331,20.0791333333333);

\path [draw=black]
(axis cs:4.12329333333333,4.08952575969141)
--(axis cs:4.12329333333333,4.91875424030859);

\path [draw=black]
(axis cs:6.92227,7.15951662420442)
--(axis cs:6.92227,7.28067004246225);

\path [draw=black]
(axis cs:15.5788666666667,15.243164177619)
--(axis cs:15.5788666666667,16.958235822381);

\path [draw=black]
(axis cs:33.2673666666667,31.1289621499671)
--(axis cs:33.2673666666667,37.9655711833663);

\path [draw=black]
(axis cs:0.811611,1.71112512002454)
--(axis cs:0.811611,1.76150821330879);

\path [draw=black]
(axis cs:0.965889333333333,1.24206723447583)
--(axis cs:0.965889333333333,1.35029943219084);

\path [draw=black]
(axis cs:23.5590333333333,20.8186854738455)
--(axis cs:23.5590333333333,28.2625145261545);

\path [draw=black]
(axis cs:94.2520333333333,79.6162268969691)
--(axis cs:94.2520333333333,108.377639769698);

\path [draw=black]
(axis cs:5.28763333333333,5.69656597614325)
--(axis cs:5.28763333333333,5.97888069052342);

\path [draw=black]
(axis cs:25.7940333333333,20.5036412952894)
--(axis cs:25.7940333333333,32.517292038044);

\path [draw=black]
(axis cs:701.2,118.888213278618)
--(axis cs:701.2,1281.32312005472);

\path [draw=black]
(axis cs:49.0621666666667,46.9883080402742)
--(axis cs:49.0621666666667,52.8287586263924);

\path [draw=black]
(axis cs:22.2417333333333,14.0948610764346)
--(axis cs:22.2417333333333,30.8840722568987);

\path [draw=black]
(axis cs:16.9769,15.8376683285096)
--(axis cs:16.9769,19.9141316714904);

\path [draw=black]
(axis cs:4.12532666666667,3.39491945586281)
--(axis cs:4.12532666666667,3.83802721080385);

\path [draw=black]
(axis cs:0.440568,0.102009017639041)
--(axis cs:0.440568,0.105882982360959);

\path [draw=black]
(axis cs:0.232507333333333,0.10074264103234)
--(axis cs:0.232507333333333,0.102116692300994);

\path [draw=black]
(axis cs:249.376666666667,176.144552308459)
--(axis cs:249.376666666667,328.710114358207);

\path [draw=black]
(axis cs:697.992,548.562633700695)
--(axis cs:697.992,861.427366299305);

\path [draw=black]
(axis cs:3.08457666666667,3.23569097542037)
--(axis cs:3.08457666666667,3.66852902457963);

\path [draw=black]
(axis cs:19.9876333333333,19.2507039988562)
--(axis cs:19.9876333333333,20.9075626678105);

\addplot [semithick, color0, dashed]
table {%
0.150507143291502 0.150507143291502
1978.7310243054 1978.7310243054
};
\addplot [black, mark=|, mark size=2, mark options={solid}, only marks]
table {%
3.71897407982745 4.50414
6.9152265018161 7.22009333333333
15.0707360644699 16.1007
31.8407551467045 34.5472666666667
0.730512875923874 1.73631666666667
0.937884508625993 1.29618333333333
19.928376969962 24.5406
79.5602855248223 93.9969333333333
4.85930352922336 5.83772333333333
19.7143315444113 26.5104666666667
116.618320188986 700.105666666667
46.2635353460228 49.9085333333333
13.9323899792797 22.4894666666667
14.8471677773955 17.8759
3.58258896412929 3.61647333333333
0.435804847822432 0.103946
0.231620311975705 0.101429666666667
174.757518121836 252.427333333333
545.142808651141 704.995
2.86104803012192 3.45211
19.1747527667336 20.0791333333333
};
\addplot [black, mark=|, mark size=2, mark options={solid}, only marks]
table {%
4.52761258683922 4.50414
6.9293134981839 7.22009333333333
16.0869972688635 16.1007
34.6939781866289 34.5472666666667
0.892709124076126 1.73631666666667
0.993894158040674 1.29618333333333
27.1896896967047 24.5406
108.943781141844 93.9969333333333
5.71596313744331 5.83772333333333
31.8737351222554 26.5104666666667
1285.78167981101 700.105666666667
51.8607979873105 49.9085333333333
30.551076687387 22.4894666666667
19.1066322226045 17.8759
4.66806436920405 3.61647333333333
0.445331152177568 0.103946
0.233394354690962 0.101429666666667
323.995815211498 252.427333333333
850.841191348859 704.995
3.30810530321142 3.45211
20.8005138999331 20.0791333333333
};
\addplot [black, mark=-, mark size=2, mark options={solid}, only marks]
table {%
4.12329333333333 4.08952575969141
6.92227 7.15951662420442
15.5788666666667 15.243164177619
33.2673666666667 31.1289621499671
0.811611 1.71112512002454
0.965889333333333 1.24206723447583
23.5590333333333 20.8186854738455
94.2520333333333 79.6162268969691
5.28763333333333 5.69656597614325
25.7940333333333 20.5036412952894
701.2 118.888213278618
49.0621666666667 46.9883080402742
22.2417333333333 14.0948610764346
16.9769 15.8376683285096
4.12532666666667 3.39491945586281
0.440568 0.102009017639041
0.232507333333333 0.10074264103234
249.376666666667 176.144552308459
697.992 548.562633700695
3.08457666666667 3.23569097542037
19.9876333333333 19.2507039988562
};
\addplot [black, mark=-, mark size=2, mark options={solid}, only marks]
table {%
4.12329333333333 4.91875424030859
6.92227 7.28067004246225
15.5788666666667 16.958235822381
33.2673666666667 37.9655711833663
0.811611 1.76150821330879
0.965889333333333 1.35029943219084
23.5590333333333 28.2625145261545
94.2520333333333 108.377639769698
5.28763333333333 5.97888069052342
25.7940333333333 32.517292038044
701.2 1281.32312005472
49.0621666666667 52.8287586263924
22.2417333333333 30.8840722568987
16.9769 19.9141316714904
4.12532666666667 3.83802721080385
0.440568 0.105882982360959
0.232507333333333 0.102116692300994
249.376666666667 328.710114358207
697.992 861.427366299305
3.08457666666667 3.66852902457963
19.9876333333333 20.9075626678105
};
\end{axis}

\end{tikzpicture}
  }
  \scalebox{0.45}{
% This file was created with tikzplotlib v0.9.17.
\begin{tikzpicture}

\definecolor{color0}{rgb}{0.12156862745098,0.466666666666667,0.705882352941177}

\begin{axis}[
log basis x={10},
log basis y={10},
tick align=outside,
tick pos=left,
title={\textsc{nauty}},
x grid style={white!69.0196078431373!black},
xlabel={time w/o preprocessor},
xmajorgrids,
xmin=0.0708851887768965, xmax=23183.4512984837,
xmode=log,
xtick style={color=black},
y grid style={white!69.0196078431373!black},
ylabel={time with preprocesser},
ymajorgrids,
ymin=0.0565463594123936, ymax=23495.4905491076,
ymode=log,
scale=0.66,
ytick style={color=black}
]
\addplot [draw=blue, fill=blue, mark=*, only marks, opacity=0.7]
table{%
x  y
1.84479 2.20797
136.995 135.778333333333
33.2747666666667 33.8612
12.3724333333333 19.6121666666667
0.405421666666667 1.20330666666667
0.513352333333333 0.826742
63.7983666666667 65.1384
2253.61333333333 2255.15666666667
12.2780433333333 11.3511
8.17919666666667 8.98363666666667
6864.31666666667 6879.60333333333
7.58314 8.22139
18.2987 18.6892666666667
0.677781 1.62396666666667
13.7445633333333 7.39199
0.192605333333333 0.104301333333333
0.127081666666667 0.101957333333333
5921.42 5928.29333333333
12981.4 13010.4666666667
1.02572633333333 1.4381
2.85391333333333 3.23043333333333
};
\path [draw=black]
(axis cs:1.76494848072588,2.20797)
--(axis cs:1.92463151927412,2.20797);

\path [draw=black]
(axis cs:136.740063406576,135.778333333333)
--(axis cs:137.249936593424,135.778333333333);

\path [draw=black]
(axis cs:32.9088409881167,33.8612)
--(axis cs:33.6406923452166,33.8612);

\path [draw=black]
(axis cs:12.2452261445078,19.6121666666667)
--(axis cs:12.4996405221588,19.6121666666667);

\path [draw=black]
(axis cs:0.367861626008778,1.20330666666667)
--(axis cs:0.442981707324555,1.20330666666667);

\path [draw=black]
(axis cs:0.472894297614551,0.826742)
--(axis cs:0.553810369052115,0.826742);

\path [draw=black]
(axis cs:63.2933125419544,65.1384)
--(axis cs:64.303420791379,65.1384);

\path [draw=black]
(axis cs:1903.2527832012,2255.15666666667)
--(axis cs:2603.97388346546,2255.15666666667);

\path [draw=black]
(axis cs:10.4251199534399,11.3511)
--(axis cs:14.1309667132267,11.3511);

\path [draw=black]
(axis cs:5.86497509513218,8.98363666666667)
--(axis cs:10.4934182382012,8.98363666666667);

\path [draw=black]
(axis cs:6850.28779089006,6879.60333333333)
--(axis cs:6878.34554244328,6879.60333333333);

\path [draw=black]
(axis cs:7.22965429722453,8.22139)
--(axis cs:7.93662570277547,8.22139);

\path [draw=black]
(axis cs:14.2534059183952,18.6892666666667)
--(axis cs:22.3439940816048,18.6892666666667);

\path [draw=black]
(axis cs:0.527473567839555,1.62396666666667)
--(axis cs:0.828088432160445,1.62396666666667);

\path [draw=black]
(axis cs:8.24477134296174,7.39199)
--(axis cs:19.2443553237049,7.39199);

\path [draw=black]
(axis cs:0.176916634208733,0.104301333333333)
--(axis cs:0.208294032457933,0.104301333333333);

\path [draw=black]
(axis cs:0.126246737394878,0.101957333333333)
--(axis cs:0.127916595938455,0.101957333333333);

\path [draw=black]
(axis cs:5071.56874991757,5928.29333333333)
--(axis cs:6771.27125008243,5928.29333333333);

\path [draw=black]
(axis cs:12945.7244247886,13010.4666666667)
--(axis cs:13017.0755752114,13010.4666666667);

\path [draw=black]
(axis cs:0.93664835295534,1.4381)
--(axis cs:1.11480431371133,1.4381);

\path [draw=black]
(axis cs:2.74842455688979,3.23043333333333)
--(axis cs:2.95940210977688,3.23043333333333);

\path [draw=black]
(axis cs:1.84479,2.12780190867517)
--(axis cs:1.84479,2.28813809132483);

\path [draw=black]
(axis cs:136.995,135.70729656664)
--(axis cs:136.995,135.849370100027);

\path [draw=black]
(axis cs:33.2747666666667,33.7095968118189)
--(axis cs:33.2747666666667,34.0128031881811);

\path [draw=black]
(axis cs:12.3724333333333,16.45268266625)
--(axis cs:12.3724333333333,22.7716506670833);

\path [draw=black]
(axis cs:0.405421666666667,1.17510361206457)
--(axis cs:0.405421666666667,1.23150972126876);

\path [draw=black]
(axis cs:0.513352333333333,0.789478260091075)
--(axis cs:0.513352333333333,0.864005739908925);

\path [draw=black]
(axis cs:63.7983666666667,64.368465634832)
--(axis cs:63.7983666666667,65.908334365168);

\path [draw=black]
(axis cs:2253.61333333333,1907.15078469211)
--(axis cs:2253.61333333333,2603.16254864122);

\path [draw=black]
(axis cs:12.2780433333333,10.5571747138427)
--(axis cs:12.2780433333333,12.1450252861573);

\path [draw=black]
(axis cs:8.17919666666667,6.56061596249961)
--(axis cs:8.17919666666667,11.4066573708337);

\path [draw=black]
(axis cs:6864.31666666667,6863.61004237882)
--(axis cs:6864.31666666667,6895.59662428784);

\path [draw=black]
(axis cs:7.58314,7.83749817435116)
--(axis cs:7.58314,8.60528182564884);

\path [draw=black]
(axis cs:18.2987,14.7110712027799)
--(axis cs:18.2987,22.6674621305534);

\path [draw=black]
(axis cs:0.677781,1.43910915499289)
--(axis cs:0.677781,1.80882417834045);

\path [draw=black]
(axis cs:13.7445633333333,6.43746637741821)
--(axis cs:13.7445633333333,8.34651362258179);

\path [draw=black]
(axis cs:0.192605333333333,0.102277318126979)
--(axis cs:0.192605333333333,0.106325348539688);

\path [draw=black]
(axis cs:0.127081666666667,0.101810985953597)
--(axis cs:0.127081666666667,0.102103680713069);

\path [draw=black]
(axis cs:5921.42,5069.62598966124)
--(axis cs:5921.42,6786.96067700543);

\path [draw=black]
(axis cs:12981.4,12971.4137680689)
--(axis cs:12981.4,13049.5195652645);

\path [draw=black]
(axis cs:1.02572633333333,1.29403254635415)
--(axis cs:1.02572633333333,1.58216745364585);

\path [draw=black]
(axis cs:2.85391333333333,3.17862286056104)
--(axis cs:2.85391333333333,3.28224380610563);

\addplot [semithick, color0, dashed]
table {%
0.0708851887768965 0.0708851887768965
23183.4512984837 23183.4512984837
};
\addplot [black, mark=|, mark size=2, mark options={solid}, only marks]
table {%
1.76494848072588 2.20797
136.740063406576 135.778333333333
32.9088409881167 33.8612
12.2452261445078 19.6121666666667
0.367861626008778 1.20330666666667
0.472894297614551 0.826742
63.2933125419544 65.1384
1903.2527832012 2255.15666666667
10.4251199534399 11.3511
5.86497509513218 8.98363666666667
6850.28779089006 6879.60333333333
7.22965429722453 8.22139
14.2534059183952 18.6892666666667
0.527473567839555 1.62396666666667
8.24477134296174 7.39199
0.176916634208733 0.104301333333333
0.126246737394878 0.101957333333333
5071.56874991757 5928.29333333333
12945.7244247886 13010.4666666667
0.93664835295534 1.4381
2.74842455688979 3.23043333333333
};
\addplot [black, mark=|, mark size=2, mark options={solid}, only marks]
table {%
1.92463151927412 2.20797
137.249936593424 135.778333333333
33.6406923452166 33.8612
12.4996405221588 19.6121666666667
0.442981707324555 1.20330666666667
0.553810369052115 0.826742
64.303420791379 65.1384
2603.97388346546 2255.15666666667
14.1309667132267 11.3511
10.4934182382012 8.98363666666667
6878.34554244328 6879.60333333333
7.93662570277547 8.22139
22.3439940816048 18.6892666666667
0.828088432160445 1.62396666666667
19.2443553237049 7.39199
0.208294032457933 0.104301333333333
0.127916595938455 0.101957333333333
6771.27125008243 5928.29333333333
13017.0755752114 13010.4666666667
1.11480431371133 1.4381
2.95940210977688 3.23043333333333
};
\addplot [black, mark=-, mark size=2, mark options={solid}, only marks]
table {%
1.84479 2.12780190867517
136.995 135.70729656664
33.2747666666667 33.7095968118189
12.3724333333333 16.45268266625
0.405421666666667 1.17510361206457
0.513352333333333 0.789478260091075
63.7983666666667 64.368465634832
2253.61333333333 1907.15078469211
12.2780433333333 10.5571747138427
8.17919666666667 6.56061596249961
6864.31666666667 6863.61004237882
7.58314 7.83749817435116
18.2987 14.7110712027799
0.677781 1.43910915499289
13.7445633333333 6.43746637741821
0.192605333333333 0.102277318126979
0.127081666666667 0.101810985953597
5921.42 5069.62598966124
12981.4 12971.4137680689
1.02572633333333 1.29403254635415
2.85391333333333 3.17862286056104
};
\addplot [black, mark=-, mark size=2, mark options={solid}, only marks]
table {%
1.84479 2.28813809132483
136.995 135.849370100027
33.2747666666667 34.0128031881811
12.3724333333333 22.7716506670833
0.405421666666667 1.23150972126876
0.513352333333333 0.864005739908925
63.7983666666667 65.908334365168
2253.61333333333 2603.16254864122
12.2780433333333 12.1450252861573
8.17919666666667 11.4066573708337
6864.31666666667 6895.59662428784
7.58314 8.60528182564884
18.2987 22.6674621305534
0.677781 1.80882417834045
13.7445633333333 8.34651362258179
0.192605333333333 0.106325348539688
0.127081666666667 0.102103680713069
5921.42 6786.96067700543
12981.4 13049.5195652645
1.02572633333333 1.58216745364585
2.85391333333333 3.28224380610563
};
\end{axis}

\end{tikzpicture}
  }
  \scalebox{0.45}{
% This file was created with tikzplotlib v0.9.17.
\begin{tikzpicture}

\definecolor{color0}{rgb}{0.12156862745098,0.466666666666667,0.705882352941177}

\begin{axis}[
log basis x={10},
log basis y={10},
tick align=outside,
tick pos=left,
title={\textsc{Traces}},
x grid style={white!69.0196078431373!black},
xlabel={time w/o preprocessor},
xmajorgrids,
xmin=0.0874703721270796, xmax=1119.22484152663,
xmode=log,
xtick style={color=black},
y grid style={white!69.0196078431373!black},
ylabel={time with preprocesser},
ymajorgrids,
ymin=0.0693353344646746, ymax=222.101430245931,
ymode=log,
scale=0.66,
ytick style={color=black}
]
\addplot [draw=blue, fill=blue, mark=*, only marks, opacity=0.7]
table{%
x  y
1.68951666666667 2.09299333333333
1.01951566666667 1.41394333333333
2.62057333333333 2.86683333333333
717.785333333333 120.2808
0.439288333333333 1.25686
0.635226333333333 0.900882
3.29895333333333 4.24939666666667
24.0630666666667 24.0464333333333
12.0094333333333 12.0178733333333
5.03127 5.79541333333333
74.7155 74.7064
8.41767 9.23802666666667
1.07356333333333 1.34178
0.822307333333333 1.73045666666667
2.05837 2.01811333333333
0.163833333333333 0.102852666666667
0.136595333333333 0.100468666666667
84.9377333333333 85.3263
93.4934333333333 93.9564666666667
1.03002833333333 1.36231
2.38997 2.65818333333333
};
\path [draw=black]
(axis cs:1.54925906310966,2.09299333333333)
--(axis cs:1.82977427022368,2.09299333333333);

\path [draw=black]
(axis cs:0.981903268743169,1.41394333333333)
--(axis cs:1.05712806459016,1.41394333333333);

\path [draw=black]
(axis cs:2.49550722539215,2.86683333333333)
--(axis cs:2.74563944127452,2.86683333333333);

\path [draw=black]
(axis cs:707.400367907831,120.2808)
--(axis cs:728.170298758836,120.2808);

\path [draw=black]
(axis cs:0.40544736529819,1.25686)
--(axis cs:0.473129301368476,1.25686);

\path [draw=black]
(axis cs:0.592038919076142,0.900882)
--(axis cs:0.678413747590524,0.900882);

\path [draw=black]
(axis cs:3.16477991809933,4.24939666666667)
--(axis cs:3.43312674856733,4.24939666666667);

\path [draw=black]
(axis cs:19.7135248330683,24.0464333333333)
--(axis cs:28.412608500265,24.0464333333333);

\path [draw=black]
(axis cs:10.7106809978633,12.0178733333333)
--(axis cs:13.3081856688034,12.0178733333333);

\path [draw=black]
(axis cs:4.45721975405748,5.79541333333333)
--(axis cs:5.60532024594252,5.79541333333333);

\path [draw=black]
(axis cs:74.1923932804867,74.7064)
--(axis cs:75.2386067195133,74.7064);

\path [draw=black]
(axis cs:5.93710360313281,9.23802666666667)
--(axis cs:10.8982363968672,9.23802666666667);

\path [draw=black]
(axis cs:0.776196965461048,1.34178)
--(axis cs:1.37092970120562,1.34178);

\path [draw=black]
(axis cs:0.750816258401641,1.73045666666667)
--(axis cs:0.893798408265026,1.73045666666667);

\path [draw=black]
(axis cs:0.999164686443338,2.01811333333333)
--(axis cs:3.11757531355666,2.01811333333333);

\path [draw=black]
(axis cs:0.162476124842022,0.102852666666667)
--(axis cs:0.165190541824645,0.102852666666667);

\path [draw=black]
(axis cs:0.13444521638561,0.100468666666667)
--(axis cs:0.138745450281057,0.100468666666667);

\path [draw=black]
(axis cs:83.8591619191212,85.3263)
--(axis cs:86.0163047475455,85.3263);

\path [draw=black]
(axis cs:92.0796798981278,93.9564666666667)
--(axis cs:94.9071867685389,93.9564666666667);

\path [draw=black]
(axis cs:0.988254962163545,1.36231)
--(axis cs:1.07180170450312,1.36231);

\path [draw=black]
(axis cs:2.31677593330057,2.65818333333333)
--(axis cs:2.46316406669943,2.65818333333333);

\path [draw=black]
(axis cs:1.68951666666667,1.94142905646632)
--(axis cs:1.68951666666667,2.24455761020035);

\path [draw=black]
(axis cs:1.01951566666667,1.33982054676342)
--(axis cs:1.01951566666667,1.48806611990325);

\path [draw=black]
(axis cs:2.62057333333333,2.71410855050851)
--(axis cs:2.62057333333333,3.01955811615816);

\path [draw=black]
(axis cs:717.785333333333,86.6731416711806)
--(axis cs:717.785333333333,153.888458328819);

\path [draw=black]
(axis cs:0.439288333333333,1.20871020041579)
--(axis cs:0.439288333333333,1.30500979958421);

\path [draw=black]
(axis cs:0.635226333333333,0.88383235400172)
--(axis cs:0.635226333333333,0.91793164599828);

\path [draw=black]
(axis cs:3.29895333333333,4.13268200196871)
--(axis cs:3.29895333333333,4.36611133136462);

\path [draw=black]
(axis cs:24.0630666666667,19.9140265008822)
--(axis cs:24.0630666666667,28.1788401657845);

\path [draw=black]
(axis cs:12.0094333333333,9.78224617268641)
--(axis cs:12.0094333333333,14.2535004939803);

\path [draw=black]
(axis cs:5.03127,5.28689226839218)
--(axis cs:5.03127,6.30393439827449);

\path [draw=black]
(axis cs:74.7155,73.7782426606083)
--(axis cs:74.7155,75.6345573393917);

\path [draw=black]
(axis cs:8.41767,6.86071472535798)
--(axis cs:8.41767,11.6153386079754);

\path [draw=black]
(axis cs:1.07356333333333,1.05048940557581)
--(axis cs:1.07356333333333,1.63307059442419);

\path [draw=black]
(axis cs:0.822307333333333,1.69273222601843)
--(axis cs:0.822307333333333,1.7681811073149);

\path [draw=black]
(axis cs:2.05837,1.71061759804222)
--(axis cs:2.05837,2.32560906862445);

\path [draw=black]
(axis cs:0.163833333333333,0.101437567062791)
--(axis cs:0.163833333333333,0.104267766270542);

\path [draw=black]
(axis cs:0.136595333333333,0.100069083272506)
--(axis cs:0.136595333333333,0.100868250060827);

\path [draw=black]
(axis cs:84.9377333333333,84.0214108833826)
--(axis cs:84.9377333333333,86.6311891166174);

\path [draw=black]
(axis cs:93.4934333333333,93.2299296644737)
--(axis cs:93.4934333333333,94.6830036688596);

\path [draw=black]
(axis cs:1.03002833333333,1.2969902597474)
--(axis cs:1.03002833333333,1.4276297402526);

\path [draw=black]
(axis cs:2.38997,2.55816712986832)
--(axis cs:2.38997,2.75819953679835);

\addplot [semithick, color0, dashed]
table {%
0.0874703721270796 0.0874703721270796
222.101430245931 222.101430245931
};
\addplot [black, mark=|, mark size=2, mark options={solid}, only marks]
table {%
1.54925906310966 2.09299333333333
0.981903268743169 1.41394333333333
2.49550722539215 2.86683333333333
707.400367907831 120.2808
0.40544736529819 1.25686
0.592038919076142 0.900882
3.16477991809933 4.24939666666667
19.7135248330683 24.0464333333333
10.7106809978633 12.0178733333333
4.45721975405748 5.79541333333333
74.1923932804867 74.7064
5.93710360313281 9.23802666666667
0.776196965461048 1.34178
0.750816258401641 1.73045666666667
0.999164686443338 2.01811333333333
0.162476124842022 0.102852666666667
0.13444521638561 0.100468666666667
83.8591619191212 85.3263
92.0796798981278 93.9564666666667
0.988254962163545 1.36231
2.31677593330057 2.65818333333333
};
\addplot [black, mark=|, mark size=2, mark options={solid}, only marks]
table {%
1.82977427022368 2.09299333333333
1.05712806459016 1.41394333333333
2.74563944127452 2.86683333333333
728.170298758836 120.2808
0.473129301368476 1.25686
0.678413747590524 0.900882
3.43312674856733 4.24939666666667
28.412608500265 24.0464333333333
13.3081856688034 12.0178733333333
5.60532024594252 5.79541333333333
75.2386067195133 74.7064
10.8982363968672 9.23802666666667
1.37092970120562 1.34178
0.893798408265026 1.73045666666667
3.11757531355666 2.01811333333333
0.165190541824645 0.102852666666667
0.138745450281057 0.100468666666667
86.0163047475455 85.3263
94.9071867685389 93.9564666666667
1.07180170450312 1.36231
2.46316406669943 2.65818333333333
};
\addplot [black, mark=-, mark size=2, mark options={solid}, only marks]
table {%
1.68951666666667 1.94142905646632
1.01951566666667 1.33982054676342
2.62057333333333 2.71410855050851
717.785333333333 86.6731416711806
0.439288333333333 1.20871020041579
0.635226333333333 0.88383235400172
3.29895333333333 4.13268200196871
24.0630666666667 19.9140265008822
12.0094333333333 9.78224617268641
5.03127 5.28689226839218
74.7155 73.7782426606083
8.41767 6.86071472535798
1.07356333333333 1.05048940557581
0.822307333333333 1.69273222601843
2.05837 1.71061759804222
0.163833333333333 0.101437567062791
0.136595333333333 0.100069083272506
84.9377333333333 84.0214108833826
93.4934333333333 93.2299296644737
1.03002833333333 1.2969902597474
2.38997 2.55816712986832
};
\addplot [black, mark=-, mark size=2, mark options={solid}, only marks]
table {%
1.68951666666667 2.24455761020035
1.01951566666667 1.48806611990325
2.62057333333333 3.01955811615816
717.785333333333 153.888458328819
0.439288333333333 1.30500979958421
0.635226333333333 0.91793164599828
3.29895333333333 4.36611133136462
24.0630666666667 28.1788401657845
12.0094333333333 14.2535004939803
5.03127 6.30393439827449
74.7155 75.6345573393917
8.41767 11.6153386079754
1.07356333333333 1.63307059442419
0.822307333333333 1.7681811073149
2.05837 2.32560906862445
0.163833333333333 0.104267766270542
0.136595333333333 0.100868250060827
84.9377333333333 86.6311891166174
93.4934333333333 94.6830036688596
1.03002833333333 1.4276297402526
2.38997 2.75819953679835
};
\end{axis}

\end{tikzpicture}
  }
  \scalebox{0.45}{
% This file was created with tikzplotlib v0.9.17.
\begin{tikzpicture}

\definecolor{color0}{rgb}{0.12156862745098,0.466666666666667,0.705882352941177}

\begin{axis}[
log basis x={10},
log basis y={10},
tick align=outside,
tick pos=left,
title={\textsc{dejavu}},
x grid style={white!69.0196078431373!black},
xlabel={time w/o preprocessor},
xmajorgrids,
xmin=0.0192960976056658, xmax=23570.7322565458,
xmode=log,
xtick style={color=black},
y grid style={white!69.0196078431373!black},
ylabel={time with preprocesser},
ymajorgrids,
ymin=0.062340526550539, ymax=2443.51541253778,
ymode=log,
scale=0.66,
ytick style={color=black}
]
\addplot [draw=blue, fill=blue, mark=*, only marks, opacity=0.7]
table{%
x  y
1.53489 2.05258
1.08553 1.1255
4.29854333333333 2.54555
13.6454133333333 10.0492266666667
0.834899333333333 1.68718
1.27229266666667 1.38504333333333
8.17293333333333 9.37468
7.95658333333333 7.07892666666667
5994.23666666667 1458.36333333333
6.81706666666667 7.52767666666667
24.3426333333333 25.4594666666667
4.81827 3.30319
6.05987333333333 4.42499
0.983014333333333 1.73871333333333
3.20145333333333 2.11205
0.102885666666667 0.104540666666667
0.268018 0.103705666666667
24.7583333333333 26.5050333333333
35.1390666666667 25.1113666666667
1.44395666666667 1.759
1.09907333333333 1.45867333333333
};
\path [draw=black]
(axis cs:1.43757700258102,2.05258)
--(axis cs:1.63220299741898,2.05258);

\path [draw=black]
(axis cs:1.04246663273733,1.1255)
--(axis cs:1.12859336726267,1.1255);

\path [draw=black]
(axis cs:1.95120272561953,2.54555)
--(axis cs:6.64588394104714,2.54555);

\path [draw=black]
(axis cs:4.84228979758258,10.0492266666667)
--(axis cs:22.4485368690841,10.0492266666667);

\path [draw=black]
(axis cs:0.798057439785061,1.68718)
--(axis cs:0.871741226881605,1.68718);

\path [draw=black]
(axis cs:0.994983773358941,1.38504333333333)
--(axis cs:1.54960155997439,1.38504333333333);

\path [draw=black]
(axis cs:7.59516145964769,9.37468)
--(axis cs:8.75070520701898,9.37468);

\path [draw=black]
(axis cs:6.85074161294313,7.07892666666667)
--(axis cs:9.06242505372353,7.07892666666667);

\path [draw=black]
(axis cs:0.001,1458.36333333333)
--(axis cs:12465.0963246633,1458.36333333333);
%-476.62299133001

\path [draw=black]
(axis cs:4.7095268061289,7.52767666666667)
--(axis cs:8.92460652720443,7.52767666666667);

\path [draw=black]
(axis cs:23.3176873044019,25.4594666666667)
--(axis cs:25.3675793622647,25.4594666666667);

\path [draw=black]
(axis cs:3.28537899269822,3.30319)
--(axis cs:6.35116100730178,3.30319);

\path [draw=black]
(axis cs:4.46190007554013,4.42499)
--(axis cs:7.65784659112654,4.42499);

\path [draw=black]
(axis cs:0.856437084471006,1.73871333333333)
--(axis cs:1.10959158219566,1.73871333333333);

\path [draw=black]
(axis cs:1.60343604547663,2.11205)
--(axis cs:4.79947062119003,2.11205);

\path [draw=black]
(axis cs:0.102484009820314,0.104540666666667)
--(axis cs:0.10328732351302,0.104540666666667);

\path [draw=black]
(axis cs:0.0364877365094576,0.103705666666667)
--(axis cs:0.499548263490542,0.103705666666667);

\path [draw=black]
(axis cs:23.9387220175934,26.5050333333333)
--(axis cs:25.5779446490732,26.5050333333333);

\path [draw=black]
(axis cs:19.5985997870792,25.1113666666667)
--(axis cs:50.6795335462542,25.1113666666667);

\path [draw=black]
(axis cs:1.20802005691264,1.759)
--(axis cs:1.67989327642069,1.759);

\path [draw=black]
(axis cs:1.06883966960857,1.45867333333333)
--(axis cs:1.1293069970581,1.45867333333333);

\path [draw=black]
(axis cs:1.53489,1.75461587922928)
--(axis cs:1.53489,2.35054412077072);

\path [draw=black]
(axis cs:1.08553,1.05097729294593)
--(axis cs:1.08553,1.20002270705407);

\path [draw=black]
(axis cs:4.29854333333333,2.51073672571178)
--(axis cs:4.29854333333333,2.58036327428822);

\path [draw=black]
(axis cs:13.6454133333333,8.62741516482982)
--(axis cs:13.6454133333333,11.4710381685035);

\path [draw=black]
(axis cs:0.834899333333333,1.6431356011885)
--(axis cs:0.834899333333333,1.7312243988115);

\path [draw=black]
(axis cs:1.27229266666667,1.10256931726184)
--(axis cs:1.27229266666667,1.66751734940482);

\path [draw=black]
(axis cs:8.17293333333333,9.19784269548914)
--(axis cs:8.17293333333333,9.55151730451086);

\path [draw=black]
(axis cs:7.95658333333333,6.12761006446621)
--(axis cs:7.95658333333333,8.03024326886712);

\path [draw=black]
(axis cs:5994.23666666667,1405.84226877613)
--(axis cs:5994.23666666667,1510.88439789054);

\path [draw=black]
(axis cs:6.81706666666667,5.7224804997394)
--(axis cs:6.81706666666667,9.33287283359393);

\path [draw=black]
(axis cs:24.3426333333333,24.4671815788237)
--(axis cs:24.3426333333333,26.4517517545097);

\path [draw=black]
(axis cs:4.81827,3.09758778551776)
--(axis cs:4.81827,3.50879221448224);

\path [draw=black]
(axis cs:6.05987333333333,3.81654495591905)
--(axis cs:6.05987333333333,5.03343504408095);

\path [draw=black]
(axis cs:0.983014333333333,1.70860477382685)
--(axis cs:0.983014333333333,1.76882189283982);

\path [draw=black]
(axis cs:3.20145333333333,1.96177407223156)
--(axis cs:3.20145333333333,2.26232592776844);

\path [draw=black]
(axis cs:0.102885666666667,0.102386009643173)
--(axis cs:0.102885666666667,0.10669532369016);

\path [draw=black]
(axis cs:0.268018,0.100821768802856)
--(axis cs:0.268018,0.106589564530477);

\path [draw=black]
(axis cs:24.7583333333333,25.3921531471098)
--(axis cs:24.7583333333333,27.6179135195569);

\path [draw=black]
(axis cs:35.1390666666667,24.6259853228875)
--(axis cs:35.1390666666667,25.5967480104459);

\path [draw=black]
(axis cs:1.44395666666667,1.73822199720859)
--(axis cs:1.44395666666667,1.77977800279141);

\path [draw=black]
(axis cs:1.09907333333333,1.43966517835828)
--(axis cs:1.09907333333333,1.47768148830838);

\addplot [semithick, color0, dashed]
table {%
0.062340526550539 0.062340526550539
2443.51541253778 2443.51541253778
};
\addplot [black, mark=|, mark size=2, mark options={solid}, only marks]
table {%
1.43757700258102 2.05258
1.04246663273733 1.1255
1.95120272561953 2.54555
4.84228979758258 10.0492266666667
0.798057439785061 1.68718
0.994983773358941 1.38504333333333
7.59516145964769 9.37468
6.85074161294313 7.07892666666667
-476.62299133001 1458.36333333333
4.7095268061289 7.52767666666667
23.3176873044019 25.4594666666667
3.28537899269822 3.30319
4.46190007554013 4.42499
0.856437084471006 1.73871333333333
1.60343604547663 2.11205
0.102484009820314 0.104540666666667
0.0364877365094576 0.103705666666667
23.9387220175934 26.5050333333333
19.5985997870792 25.1113666666667
1.20802005691264 1.759
1.06883966960857 1.45867333333333
};
\addplot [black, mark=|, mark size=2, mark options={solid}, only marks]
table {%
1.63220299741898 2.05258
1.12859336726267 1.1255
6.64588394104714 2.54555
22.4485368690841 10.0492266666667
0.871741226881605 1.68718
1.54960155997439 1.38504333333333
8.75070520701898 9.37468
9.06242505372353 7.07892666666667
12465.0963246633 1458.36333333333
8.92460652720443 7.52767666666667
25.3675793622647 25.4594666666667
6.35116100730178 3.30319
7.65784659112654 4.42499
1.10959158219566 1.73871333333333
4.79947062119003 2.11205
0.10328732351302 0.104540666666667
0.499548263490542 0.103705666666667
25.5779446490732 26.5050333333333
50.6795335462542 25.1113666666667
1.67989327642069 1.759
1.1293069970581 1.45867333333333
};
\addplot [black, mark=-, mark size=2, mark options={solid}, only marks]
table {%
1.53489 1.75461587922928
1.08553 1.05097729294593
4.29854333333333 2.51073672571178
13.6454133333333 8.62741516482982
0.834899333333333 1.6431356011885
1.27229266666667 1.10256931726184
8.17293333333333 9.19784269548914
7.95658333333333 6.12761006446621
5994.23666666667 1405.84226877613
6.81706666666667 5.7224804997394
24.3426333333333 24.4671815788237
4.81827 3.09758778551776
6.05987333333333 3.81654495591905
0.983014333333333 1.70860477382685
3.20145333333333 1.96177407223156
0.102885666666667 0.102386009643173
0.268018 0.100821768802856
24.7583333333333 25.3921531471098
35.1390666666667 24.6259853228875
1.44395666666667 1.73822199720859
1.09907333333333 1.43966517835828
};
\addplot [black, mark=-, mark size=2, mark options={solid}, only marks]
table {%
1.53489 2.35054412077072
1.08553 1.20002270705407
4.29854333333333 2.58036327428822
13.6454133333333 11.4710381685035
0.834899333333333 1.7312243988115
1.27229266666667 1.66751734940482
8.17293333333333 9.55151730451086
7.95658333333333 8.03024326886712
5994.23666666667 1510.88439789054
6.81706666666667 9.33287283359393
24.3426333333333 26.4517517545097
4.81827 3.50879221448224
6.05987333333333 5.03343504408095
0.983014333333333 1.76882189283982
3.20145333333333 2.26232592776844
0.102885666666667 0.10669532369016
0.268018 0.106589564530477
24.7583333333333 27.6179135195569
35.1390666666667 25.5967480104459
1.44395666666667 1.77977800279141
1.09907333333333 1.47768148830838
};
\end{axis}

\end{tikzpicture}
  }
  \scalebox{0.45}{
% This file was created with tikzplotlib v0.9.17.
\begin{tikzpicture}

\definecolor{color0}{rgb}{0.12156862745098,0.466666666666667,0.705882352941177}

\begin{axis}[
log basis x={10},
log basis y={10},
tick align=outside,
tick pos=left,
title={\textsc{saucy}},
x grid style={white!69.0196078431373!black},
xlabel={time w/o preprocessor},
xmajorgrids,
xmin=0.0549805177414166, xmax=133203.52789875,
xmode=log,
xtick style={color=black},
y grid style={white!69.0196078431373!black},
ylabel={time with preprocesser},
ymajorgrids,
ymin=0.0506822865076788, ymax=133720.783034406,
ymode=log,
scale=0.66,
ytick style={color=black}
]
\addplot [draw=blue, fill=blue, mark=*, only marks, opacity=0.7]
table{%
x  y
2.43990666666667 2.83570666666667
1.175131 1.52598466666667
2.29419666666667 2.57523333333333
305.642333333333 108.046233333333
0.830165 1.32189
0.642145333333333 0.917564666666667
1448.53233333333 1460.05533333333
3392.49666666667 3346.76
0.874596333333333 1.24847333333333
4.94437666666667 5.18200333333333
4776.38666666667 4707.82
16.4581 16.826
40000.273236 40000.3727333333
1.43879666666667 2.36738666666667
2.07004333333333 1.77313666666667
0.193207333333333 0.102645333333333
0.108026666666667 0.100083666666667
4333.6 4355.47666666667
7667.01666666667 7677.43666666667
1.44170666666667 1.75363
6.43794666666667 6.66779666666667
};
\path [draw=black]
(axis cs:2.42190938292099,2.83570666666667)
--(axis cs:2.45790395041234,2.83570666666667);

\path [draw=black]
(axis cs:0.678476454420614,1.52598466666667)
--(axis cs:1.67178554557939,1.52598466666667);

\path [draw=black]
(axis cs:2.1357862868861,2.57523333333333)
--(axis cs:2.45260704644723,2.57523333333333);

\path [draw=black]
(axis cs:298.455733936617,108.046233333333)
--(axis cs:312.82893273005,108.046233333333);

\path [draw=black]
(axis cs:0.818564186723912,1.32189)
--(axis cs:0.841765813276088,1.32189);

\path [draw=black]
(axis cs:0.624403343697945,0.917564666666667)
--(axis cs:0.659887322968722,0.917564666666667);

\path [draw=black]
(axis cs:1104.84287094387,1460.05533333333)
--(axis cs:1792.22179572279,1460.05533333333);

\path [draw=black]
(axis cs:3071.16375753832,3346.76)
--(axis cs:3713.82957579502,3346.76);

\path [draw=black]
(axis cs:0.851367447374491,1.24847333333333)
--(axis cs:0.897825219292175,1.24847333333333);

\path [draw=black]
(axis cs:2.45528308351458,5.18200333333333)
--(axis cs:7.43347024981875,5.18200333333333);

\path [draw=black]
(axis cs:4673.1155302237,4707.82)
--(axis cs:4879.65780310963,4707.82);

\path [draw=black]
(axis cs:15.3314740165136,16.826)
--(axis cs:17.5847259834864,16.826);

\path [draw=black]
(axis cs:11716.388402595,40000.3727333333)
--(axis cs:68284.158069405,40000.3727333333);

\path [draw=black]
(axis cs:1.42504592042429,2.36738666666667)
--(axis cs:1.45254741290904,2.36738666666667);

\path [draw=black]
(axis cs:1.17339170505795,1.77313666666667)
--(axis cs:2.96669496160871,1.77313666666667);

\path [draw=black]
(axis cs:0.192327213518258,0.102645333333333)
--(axis cs:0.194087453148409,0.102645333333333);

\path [draw=black]
(axis cs:0.107251800943532,0.100083666666667)
--(axis cs:0.108801532389801,0.100083666666667);

\path [draw=black]
(axis cs:3116.44394777553,4355.47666666667)
--(axis cs:5550.75605222447,4355.47666666667);

\path [draw=black]
(axis cs:7618.65579657983,7677.43666666667)
--(axis cs:7715.3775367535,7677.43666666667);

\path [draw=black]
(axis cs:1.26631973417485,1.75363)
--(axis cs:1.61709359915848,1.75363);

\path [draw=black]
(axis cs:6.38861112820244,6.66779666666667)
--(axis cs:6.4872822051309,6.66779666666667);

\path [draw=black]
(axis cs:2.43990666666667,2.7695326151896)
--(axis cs:2.43990666666667,2.90188071814373);

\path [draw=black]
(axis cs:1.175131,0.977465807005893)
--(axis cs:1.175131,2.07450352632744);

\path [draw=black]
(axis cs:2.29419666666667,2.48912970726438)
--(axis cs:2.29419666666667,2.66133695940229);

\path [draw=black]
(axis cs:305.642333333333,65.4234875321899)
--(axis cs:305.642333333333,150.668979134477);

\path [draw=black]
(axis cs:0.830165,1.24212594627486)
--(axis cs:0.830165,1.40165405372514);

\path [draw=black]
(axis cs:0.642145333333333,0.881899530213511)
--(axis cs:0.642145333333333,0.953229803119823);

\path [draw=black]
(axis cs:1448.53233333333,1115.89246074214)
--(axis cs:1448.53233333333,1804.21820592453);

\path [draw=black]
(axis cs:3392.49666666667,3077.99576713657)
--(axis cs:3392.49666666667,3615.52423286343);

\path [draw=black]
(axis cs:0.874596333333333,1.1936169867948)
--(axis cs:0.874596333333333,1.30332967987187);

\path [draw=black]
(axis cs:4.94437666666667,3.59222696465055)
--(axis cs:4.94437666666667,6.77177970201612);

\path [draw=black]
(axis cs:4776.38666666667,4644.54352043084)
--(axis cs:4776.38666666667,4771.09647956916);

\path [draw=black]
(axis cs:16.4581,16.2316594747117)
--(axis cs:16.4581,17.4203405252883);

\path [draw=black]
(axis cs:40000.273236,11716.6286104066)
--(axis cs:40000.273236,68284.1168562601);

\path [draw=black]
(axis cs:1.43879666666667,2.30587213378577)
--(axis cs:1.43879666666667,2.42890119954757);

\path [draw=black]
(axis cs:2.07004333333333,1.65121725603452)
--(axis cs:2.07004333333333,1.89505607729881);

\path [draw=black]
(axis cs:0.193207333333333,0.100596347188377)
--(axis cs:0.193207333333333,0.104694319478289);

\path [draw=black]
(axis cs:0.108026666666667,0.0992511194374425)
--(axis cs:0.108026666666667,0.100916213895891);

\path [draw=black]
(axis cs:4333.6,3193.15250951906)
--(axis cs:4333.6,5517.80082381427);

\path [draw=black]
(axis cs:7667.01666666667,7665.65577441598)
--(axis cs:7667.01666666667,7689.21755891735);

\path [draw=black]
(axis cs:1.44170666666667,1.62611858691605)
--(axis cs:1.44170666666667,1.88114141308395);

\path [draw=black]
(axis cs:6.43794666666667,6.58101622557793)
--(axis cs:6.43794666666667,6.7545771077554);

\addplot [semithick, color0, dashed]
table {%
0.0549805177414166 0.0549805177414166
133203.52789875 133203.52789875
};
\addplot [black, mark=|, mark size=2, mark options={solid}, only marks]
table {%
2.42190938292099 2.83570666666667
0.678476454420614 1.52598466666667
2.1357862868861 2.57523333333333
298.455733936617 108.046233333333
0.818564186723912 1.32189
0.624403343697945 0.917564666666667
1104.84287094387 1460.05533333333
3071.16375753832 3346.76
0.851367447374491 1.24847333333333
2.45528308351458 5.18200333333333
4673.1155302237 4707.82
15.3314740165136 16.826
11716.388402595 40000.3727333333
1.42504592042429 2.36738666666667
1.17339170505795 1.77313666666667
0.192327213518258 0.102645333333333
0.107251800943532 0.100083666666667
3116.44394777553 4355.47666666667
7618.65579657983 7677.43666666667
1.26631973417485 1.75363
6.38861112820244 6.66779666666667
};
\addplot [black, mark=|, mark size=2, mark options={solid}, only marks]
table {%
2.45790395041234 2.83570666666667
1.67178554557939 1.52598466666667
2.45260704644723 2.57523333333333
312.82893273005 108.046233333333
0.841765813276088 1.32189
0.659887322968722 0.917564666666667
1792.22179572279 1460.05533333333
3713.82957579502 3346.76
0.897825219292175 1.24847333333333
7.43347024981875 5.18200333333333
4879.65780310963 4707.82
17.5847259834864 16.826
68284.158069405 40000.3727333333
1.45254741290904 2.36738666666667
2.96669496160871 1.77313666666667
0.194087453148409 0.102645333333333
0.108801532389801 0.100083666666667
5550.75605222447 4355.47666666667
7715.3775367535 7677.43666666667
1.61709359915848 1.75363
6.4872822051309 6.66779666666667
};
\addplot [black, mark=-, mark size=2, mark options={solid}, only marks]
table {%
2.43990666666667 2.7695326151896
1.175131 0.977465807005893
2.29419666666667 2.48912970726438
305.642333333333 65.4234875321899
0.830165 1.24212594627486
0.642145333333333 0.881899530213511
1448.53233333333 1115.89246074214
3392.49666666667 3077.99576713657
0.874596333333333 1.1936169867948
4.94437666666667 3.59222696465055
4776.38666666667 4644.54352043084
16.4581 16.2316594747117
40000.273236 11716.6286104066
1.43879666666667 2.30587213378577
2.07004333333333 1.65121725603452
0.193207333333333 0.100596347188377
0.108026666666667 0.0992511194374425
4333.6 3193.15250951906
7667.01666666667 7665.65577441598
1.44170666666667 1.62611858691605
6.43794666666667 6.58101622557793
};
\addplot [black, mark=-, mark size=2, mark options={solid}, only marks]
table {%
2.43990666666667 2.90188071814373
1.175131 2.07450352632744
2.29419666666667 2.66133695940229
305.642333333333 150.668979134477
0.830165 1.40165405372514
0.642145333333333 0.953229803119823
1448.53233333333 1804.21820592453
3392.49666666667 3615.52423286343
0.874596333333333 1.30332967987187
4.94437666666667 6.77177970201612
4776.38666666667 4771.09647956916
16.4581 17.4203405252883
40000.273236 68284.1168562601
1.43879666666667 2.42890119954757
2.07004333333333 1.89505607729881
0.193207333333333 0.104694319478289
0.108026666666667 0.100916213895891
4333.6 5517.80082381427
7667.01666666667 7689.21755891735
1.44170666666667 1.88114141308395
6.43794666666667 6.7545771077554
};
\end{axis}

\end{tikzpicture}
  }
      \caption{Solvers with \textsc{sassy} vs. solvers without \textsc{sassy} on \textbf{portfolio\_comb (p)}.} \label{fig:rescomb_permute}
\end{figure*}
\begin{figure*}[t!]
  \centering
  \scalebox{0.45}{
\input{results/bli_portfolio_prep_permute.tex}
  }
  \scalebox{0.45}{
\input{results/nau_portfolio_prep_permute.tex}
  }
  \scalebox{0.45}{
\input{results/tra_portfolio_prep_permute.tex}
  }
  \scalebox{0.45}{
\input{results/dej_portfolio_prep_permute.tex}
  }
  \scalebox{0.45}{
\input{results/sau_portfolio_prep_permute.tex}
  }
      \caption{Solvers with \textsc{sassy} vs. solvers without \textsc{sassy} on \textbf{portfolio\_pract (p)}. Timeout is $60s$. The green bar shows instances that timed out without the preprocessor. With the preprocessor enabled no instance timed out.} \label{fig:respract_permute} 
\end{figure*}
\begin{figure*}[t!]
  \centering
  \scalebox{0.45}{ 
% This file was created with tikzplotlib v0.9.17.
\begin{tikzpicture}

\definecolor{color0}{rgb}{0.12156862745098,0.466666666666667,0.705882352941177}

\begin{axis}[
log basis x={10},
log basis y={10},
tick align=outside,
tick pos=left,
title={\textsc{bliss}},
x grid style={white!69.0196078431373!black},
xlabel={time w/o preprocessor},
xmajorgrids,
xmin=0.13912708659536, xmax=5451.75900667369,
xmode=log,
xtick style={color=black},
y grid style={white!69.0196078431373!black},
ylabel={time with preprocesser},
ymajorgrids,
ymin=0.0584664572403816, ymax=5683.74062285528,
ymode=log,
scale=0.66,
ytick style={color=black}
]
\addplot [draw=blue, fill=blue, mark=*, only marks, opacity=0.7]
table{%
x  y
3.97799666666667 4.31731
7.01894 7.42769
13.3871 13.6155
31.0705333333333 9.11755333333333
0.829833666666667 1.45711
0.850395666666667 1.18999333333333
29.5259666666667 30.7029333333333
173.235333333333 171.204333333333
5.65273333333333 5.53567
27.0141333333333 27.8269
3362.98333333333 3363.31333333333
48.5764 49.7409666666667
10.4352666666667 10.6905
15.4289 16.3799666666667
3.10139333333333 3.38825666666667
0.453723333333333 0.109101
0.230973666666667 0.0999606666666667
219.407 220.604666666667
985.766333333333 988.685666666667
3.20712333333333 3.39186666666667
12.1395 12.6061
};
\path [draw=black]
(axis cs:3.78448149407748,4.31731)
--(axis cs:4.17151183925585,4.31731);

\path [draw=black]
(axis cs:7.00962119821722,7.42769)
--(axis cs:7.02825880178278,7.42769);

\path [draw=black]
(axis cs:13.2969586665286,13.6155)
--(axis cs:13.4772413334714,13.6155);

\path [draw=black]
(axis cs:30.8545905271194,9.11755333333333)
--(axis cs:31.2864761395473,9.11755333333333);

\path [draw=black]
(axis cs:0.80200303126449,1.45711)
--(axis cs:0.857664302068843,1.45711);

\path [draw=black]
(axis cs:0.821300347451676,1.18999333333333)
--(axis cs:0.879490985881658,1.18999333333333);

\path [draw=black]
(axis cs:29.45305314102,30.7029333333333)
--(axis cs:29.5988801923133,30.7029333333333);

\path [draw=black]
(axis cs:170.026405128742,171.204333333333)
--(axis cs:176.444261537925,171.204333333333);

\path [draw=black]
(axis cs:4.88886770499576,5.53567)
--(axis cs:6.41659896167091,5.53567);

\path [draw=black]
(axis cs:26.6380269269948,27.8269)
--(axis cs:27.3902397396719,27.8269);

\path [draw=black]
(axis cs:3354.97062537105,3363.31333333333)
--(axis cs:3370.99604129562,3363.31333333333);

\path [draw=black]
(axis cs:48.2627536280033,49.7409666666667)
--(axis cs:48.8900463719967,49.7409666666667);

\path [draw=black]
(axis cs:10.3533546271229,10.6905)
--(axis cs:10.5171787062104,10.6905);

\path [draw=black]
(axis cs:15.3620326188141,16.3799666666667)
--(axis cs:15.4957673811859,16.3799666666667);

\path [draw=black]
(axis cs:3.08644718927467,3.38825666666667)
--(axis cs:3.11633947739199,3.38825666666667);

\path [draw=black]
(axis cs:0.434822251806371,0.109101)
--(axis cs:0.472624414860296,0.109101);

\path [draw=black]
(axis cs:0.225003927066911,0.0999606666666667)
--(axis cs:0.236943406266422,0.0999606666666667);

\path [draw=black]
(axis cs:218.811365324493,220.604666666667)
--(axis cs:220.002634675507,220.604666666667);

\path [draw=black]
(axis cs:982.395457944226,988.685666666667)
--(axis cs:989.137208722441,988.685666666667);

\path [draw=black]
(axis cs:2.85931153649168,3.39186666666667)
--(axis cs:3.55493513017499,3.39186666666667);

\path [draw=black]
(axis cs:12.0011723936928,12.6061)
--(axis cs:12.2778276063072,12.6061);

\path [draw=black]
(axis cs:3.97799666666667,4.27534044833851)
--(axis cs:3.97799666666667,4.35927955166149);

\path [draw=black]
(axis cs:7.01894,7.39612267195343)
--(axis cs:7.01894,7.45925732804657);

\path [draw=black]
(axis cs:13.3871,13.4687066531935)
--(axis cs:13.3871,13.7622933468065);

\path [draw=black]
(axis cs:31.0705333333333,9.10443671198652)
--(axis cs:31.0705333333333,9.13066995468014);

\path [draw=black]
(axis cs:0.829833666666667,1.45388670975555)
--(axis cs:0.829833666666667,1.46033329024446);

\path [draw=black]
(axis cs:0.850395666666667,1.12086907597345)
--(axis cs:0.850395666666667,1.25911759069322);

\path [draw=black]
(axis cs:29.5259666666667,30.6146600707399)
--(axis cs:29.5259666666667,30.7912065959268);

\path [draw=black]
(axis cs:173.235333333333,170.709156372821)
--(axis cs:173.235333333333,171.699510293846);

\path [draw=black]
(axis cs:5.65273333333333,5.47651375096408)
--(axis cs:5.65273333333333,5.59482624903592);

\path [draw=black]
(axis cs:27.0141333333333,27.6551296688404)
--(axis cs:27.0141333333333,27.9986703311596);

\path [draw=black]
(axis cs:3362.98333333333,3354.37975497848)
--(axis cs:3362.98333333333,3372.24691168818);

\path [draw=black]
(axis cs:48.5764,49.2858481741593)
--(axis cs:48.5764,50.1960851591741);

\path [draw=black]
(axis cs:10.4352666666667,10.5941404303317)
--(axis cs:10.4352666666667,10.7868595696683);

\path [draw=black]
(axis cs:15.4289,16.3070889456388)
--(axis cs:15.4289,16.4528443876946);

\path [draw=black]
(axis cs:3.10139333333333,3.37160098828894)
--(axis cs:3.10139333333333,3.40491234504439);

\path [draw=black]
(axis cs:0.453723333333333,0.102211988605032)
--(axis cs:0.453723333333333,0.115990011394968);

\path [draw=black]
(axis cs:0.230973666666667,0.098542066104297)
--(axis cs:0.230973666666667,0.101379267229036);

\path [draw=black]
(axis cs:219.407,219.188739094759)
--(axis cs:219.407,222.020594238575);

\path [draw=black]
(axis cs:985.766333333333,987.425704145178)
--(axis cs:985.766333333333,989.945629188155);

\path [draw=black]
(axis cs:3.20712333333333,3.34965598765978)
--(axis cs:3.20712333333333,3.43407734567355);

\path [draw=black]
(axis cs:12.1395,12.4757809300217)
--(axis cs:12.1395,12.7364190699783);

\addplot [semithick, color0, dashed]
table {%
0.13912708659536 0.13912708659536
5451.75900667369 5451.75900667369
};
\addplot [black, mark=|, mark size=2, mark options={solid}, only marks]
table {%
3.78448149407748 4.31731
7.00962119821722 7.42769
13.2969586665286 13.6155
30.8545905271194 9.11755333333333
0.80200303126449 1.45711
0.821300347451676 1.18999333333333
29.45305314102 30.7029333333333
170.026405128742 171.204333333333
4.88886770499576 5.53567
26.6380269269948 27.8269
3354.97062537105 3363.31333333333
48.2627536280033 49.7409666666667
10.3533546271229 10.6905
15.3620326188141 16.3799666666667
3.08644718927467 3.38825666666667
0.434822251806371 0.109101
0.225003927066911 0.0999606666666667
218.811365324493 220.604666666667
982.395457944226 988.685666666667
2.85931153649168 3.39186666666667
12.0011723936928 12.6061
};
\addplot [black, mark=|, mark size=2, mark options={solid}, only marks]
table {%
4.17151183925585 4.31731
7.02825880178278 7.42769
13.4772413334714 13.6155
31.2864761395473 9.11755333333333
0.857664302068843 1.45711
0.879490985881658 1.18999333333333
29.5988801923133 30.7029333333333
176.444261537925 171.204333333333
6.41659896167091 5.53567
27.3902397396719 27.8269
3370.99604129562 3363.31333333333
48.8900463719967 49.7409666666667
10.5171787062104 10.6905
15.4957673811859 16.3799666666667
3.11633947739199 3.38825666666667
0.472624414860296 0.109101
0.236943406266422 0.0999606666666667
220.002634675507 220.604666666667
989.137208722441 988.685666666667
3.55493513017499 3.39186666666667
12.2778276063072 12.6061
};
\addplot [black, mark=-, mark size=2, mark options={solid}, only marks]
table {%
3.97799666666667 4.27534044833851
7.01894 7.39612267195343
13.3871 13.4687066531935
31.0705333333333 9.10443671198652
0.829833666666667 1.45388670975555
0.850395666666667 1.12086907597345
29.5259666666667 30.6146600707399
173.235333333333 170.709156372821
5.65273333333333 5.47651375096408
27.0141333333333 27.6551296688404
3362.98333333333 3354.37975497848
48.5764 49.2858481741593
10.4352666666667 10.5941404303317
15.4289 16.3070889456388
3.10139333333333 3.37160098828894
0.453723333333333 0.102211988605032
0.230973666666667 0.098542066104297
219.407 219.188739094759
985.766333333333 987.425704145178
3.20712333333333 3.34965598765978
12.1395 12.4757809300217
};
\addplot [black, mark=-, mark size=2, mark options={solid}, only marks]
table {%
3.97799666666667 4.35927955166149
7.01894 7.45925732804657
13.3871 13.7622933468065
31.0705333333333 9.13066995468014
0.829833666666667 1.46033329024446
0.850395666666667 1.25911759069322
29.5259666666667 30.7912065959268
173.235333333333 171.699510293846
5.65273333333333 5.59482624903592
27.0141333333333 27.9986703311596
3362.98333333333 3372.24691168818
48.5764 50.1960851591741
10.4352666666667 10.7868595696683
15.4289 16.4528443876946
3.10139333333333 3.40491234504439
0.453723333333333 0.115990011394968
0.230973666666667 0.101379267229036
219.407 222.020594238575
985.766333333333 989.945629188155
3.20712333333333 3.43407734567355
12.1395 12.7364190699783
};
\end{axis}

\end{tikzpicture}
  }
  \scalebox{0.45}{
% This file was created with tikzplotlib v0.9.17.
\begin{tikzpicture}

\definecolor{color0}{rgb}{0.12156862745098,0.466666666666667,0.705882352941177}

\begin{axis}[
log basis x={10},
log basis y={10},
tick align=outside,
tick pos=left,
title={\textsc{nauty}},
x grid style={white!69.0196078431373!black},
xlabel={time w/o preprocessor},
xmajorgrids,
xmin=0.0704474534765265, xmax=23198.3347920468,
xmode=log,
xtick style={color=black},
y grid style={white!69.0196078431373!black},
ylabel={time with preprocesser},
ymajorgrids,
ymin=0.0546865757864616, ymax=23337.9406089934,
ymode=log,
scale=0.66,
ytick style={color=black}
]
\addplot [draw=blue, fill=blue, mark=*, only marks, opacity=0.7]
table{%
x  y
1.42956333333333 1.80861333333333
137.606666666667 135.663
64.0416666666667 64.9317
12.277 7.96701333333333
0.390820666666667 1.12694666666667
0.408222333333333 0.725481333333333
20.0352666666667 20.8515666666667
1122.47666666667 1124.39333333333
11.27515 9.75324
6.05242666666667 6.88715666666667
6830.76 6813.88
9.25754666666667 9.98488333333333
19.6520666666667 20.2006666666667
0.622079333333333 1.47838
4.41635666666667 4.56986
0.190954666666667 0.103065
0.125804333333333 0.101248666666667
4371.28333333333 4370.44666666667
12957.8 12936.5333333333
1.1796 1.44106333333333
2.32415666666667 2.66743333333333
};
\path [draw=black]
(axis cs:1.39143385188452,1.80861333333333)
--(axis cs:1.46769281478215,1.80861333333333);

\path [draw=black]
(axis cs:136.788986071596,135.663)
--(axis cs:138.424347261738,135.663);

\path [draw=black]
(axis cs:63.7167478426164,64.9317)
--(axis cs:64.366585490717,64.9317);

\path [draw=black]
(axis cs:12.1624709352755,7.96701333333333)
--(axis cs:12.3915290647245,7.96701333333333);

\path [draw=black]
(axis cs:0.384539152128843,1.12694666666667)
--(axis cs:0.39710218120449,1.12694666666667);

\path [draw=black]
(axis cs:0.389634236885457,0.725481333333333)
--(axis cs:0.42681042978121,0.725481333333333);

\path [draw=black]
(axis cs:19.7220766675359,20.8515666666667)
--(axis cs:20.3484566657975,20.8515666666667);

\path [draw=black]
(axis cs:1121.70637090817,1124.39333333333)
--(axis cs:1123.24696242516,1124.39333333333);

\path [draw=black]
(axis cs:8.34104094475569,9.75324)
--(axis cs:14.2092590552443,9.75324);

\path [draw=black]
(axis cs:5.95286170049438,6.88715666666667)
--(axis cs:6.15199163283895,6.88715666666667);

\path [draw=black]
(axis cs:6804.48276523934,6813.88)
--(axis cs:6857.03723476066,6813.88);

\path [draw=black]
(axis cs:9.2037264063345,9.98488333333333)
--(axis cs:9.31136692699883,9.98488333333333);

\path [draw=black]
(axis cs:19.6394106184858,20.2006666666667)
--(axis cs:19.6647227148475,20.2006666666667);

\path [draw=black]
(axis cs:0.604229371067578,1.47838)
--(axis cs:0.639929295599089,1.47838);

\path [draw=black]
(axis cs:4.33313781450229,4.56986)
--(axis cs:4.49957551883104,4.56986);

\path [draw=black]
(axis cs:0.190187245733352,0.103065)
--(axis cs:0.191722087599981,0.103065);

\path [draw=black]
(axis cs:0.125506122744059,0.101248666666667)
--(axis cs:0.126102543922608,0.101248666666667);

\path [draw=black]
(axis cs:4355.44124943562,4370.44666666667)
--(axis cs:4387.12541723105,4370.44666666667);

\path [draw=black]
(axis cs:12894.2144670542,12936.5333333333)
--(axis cs:13021.3855329458,12936.5333333333);

\path [draw=black]
(axis cs:0.932861980230042,1.44106333333333)
--(axis cs:1.42633801976996,1.44106333333333);

\path [draw=black]
(axis cs:2.28135156390751,2.66743333333333)
--(axis cs:2.36696176942583,2.66743333333333);

\path [draw=black]
(axis cs:1.42956333333333,1.74205167253007)
--(axis cs:1.42956333333333,1.8751749941366);

\path [draw=black]
(axis cs:137.606666666667,135.150839217953)
--(axis cs:137.606666666667,136.175160782047);

\path [draw=black]
(axis cs:64.0416666666667,64.3066063216872)
--(axis cs:64.0416666666667,65.5567936783128);

\path [draw=black]
(axis cs:12.277,7.94673245509243)
--(axis cs:12.277,7.98729421157423);

\path [draw=black]
(axis cs:0.390820666666667,1.10808219152893)
--(axis cs:0.390820666666667,1.1458111418044);

\path [draw=black]
(axis cs:0.408222333333333,0.688150855505038)
--(axis cs:0.408222333333333,0.762811811161629);

\path [draw=black]
(axis cs:20.0352666666667,20.8175219575003)
--(axis cs:20.0352666666667,20.885611375833);

\path [draw=black]
(axis cs:1122.47666666667,1122.84995320445)
--(axis cs:1122.47666666667,1125.93671346221);

\path [draw=black]
(axis cs:11.27515,9.66987232960754)
--(axis cs:11.27515,9.83660767039246);

\path [draw=black]
(axis cs:6.05242666666667,6.8408427016501)
--(axis cs:6.05242666666667,6.93347063168323);

\path [draw=black]
(axis cs:6830.76,6807.36226521967)
--(axis cs:6830.76,6820.39773478033);

\path [draw=black]
(axis cs:9.25754666666667,9.90029253228368)
--(axis cs:9.25754666666667,10.069474134383);

\path [draw=black]
(axis cs:19.6520666666667,20.0934851902405)
--(axis cs:19.6520666666667,20.3078481430928);

\path [draw=black]
(axis cs:0.622079333333333,1.46189860240554)
--(axis cs:0.622079333333333,1.49486139759446);

\path [draw=black]
(axis cs:4.41635666666667,4.56177642405862)
--(axis cs:4.41635666666667,4.57794357594138);

\path [draw=black]
(axis cs:0.190954666666667,0.102307205392823)
--(axis cs:0.190954666666667,0.103822794607177);

\path [draw=black]
(axis cs:0.125804333333333,0.0985821045437159)
--(axis cs:0.125804333333333,0.103915228789617);

\path [draw=black]
(axis cs:4371.28333333333,4353.89098016457)
--(axis cs:4371.28333333333,4387.00235316876);

\path [draw=black]
(axis cs:12957.8,12926.781296509)
--(axis cs:12957.8,12946.2853701576);

\path [draw=black]
(axis cs:1.1796,1.30807952421023)
--(axis cs:1.1796,1.57404714245643);

\path [draw=black]
(axis cs:2.32415666666667,2.64303214165023)
--(axis cs:2.32415666666667,2.69183452501644);

\addplot [semithick, color0, dashed]
table {%
0.0704474534765265 0.0704474534765265
23198.3347920468 23198.3347920468
};
\addplot [black, mark=|, mark size=2, mark options={solid}, only marks]
table {%
1.39143385188452 1.80861333333333
136.788986071596 135.663
63.7167478426164 64.9317
12.1624709352755 7.96701333333333
0.384539152128843 1.12694666666667
0.389634236885457 0.725481333333333
19.7220766675359 20.8515666666667
1121.70637090817 1124.39333333333
8.34104094475569 9.75324
5.95286170049438 6.88715666666667
6804.48276523934 6813.88
9.2037264063345 9.98488333333333
19.6394106184858 20.2006666666667
0.604229371067578 1.47838
4.33313781450229 4.56986
0.190187245733352 0.103065
0.125506122744059 0.101248666666667
4355.44124943562 4370.44666666667
12894.2144670542 12936.5333333333
0.932861980230042 1.44106333333333
2.28135156390751 2.66743333333333
};
\addplot [black, mark=|, mark size=2, mark options={solid}, only marks]
table {%
1.46769281478215 1.80861333333333
138.424347261738 135.663
64.366585490717 64.9317
12.3915290647245 7.96701333333333
0.39710218120449 1.12694666666667
0.42681042978121 0.725481333333333
20.3484566657975 20.8515666666667
1123.24696242516 1124.39333333333
14.2092590552443 9.75324
6.15199163283895 6.88715666666667
6857.03723476066 6813.88
9.31136692699883 9.98488333333333
19.6647227148475 20.2006666666667
0.639929295599089 1.47838
4.49957551883104 4.56986
0.191722087599981 0.103065
0.126102543922608 0.101248666666667
4387.12541723105 4370.44666666667
13021.3855329458 12936.5333333333
1.42633801976996 1.44106333333333
2.36696176942583 2.66743333333333
};
\addplot [black, mark=-, mark size=2, mark options={solid}, only marks]
table {%
1.42956333333333 1.74205167253007
137.606666666667 135.150839217953
64.0416666666667 64.3066063216872
12.277 7.94673245509243
0.390820666666667 1.10808219152893
0.408222333333333 0.688150855505038
20.0352666666667 20.8175219575003
1122.47666666667 1122.84995320445
11.27515 9.66987232960754
6.05242666666667 6.8408427016501
6830.76 6807.36226521967
9.25754666666667 9.90029253228368
19.6520666666667 20.0934851902405
0.622079333333333 1.46189860240554
4.41635666666667 4.56177642405862
0.190954666666667 0.102307205392823
0.125804333333333 0.0985821045437159
4371.28333333333 4353.89098016457
12957.8 12926.781296509
1.1796 1.30807952421023
2.32415666666667 2.64303214165023
};
\addplot [black, mark=-, mark size=2, mark options={solid}, only marks]
table {%
1.42956333333333 1.8751749941366
137.606666666667 136.175160782047
64.0416666666667 65.5567936783128
12.277 7.98729421157423
0.390820666666667 1.1458111418044
0.408222333333333 0.762811811161629
20.0352666666667 20.885611375833
1122.47666666667 1125.93671346221
11.27515 9.83660767039246
6.05242666666667 6.93347063168323
6830.76 6820.39773478033
9.25754666666667 10.069474134383
19.6520666666667 20.3078481430928
0.622079333333333 1.49486139759446
4.41635666666667 4.57794357594138
0.190954666666667 0.103822794607177
0.125804333333333 0.103915228789617
4371.28333333333 4387.00235316876
12957.8 12946.2853701576
1.1796 1.57404714245643
2.32415666666667 2.69183452501644
};
\end{axis}

\end{tikzpicture}
  }
  \scalebox{0.45}{
% This file was created with tikzplotlib v0.9.17.
\begin{tikzpicture}

\definecolor{color0}{rgb}{0.12156862745098,0.466666666666667,0.705882352941177}

\begin{axis}[
log basis x={10},
log basis y={10},
tick align=outside,
tick pos=left,
title={\textsc{Traces}},
x grid style={white!69.0196078431373!black},
xlabel={time w/o preprocessor},
xmajorgrids,
xmin=0.0832456217704071, xmax=1125.53737017277,
xmode=log,
xtick style={color=black},
y grid style={white!69.0196078431373!black},
ylabel={time with preprocesser},
ymajorgrids,
ymin=0.0577641033515621, ymax=10621.4890613609,
ymode=log,
scale=0.66,
ytick style={color=black}
]
\addplot [draw=blue, fill=blue, mark=*, only marks, opacity=0.7]
table{%
x  y
1.46449666666667 1.79343666666667
0.890488 1.26379666666667
2.29443 2.52068666666667
726.657 5988.13333333333
0.388858333333333 1.22604
0.533303666666667 0.851815
5.28178 6.33977666666667
17.5761666666667 17.747
13.3055666666667 11.2509
4.67058 5.65018333333333
73.9695333333333 73.9329666666667
8.26151666666667 8.88268666666667
0.593942666666667 0.840217333333333
0.740447 1.75039333333333
1.85368 2.11668333333333
0.190915666666667 0.100928
0.152851333333333 0.101779333333333
81.0452666666667 82.6680666666667
91.9845666666667 92.3530333333333
1.266249 1.39550333333333
2.03953666666667 2.41675
};
\path [draw=black]
(axis cs:1.44331289562477,1.79343666666667)
--(axis cs:1.48568043770856,1.79343666666667);

\path [draw=black]
(axis cs:0.868021058048175,1.26379666666667)
--(axis cs:0.912954941951825,1.26379666666667);

\path [draw=black]
(axis cs:2.21423013798848,2.52068666666667)
--(axis cs:2.37462986201152,2.52068666666667);

\path [draw=black]
(axis cs:722.869439747454,5988.13333333333)
--(axis cs:730.444560252546,5988.13333333333);

\path [draw=black]
(axis cs:0.387560063373391,1.22604)
--(axis cs:0.390156603293276,1.22604);

\path [draw=black]
(axis cs:0.5101738427686,0.851815)
--(axis cs:0.556433490564733,0.851815);

\path [draw=black]
(axis cs:5.22876608798941,6.33977666666667)
--(axis cs:5.33479391201059,6.33977666666667);

\path [draw=black]
(axis cs:17.5374152939311,17.747)
--(axis cs:17.6149180394022,17.747);

\path [draw=black]
(axis cs:9.57291272223398,11.2509)
--(axis cs:17.0382206110994,11.2509);

\path [draw=black]
(axis cs:4.58022514364647,5.65018333333333)
--(axis cs:4.76093485635353,5.65018333333333);

\path [draw=black]
(axis cs:73.6884241453733,73.9329666666667)
--(axis cs:74.2506425212934,73.9329666666667);

\path [draw=black]
(axis cs:8.08534802902396,8.88268666666667)
--(axis cs:8.43768530430938,8.88268666666667);

\path [draw=black]
(axis cs:0.591575769072299,0.840217333333333)
--(axis cs:0.596309564261034,0.840217333333333);

\path [draw=black]
(axis cs:0.736479000924059,1.75039333333333)
--(axis cs:0.744414999075941,1.75039333333333);

\path [draw=black]
(axis cs:1.8311859830177,2.11668333333333)
--(axis cs:1.8761740169823,2.11668333333333);

\path [draw=black]
(axis cs:0.171818691821885,0.100928)
--(axis cs:0.210012641511448,0.100928);

\path [draw=black]
(axis cs:0.12827264833551,0.101779333333333)
--(axis cs:0.177430018331157,0.101779333333333);

\path [draw=black]
(axis cs:80.8393483279442,82.6680666666667)
--(axis cs:81.2511850053891,82.6680666666667);

\path [draw=black]
(axis cs:91.4772357665411,92.3530333333333)
--(axis cs:92.4918975667922,92.3530333333333);

\path [draw=black]
(axis cs:0.923506241464292,1.39550333333333)
--(axis cs:1.60899175853571,1.39550333333333);

\path [draw=black]
(axis cs:2.020959787967,2.41675)
--(axis cs:2.05811354536634,2.41675);

\path [draw=black]
(axis cs:1.46449666666667,1.78287648253528)
--(axis cs:1.46449666666667,1.80399685079806);

\path [draw=black]
(axis cs:0.890488,1.18044868915338)
--(axis cs:0.890488,1.34714464417995);

\path [draw=black]
(axis cs:2.29443,2.51084270969353)
--(axis cs:2.29443,2.5305306236398);

\path [draw=black]
(axis cs:726.657,5854.33175951764)
--(axis cs:726.657,6121.93490714902);

\path [draw=black]
(axis cs:0.388858333333333,1.19934403650986)
--(axis cs:0.388858333333333,1.25273596349014);

\path [draw=black]
(axis cs:0.533303666666667,0.825533689061109)
--(axis cs:0.533303666666667,0.878096310938891);

\path [draw=black]
(axis cs:5.28178,6.29389911272625)
--(axis cs:5.28178,6.38565422060708);

\path [draw=black]
(axis cs:17.5761666666667,17.5775337595075)
--(axis cs:17.5761666666667,17.9164662404925);

\path [draw=black]
(axis cs:13.3055666666667,11.1836475898821)
--(axis cs:13.3055666666667,11.3181524101179);

\path [draw=black]
(axis cs:4.67058,5.59288935090741)
--(axis cs:4.67058,5.70747731575926);

\path [draw=black]
(axis cs:73.9695333333333,73.6299687862194)
--(axis cs:73.9695333333333,74.235964547114);

\path [draw=black]
(axis cs:8.26151666666667,8.8458151724077)
--(axis cs:8.26151666666667,8.91955816092563);

\path [draw=black]
(axis cs:0.593942666666667,0.821518213775926)
--(axis cs:0.593942666666667,0.858916452890741);

\path [draw=black]
(axis cs:0.740447,1.65154824229578)
--(axis cs:0.740447,1.84923842437089);

\path [draw=black]
(axis cs:1.85368,2.04291324620071)
--(axis cs:1.85368,2.19045342046596);

\path [draw=black]
(axis cs:0.190915666666667,0.100680757878454)
--(axis cs:0.190915666666667,0.101175242121546);

\path [draw=black]
(axis cs:0.152851333333333,0.100220077670454)
--(axis cs:0.152851333333333,0.103338588996213);

\path [draw=black]
(axis cs:81.0452666666667,82.3786056234233)
--(axis cs:81.0452666666667,82.9575277099101);

\path [draw=black]
(axis cs:91.9845666666667,91.7051384327343)
--(axis cs:91.9845666666667,93.0009282339324);

\path [draw=black]
(axis cs:1.266249,1.35622570583087)
--(axis cs:1.266249,1.4347809608358);

\path [draw=black]
(axis cs:2.03953666666667,2.40486257527188)
--(axis cs:2.03953666666667,2.42863742472812);

\addplot [semithick, color0, dashed]
table {%
0.0832456217704071 0.0832456217704071
1125.53737017277 1125.53737017277
};
\addplot [black, mark=|, mark size=2, mark options={solid}, only marks]
table {%
1.44331289562477 1.79343666666667
0.868021058048175 1.26379666666667
2.21423013798848 2.52068666666667
722.869439747454 5988.13333333333
0.387560063373391 1.22604
0.5101738427686 0.851815
5.22876608798941 6.33977666666667
17.5374152939311 17.747
9.57291272223398 11.2509
4.58022514364647 5.65018333333333
73.6884241453733 73.9329666666667
8.08534802902396 8.88268666666667
0.591575769072299 0.840217333333333
0.736479000924059 1.75039333333333
1.8311859830177 2.11668333333333
0.171818691821885 0.100928
0.12827264833551 0.101779333333333
80.8393483279442 82.6680666666667
91.4772357665411 92.3530333333333
0.923506241464292 1.39550333333333
2.020959787967 2.41675
};
\addplot [black, mark=|, mark size=2, mark options={solid}, only marks]
table {%
1.48568043770856 1.79343666666667
0.912954941951825 1.26379666666667
2.37462986201152 2.52068666666667
730.444560252546 5988.13333333333
0.390156603293276 1.22604
0.556433490564733 0.851815
5.33479391201059 6.33977666666667
17.6149180394022 17.747
17.0382206110994 11.2509
4.76093485635353 5.65018333333333
74.2506425212934 73.9329666666667
8.43768530430938 8.88268666666667
0.596309564261034 0.840217333333333
0.744414999075941 1.75039333333333
1.8761740169823 2.11668333333333
0.210012641511448 0.100928
0.177430018331157 0.101779333333333
81.2511850053891 82.6680666666667
92.4918975667922 92.3530333333333
1.60899175853571 1.39550333333333
2.05811354536634 2.41675
};
\addplot [black, mark=-, mark size=2, mark options={solid}, only marks]
table {%
1.46449666666667 1.78287648253528
0.890488 1.18044868915338
2.29443 2.51084270969353
726.657 5854.33175951764
0.388858333333333 1.19934403650986
0.533303666666667 0.825533689061109
5.28178 6.29389911272625
17.5761666666667 17.5775337595075
13.3055666666667 11.1836475898821
4.67058 5.59288935090741
73.9695333333333 73.6299687862194
8.26151666666667 8.8458151724077
0.593942666666667 0.821518213775926
0.740447 1.65154824229578
1.85368 2.04291324620071
0.190915666666667 0.100680757878454
0.152851333333333 0.100220077670454
81.0452666666667 82.3786056234233
91.9845666666667 91.7051384327343
1.266249 1.35622570583087
2.03953666666667 2.40486257527188
};
\addplot [black, mark=-, mark size=2, mark options={solid}, only marks]
table {%
1.46449666666667 1.80399685079806
0.890488 1.34714464417995
2.29443 2.5305306236398
726.657 6121.93490714902
0.388858333333333 1.25273596349014
0.533303666666667 0.878096310938891
5.28178 6.38565422060708
17.5761666666667 17.9164662404925
13.3055666666667 11.3181524101179
4.67058 5.70747731575926
73.9695333333333 74.235964547114
8.26151666666667 8.91955816092563
0.593942666666667 0.858916452890741
0.740447 1.84923842437089
1.85368 2.19045342046596
0.190915666666667 0.101175242121546
0.152851333333333 0.103338588996213
81.0452666666667 82.9575277099101
91.9845666666667 93.0009282339324
1.266249 1.4347809608358
2.03953666666667 2.42863742472812
};
\end{axis}

\end{tikzpicture}
  }
  \scalebox{0.45}{
% This file was created with tikzplotlib v0.9.17.
\begin{tikzpicture}

\definecolor{color0}{rgb}{0.12156862745098,0.466666666666667,0.705882352941177}

\begin{axis}[
log basis x={10},
log basis y={10},
tick align=outside,
tick pos=left,
title={\textsc{dejavu}},
x grid style={white!69.0196078431373!black},
xlabel={time w/o preprocessor},
xmajorgrids,
xmin=0.0637233018755665, xmax=2396.09720542543,
xmode=log,
xtick style={color=black},
y grid style={white!69.0196078431373!black},
ylabel={time with preprocesser},
ymajorgrids,
ymin=0.0553954583247683, ymax=22111.7101243484,
ymode=log,
scale=0.66,
ytick style={color=black}
]
\addplot [draw=blue, fill=blue, mark=*, only marks, opacity=0.7]
table{%
x  y
1.73028666666667 2.08859
0.869252333333333 1.41726
4.36094333333333 2.28884
17.5044166666667 16.9607
1.3341 1.62675666666667
1.09240233333333 1.44702333333333
9.54251666666667 9.87102
8.94384333333333 10.79528
1456.35666666667 5925.72
6.57061 6.36546666666667
29.4844666666667 30.6833666666667
3.69778 3.16819666666667
3.92745666666667 3.46716
0.926254333333333 1.69687
1.85601666666667 2.07699333333333
0.103552 0.102502333333333
0.103776 0.100246333333333
25.0119 39.1488
23.9352666666667 24.7750666666667
2.39057666666667 1.79723333333333
1.18379333333333 2.79124
};
\path [draw=black]
(axis cs:1.5328115282596,2.08859)
--(axis cs:1.92776180507373,2.08859);

\path [draw=black]
(axis cs:0.728100143220832,1.41726)
--(axis cs:1.01040452344583,1.41726);

\path [draw=black]
(axis cs:1.64843425456102,2.28884)
--(axis cs:7.07345241210565,2.28884);

\path [draw=black]
(axis cs:2.91251483502427,16.9607)
--(axis cs:32.0963184983091,16.9607);

\path [draw=black]
(axis cs:0.569432067321594,1.62675666666667)
--(axis cs:2.09876793267841,1.62675666666667);

\path [draw=black]
(axis cs:0.866046448673309,1.44702333333333)
--(axis cs:1.31875821799336,1.44702333333333);

\path [draw=black]
(axis cs:9.08332057330833,9.87102)
--(axis cs:10.001712760025,9.87102);

\path [draw=black]
(axis cs:7.81992262283419,10.79528)
--(axis cs:10.0677640438325,10.79528);

\path [draw=black]
(axis cs:1428.34898042005,5925.72)
--(axis cs:1484.36435291328,5925.72);

\path [draw=black]
(axis cs:5.25867809315422,6.36546666666667)
--(axis cs:7.88254190684578,6.36546666666667);

\path [draw=black]
(axis cs:29.3128272319334,30.6833666666667)
--(axis cs:29.6561061013999,30.6833666666667);

\path [draw=black]
(axis cs:2.32262716010668,3.16819666666667)
--(axis cs:5.07293283989332,3.16819666666667);

\path [draw=black]
(axis cs:3.41233417208676,3.46716)
--(axis cs:4.44257916124657,3.46716);

\path [draw=black]
(axis cs:0.887986759798783,1.69687)
--(axis cs:0.964521906867884,1.69687);

\path [draw=black]
(axis cs:1.62592939607377,2.07699333333333)
--(axis cs:2.08610393725956,2.07699333333333);

\path [draw=black]
(axis cs:0.102863710816589,0.102502333333333)
--(axis cs:0.104240289183411,0.102502333333333);

\path [draw=black]
(axis cs:0.102925333986416,0.100246333333333)
--(axis cs:0.104626666013584,0.100246333333333);

\path [draw=black]
(axis cs:24.8978426167814,39.1488)
--(axis cs:25.1259573832186,39.1488);

\path [draw=black]
(axis cs:23.7654394145675,24.7750666666667)
--(axis cs:24.1050939187658,24.7750666666667);

\path [draw=black]
(axis cs:0.82761974117078,1.79723333333333)
--(axis cs:3.95353359216255,1.79723333333333);

\path [draw=black]
(axis cs:1.12069557091483,2.79124)
--(axis cs:1.24689109575184,2.79124);

\path [draw=black]
(axis cs:1.73028666666667,1.80920596072789)
--(axis cs:1.73028666666667,2.36797403927211);

\path [draw=black]
(axis cs:0.869252333333333,1.3555880528387)
--(axis cs:0.869252333333333,1.4789319471613);

\path [draw=black]
(axis cs:4.36094333333333,2.12535189136005)
--(axis cs:4.36094333333333,2.45232810863995);

\path [draw=black]
(axis cs:17.5044166666667,16.803180710176)
--(axis cs:17.5044166666667,17.118219289824);

\path [draw=black]
(axis cs:1.3341,1.5922810807801)
--(axis cs:1.3341,1.66123225255324);

\path [draw=black]
(axis cs:1.09240233333333,1.23430181705233)
--(axis cs:1.09240233333333,1.65974484961434);

\path [draw=black]
(axis cs:9.54251666666667,8.60803097645308)
--(axis cs:9.54251666666667,11.1340090235469);

\path [draw=black]
(axis cs:8.94384333333333,7.72838952027954)
--(axis cs:8.94384333333333,13.8621704797205);

\path [draw=black]
(axis cs:1456.35666666667,0.001) 
--(axis cs:1456.35666666667,12303.3870923204);
%-451.94709232041
\path [draw=black]
(axis cs:6.57061,6.06308395548393)
--(axis cs:6.57061,6.6678493778494);

\path [draw=black]
(axis cs:29.4844666666667,30.5044843155094)
--(axis cs:29.4844666666667,30.8622490178239);

\path [draw=black]
(axis cs:3.69778,2.99297040525335)
--(axis cs:3.69778,3.34342292807999);

\path [draw=black]
(axis cs:3.92745666666667,3.37923071136419)
--(axis cs:3.92745666666667,3.55508928863581);

\path [draw=black]
(axis cs:0.926254333333333,1.6579504256618)
--(axis cs:0.926254333333333,1.7357895743382);

\path [draw=black]
(axis cs:1.85601666666667,1.90992878346646)
--(axis cs:1.85601666666667,2.24405788320021);

\path [draw=black]
(axis cs:0.103552,0.101326612183476)
--(axis cs:0.103552,0.103678054483191);

\path [draw=black]
(axis cs:0.103776,0.0995570006447456)
--(axis cs:0.103776,0.100935666021921);

\path [draw=black]
(axis cs:25.0119,21.3520326124471)
--(axis cs:25.0119,56.9455673875529);

\path [draw=black]
(axis cs:23.9352666666667,24.5465349134633)
--(axis cs:23.9352666666667,25.0035984198701);

\path [draw=black]
(axis cs:2.39057666666667,1.64170658872364)
--(axis cs:2.39057666666667,1.95276007794303);

\path [draw=black]
(axis cs:1.18379333333333,0.889074279336665)
--(axis cs:1.18379333333333,4.69340572066334);

\addplot [semithick, color0, dashed]
table {%
0.0637233018755665 0.0637233018755665
2396.09720542543 2396.09720542543
};
\addplot [black, mark=|, mark size=2, mark options={solid}, only marks]
table {%
1.5328115282596 2.08859
0.728100143220832 1.41726
1.64843425456102 2.28884
2.91251483502427 16.9607
0.569432067321594 1.62675666666667
0.866046448673309 1.44702333333333
9.08332057330833 9.87102
7.81992262283419 10.79528
1428.34898042005 5925.72
5.25867809315422 6.36546666666667
29.3128272319334 30.6833666666667
2.32262716010668 3.16819666666667
3.41233417208676 3.46716
0.887986759798783 1.69687
1.62592939607377 2.07699333333333
0.102863710816589 0.102502333333333
0.102925333986416 0.100246333333333
24.8978426167814 39.1488
23.7654394145675 24.7750666666667
0.82761974117078 1.79723333333333
1.12069557091483 2.79124
};
\addplot [black, mark=|, mark size=2, mark options={solid}, only marks]
table {%
1.92776180507373 2.08859
1.01040452344583 1.41726
7.07345241210565 2.28884
32.0963184983091 16.9607
2.09876793267841 1.62675666666667
1.31875821799336 1.44702333333333
10.001712760025 9.87102
10.0677640438325 10.79528
1484.36435291328 5925.72
7.88254190684578 6.36546666666667
29.6561061013999 30.6833666666667
5.07293283989332 3.16819666666667
4.44257916124657 3.46716
0.964521906867884 1.69687
2.08610393725956 2.07699333333333
0.104240289183411 0.102502333333333
0.104626666013584 0.100246333333333
25.1259573832186 39.1488
24.1050939187658 24.7750666666667
3.95353359216255 1.79723333333333
1.24689109575184 2.79124
};
\addplot [black, mark=-, mark size=2, mark options={solid}, only marks]
table {%
1.73028666666667 1.80920596072789
0.869252333333333 1.3555880528387
4.36094333333333 2.12535189136005
17.5044166666667 16.803180710176
1.3341 1.5922810807801
1.09240233333333 1.23430181705233
9.54251666666667 8.60803097645308
8.94384333333333 7.72838952027954
1456.35666666667 -451.94709232041
6.57061 6.06308395548393
29.4844666666667 30.5044843155094
3.69778 2.99297040525335
3.92745666666667 3.37923071136419
0.926254333333333 1.6579504256618
1.85601666666667 1.90992878346646
0.103552 0.101326612183476
0.103776 0.0995570006447456
25.0119 21.3520326124471
23.9352666666667 24.5465349134633
2.39057666666667 1.64170658872364
1.18379333333333 0.889074279336665
};
\addplot [black, mark=-, mark size=2, mark options={solid}, only marks]
table {%
1.73028666666667 2.36797403927211
0.869252333333333 1.4789319471613
4.36094333333333 2.45232810863995
17.5044166666667 17.118219289824
1.3341 1.66123225255324
1.09240233333333 1.65974484961434
9.54251666666667 11.1340090235469
8.94384333333333 13.8621704797205
1456.35666666667 12303.3870923204
6.57061 6.6678493778494
29.4844666666667 30.8622490178239
3.69778 3.34342292807999
3.92745666666667 3.55508928863581
0.926254333333333 1.7357895743382
1.85601666666667 2.24405788320021
0.103552 0.103678054483191
0.103776 0.100935666021921
25.0119 56.9455673875529
23.9352666666667 25.0035984198701
2.39057666666667 1.95276007794303
1.18379333333333 4.69340572066334
};
\end{axis}

\end{tikzpicture}
  }
  \scalebox{0.45}{
% This file was created with tikzplotlib v0.9.17.
\begin{tikzpicture}

\definecolor{color0}{rgb}{0.12156862745098,0.466666666666667,0.705882352941177}

\begin{axis}[
log basis x={10},
log basis y={10},
tick align=outside,
tick pos=left,
title={\textsc{saucy}},
x grid style={white!69.0196078431373!black},
xlabel={time w/o preprocessor},
xmajorgrids,
xmin=0.0603949247030766, xmax=12671.6699781147,
xmode=log,
xtick style={color=black},
y grid style={white!69.0196078431373!black},
ylabel={time with preprocesser},
ymajorgrids,
ymin=0.0571641997291966, ymax=12680.3875469513,
ymode=log,
scale=0.66,
ytick style={color=black}
]
\draw[line width=4pt, green!30] (axis cs:60000,0.001) -- (axis cs:60000,100000);
\addplot [draw=blue, fill=blue, mark=*, only marks, opacity=0.7]
table{%
x  y
2.33671 2.59251666666667
1.97195333333333 2.25668
2.00062 2.20903666666667
269.982 69.2879666666667
1.07610933333333 1.23393666666667
0.584174666666667 0.884029
1250.2 1247.81333333333
3314.14 3332.01666666667
1.282857 1.26285
9.33786666666667 10.3748
4521.09 4469.13666666667
14.764 15.5680666666667
0.434026333333333 0.649473333333333
1.4253 2.38417333333333
1.46375 1.68
0.221559333333333 0.102304666666667
0.106383666666667 0.101347
2498.21333333333 2482.69666666667
7246.86666666667 7213.68666666667
1.82892 1.61945333333333
5.93936666666667 6.33997666666667
};
\path [draw=black]
(axis cs:2.20545899289275,2.59251666666667)
--(axis cs:2.46796100710725,2.59251666666667);

\path [draw=black]
(axis cs:1.93652746582259,2.25668)
--(axis cs:2.00737920084408,2.25668);

\path [draw=black]
(axis cs:1.98294713379217,2.20903666666667)
--(axis cs:2.01829286620783,2.20903666666667);

\path [draw=black]
(axis cs:269.552414152933,69.2879666666667)
--(axis cs:270.411585847067,69.2879666666667);

\path [draw=black]
(axis cs:0.701623000265994,1.23393666666667)
--(axis cs:1.45059566640067,1.23393666666667);

\path [draw=black]
(axis cs:0.577511184777377,0.884029)
--(axis cs:0.590838148555956,0.884029);

\path [draw=black]
(axis cs:1242.70116897999,1247.81333333333)
--(axis cs:1257.69883102001,1247.81333333333);

\path [draw=black]
(axis cs:3303.10838784825,3332.01666666667)
--(axis cs:3325.17161215175,3332.01666666667);

\path [draw=black]
(axis cs:0.673036782942,1.26285)
--(axis cs:1.892677217058,1.26285);

\path [draw=black]
(axis cs:9.18144358091666,10.3748)
--(axis cs:9.49428975241668,10.3748);

\path [draw=black]
(axis cs:4471.02723672562,4469.13666666667)
--(axis cs:4571.15276327438,4469.13666666667);

\path [draw=black]
(axis cs:14.7139887346024,15.5680666666667)
--(axis cs:14.8140112653976,15.5680666666667);

\path [draw=black]
(axis cs:0.433334257069071,0.649473333333333)
--(axis cs:0.434718409597596,0.649473333333333);

\path [draw=black]
(axis cs:1.4181461059555,2.38417333333333)
--(axis cs:1.4324538940445,2.38417333333333);

\path [draw=black]
(axis cs:1.44963953343554,1.68)
--(axis cs:1.47786046656446,1.68);

\path [draw=black]
(axis cs:0.201056432590051,0.102304666666667)
--(axis cs:0.242062234076616,0.102304666666667);

\path [draw=black]
(axis cs:0.105414890692104,0.101347)
--(axis cs:0.107352442641229,0.101347);

\path [draw=black]
(axis cs:2494.06953207255,2482.69666666667)
--(axis cs:2502.35713459412,2482.69666666667);

\path [draw=black]
(axis cs:7233.80497628772,7213.68666666667)
--(axis cs:7259.92835704562,7213.68666666667);

\path [draw=black]
(axis cs:1.08569809365798,1.61945333333333)
--(axis cs:2.57214190634202,1.61945333333333);

\path [draw=black]
(axis cs:5.85199670030395,6.33997666666667)
--(axis cs:6.02673663302938,6.33997666666667);

\path [draw=black]
(axis cs:2.33671,2.56912215828922)
--(axis cs:2.33671,2.61591117504411);

\path [draw=black]
(axis cs:1.97195333333333,2.18882306667702)
--(axis cs:1.97195333333333,2.32453693332298);

\path [draw=black]
(axis cs:2.00062,2.1617939456518)
--(axis cs:2.00062,2.25627938768154);

\path [draw=black]
(axis cs:269.982,69.1156326434572)
--(axis cs:269.982,69.4603006898761);

\path [draw=black]
(axis cs:1.07610933333333,1.23091644224528)
--(axis cs:1.07610933333333,1.23695689108806);

\path [draw=black]
(axis cs:0.584174666666667,0.853248786929479)
--(axis cs:0.584174666666667,0.914809213070521);

\path [draw=black]
(axis cs:1250.2,1240.70883554046)
--(axis cs:1250.2,1254.9178311262);

\path [draw=black]
(axis cs:3314.14,3290.23536276701)
--(axis cs:3314.14,3373.79797056632);

\path [draw=black]
(axis cs:1.282857,1.2254925661124)
--(axis cs:1.282857,1.3002074338876);

\path [draw=black]
(axis cs:9.33786666666667,10.1464044950822)
--(axis cs:9.33786666666667,10.6031955049178);

\path [draw=black]
(axis cs:4521.09,4450.51432822895)
--(axis cs:4521.09,4487.75900510438);

\path [draw=black]
(axis cs:14.764,15.3514807894211)
--(axis cs:14.764,15.7846525439122);

\path [draw=black]
(axis cs:0.434026333333333,0.638299224261217)
--(axis cs:0.434026333333333,0.66064744240545);

\path [draw=black]
(axis cs:1.4253,2.33538239542247)
--(axis cs:1.4253,2.4329642712442);

\path [draw=black]
(axis cs:1.46375,1.65815303072125)
--(axis cs:1.46375,1.70184696927875);

\path [draw=black]
(axis cs:0.221559333333333,0.101897395804027)
--(axis cs:0.221559333333333,0.102711937529307);

\path [draw=black]
(axis cs:0.106383666666667,0.100028673535627)
--(axis cs:0.106383666666667,0.102665326464373);

\path [draw=black]
(axis cs:2498.21333333333,2470.80389762176)
--(axis cs:2498.21333333333,2494.58943571158);

\path [draw=black]
(axis cs:7246.86666666667,7180.80911573057)
--(axis cs:7246.86666666667,7246.56421760276);

\path [draw=black]
(axis cs:1.82892,1.59029823734271)
--(axis cs:1.82892,1.64860842932395);

\path [draw=black]
(axis cs:5.93936666666667,6.31384396131707)
--(axis cs:5.93936666666667,6.36610937201626);

\addplot [semithick, color0, dashed]
table {%
0.0603949247030766 0.0603949247030766
12671.6699781147 12671.6699781147
};
\addplot [black, mark=|, mark size=2, mark options={solid}, only marks]
table {%
2.20545899289275 2.59251666666667
1.93652746582259 2.25668
1.98294713379217 2.20903666666667
269.552414152933 69.2879666666667
0.701623000265994 1.23393666666667
0.577511184777377 0.884029
1242.70116897999 1247.81333333333
3303.10838784825 3332.01666666667
0.673036782942 1.26285
9.18144358091666 10.3748
4471.02723672562 4469.13666666667
14.7139887346024 15.5680666666667
0.433334257069071 0.649473333333333
1.4181461059555 2.38417333333333
1.44963953343554 1.68
0.201056432590051 0.102304666666667
0.105414890692104 0.101347
2494.06953207255 2482.69666666667
7233.80497628772 7213.68666666667
1.08569809365798 1.61945333333333
5.85199670030395 6.33997666666667
};
\addplot [black, mark=|, mark size=2, mark options={solid}, only marks]
table {%
2.46796100710725 2.59251666666667
2.00737920084408 2.25668
2.01829286620783 2.20903666666667
270.411585847067 69.2879666666667
1.45059566640067 1.23393666666667
0.590838148555956 0.884029
1257.69883102001 1247.81333333333
3325.17161215175 3332.01666666667
1.892677217058 1.26285
9.49428975241668 10.3748
4571.15276327438 4469.13666666667
14.8140112653976 15.5680666666667
0.434718409597596 0.649473333333333
1.4324538940445 2.38417333333333
1.47786046656446 1.68
0.242062234076616 0.102304666666667
0.107352442641229 0.101347
2502.35713459412 2482.69666666667
7259.92835704562 7213.68666666667
2.57214190634202 1.61945333333333
6.02673663302938 6.33997666666667
};
\addplot [black, mark=-, mark size=2, mark options={solid}, only marks]
table {%
2.33671 2.56912215828922
1.97195333333333 2.18882306667702
2.00062 2.1617939456518
269.982 69.1156326434572
1.07610933333333 1.23091644224528
0.584174666666667 0.853248786929479
1250.2 1240.70883554046
3314.14 3290.23536276701
1.282857 1.2254925661124
9.33786666666667 10.1464044950822
4521.09 4450.51432822895
14.764 15.3514807894211
0.434026333333333 0.638299224261217
1.4253 2.33538239542247
1.46375 1.65815303072125
0.221559333333333 0.101897395804027
0.106383666666667 0.100028673535627
2498.21333333333 2470.80389762176
7246.86666666667 7180.80911573057
1.82892 1.59029823734271
5.93936666666667 6.31384396131707
};
\addplot [black, mark=-, mark size=2, mark options={solid}, only marks]
table {%
2.33671 2.61591117504411
1.97195333333333 2.32453693332298
2.00062 2.25627938768154
269.982 69.4603006898761
1.07610933333333 1.23695689108806
0.584174666666667 0.914809213070521
1250.2 1254.9178311262
3314.14 3373.79797056632
1.282857 1.3002074338876
9.33786666666667 10.6031955049178
4521.09 4487.75900510438
14.764 15.7846525439122
0.434026333333333 0.66064744240545
1.4253 2.4329642712442
1.46375 1.70184696927875
0.221559333333333 0.102711937529307
0.106383666666667 0.102665326464373
2498.21333333333 2494.58943571158
7246.86666666667 7246.56421760276
1.82892 1.64860842932395
5.93936666666667 6.36610937201626
};
\end{axis}

\end{tikzpicture}
  }
      \caption{Solvers with \textsc{sassy} vs. solvers without \textsc{sassy} on \textbf{portfolio\_comb} (not permuted).} \label{fig:rescomb_nopermute}
\end{figure*}
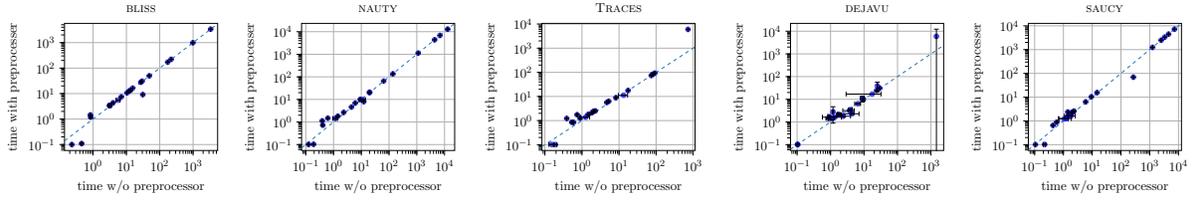
\begin{figure*}[t!]
  \centering
  \scalebox{0.45}{
\input{results/bli_portfolio_prep_nopermute.tex}
  }
  \scalebox{0.45}{
\input{results/nau_portfolio_prep_nopermute.tex}
  }
  \scalebox{0.45}{
\input{results/tra_portfolio_prep_nopermute.tex}
  }
  \scalebox{0.45}{
\input{results/dej_portfolio_prep_nopermute.tex}
  }
  \scalebox{0.45}{
\input{results/sau_portfolio_prep_nopermute.tex}
  }
      \caption{Solvers with \textsc{sassy} vs. solvers without \textsc{sassy} on \textbf{portfolio\_pract} (not permuted). Timeout is $60s$. The green bar shows instances that timed out without the preprocessor. With the preprocessor enabled no instance timed out.} \label{fig:respract_nopermute} 
\end{figure*}

\begin{figure*}[p!]
  \centering
  \resizebox{.975\hsize}{!}{
      \begin{tabular}{|l||c|c||c|c|c|c|c|}\hline
        & \multicolumn{2}{c||}{state-of-the-art} & \multicolumn{5}{c|}{this paper}\\\hline
        graph class & \textsc{saucy} & \textsc{Traces} & \textsc{sy+dejavu} & \textsc{sy+Traces} & \textsc{sy+bliss} & \textsc{sy+saucy} & \textsc{sy+nauty}\\\hline
        dac (p) & $ 0.51 \pm 0.057 $ & $ 2.49 \pm 0.072 $ & $ \textbf{0.38} \pm 0.006 $ & $ 0.91 \pm 0.016 $ & $ 0.5 \pm 0.002 $ & $ 0.41 \pm 0.058 $ & $ 0.46 \pm 0.005 $\\
states (p) & $ 7.32 \pm 0.031 $ & $ 12.55 \pm 0.281 $ & $ \textbf{6.79} \pm 0.075 $ & $ 6.8 \pm 0.069 $ & $ 6.8 \pm 0.062 $ & $ 6.83 \pm 0.085 $ & $ 6.8 \pm 0.074 $\\
internet (p) & $ 0.19 \pm 0.005 $ & $ 2.47 \pm 0.379 $ & $ \textbf{0.14} \pm 0.003 $ & $ 0.15 \pm 0.004 $ & $ \textbf{0.14} \pm 0.002 $ & $ \textbf{0.14} \pm 0.004 $ & $ \textbf{0.14} \pm 0.003 $\\
ispd (p) & $ 7.12 \pm 0.069 $ & $ 7.04 \pm 0.085 $ & $ 5.48 \pm 0.046 $ & $ \textbf{5.43} \pm 0.023 $ & $ 5.46 \pm 0.005 $ & $ 5.45 \pm 0.025 $ & $ 5.45 \pm 0.008 $\\
MIP2017 (p) & $ 22.3 \pm 0.22 $ & $ 803.63 \pm 10.469 $ & $ \textbf{14.07} \pm 0.171 $ & $ 92.76 \pm 2.796 $ & $ 28.59 \pm 0.234 $ & $ 15.52 \pm 0.324 $ & $ 26.54 \pm 0.128 $\\
SAT2021 (p) & $ 2217.61 \pm 0.627 $ & $ 3645.33 \pm 11.963 $ & $ \textbf{1701.62} \pm 10.411 $ & $ 1856.39 \pm 10.138 $ & $ 1939.54 \pm 4.926 $ & $ 1786.68 \pm 6.005 $ & $ 1763.12 \pm 6.214 $\\
SAT2021-up (p) & $ 1886.87 \pm 4.472 $ & $ 2948.5 \pm 25.254 $ & $ \textbf{1439.5} \pm 2.972 $ & $ 1538.71 \pm 8.639 $ & $ 1650.68 \pm 3.577 $ & $ 1508.91 \pm 2.327 $ & $ 1481.28 \pm 3.556 $\\\hline
      \end{tabular}}\\
      \caption{Benchmark results on various sets of large, practical graphs (\textbf{randomly permuted}), timeout is $60s$. Running out of memory also counts as a timeout. The benchmarks compare solver configurations using the preprocessor (``\textsc{sy+}'') to state of the art \saucy{} and \Traces{}. Shown values are the sum over all instances in the set in seconds. The average and standard deviation of $3$ consecutive runs is used. Bold entries indicates the fastest running time for the given set.} \label{fig:benchmarks1_permute}
    \end{figure*}
      \begin{figure*}[p!]    
        \centering
   %   \vspace{0.3cm}\hspace{0.3cm}\\
   \scalebox{0.95}{
      \includegraphics{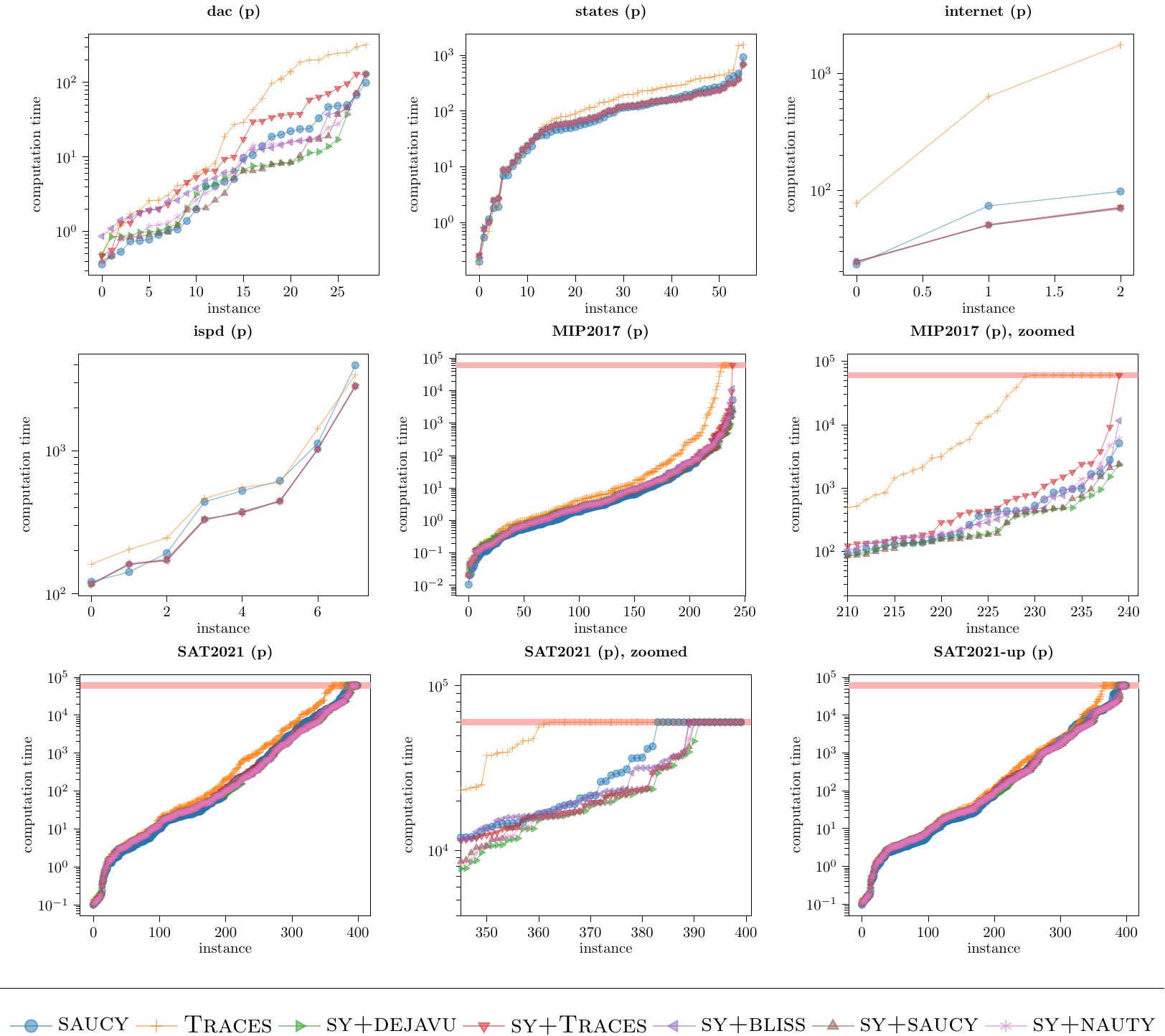}
   }
  \caption{Detailed plots for the various sets of Figure~\ref{fig:benchmarks1_permute}. The red bar illustrates timeouts. Instances are sorted according to running time.} \label{fig:benchmarks2_permute}
\end{figure*}

\begin{figure*}[p!]
  \centering
  \resizebox{.975\hsize}{!}{
      \begin{tabular}{|l||c|c||c|c|c|c|c|}\hline
                    & \multicolumn{2}{c||}{state-of-the-art} & \multicolumn{5}{c|}{this paper}\\\hline
        graph class & \textsc{saucy} & \textsc{Traces} & \textsc{sy+dejavu} & \textsc{sy+Traces} & \textsc{sy+bliss} & \textsc{sy+saucy} & \textsc{sy+nauty}\\\hline
        dac & $ 0.35 \pm 0.008 $ & $ 2.47 \pm 0.005 $ & $ 0.28 \pm 0.001 $ & $ 0.81 \pm 0.002 $ & $ 0.37 \pm 0.002 $ & $ \textbf{0.27} \pm 0.002 $ & $ 0.34 \pm 0.001 $\\
states & $ \textbf{2.89} \pm 0.054 $ & $ 7.58 \pm 0.158 $ & $ 3.85 \pm 0.048 $ & $ 3.85 \pm 0.04 $ & $ 3.85 \pm 0.041 $ & $ 3.85 \pm 0.038 $ & $ 3.85 \pm 0.043 $\\
internet & $ 0.15 \pm 0.002 $ & $ 2.23 \pm 0.022 $ & $ \textbf{0.13} \pm 0.000 $ & $ \textbf{0.13} \pm 0.001 $ & $ \textbf{0.13} \pm 0.001 $ & $ \textbf{0.13} \pm 0.000 $ & $ \textbf{0.13} \pm 0.000 $\\
ispd & $ 3.83 \pm 0.028 $ & $ 4.84 \pm 0.059 $ & $ 3.7 \pm 0.057 $ & $ 3.7 \pm 0.061 $ & $ 3.7 \pm 0.057 $ & $ 3.7 \pm 0.054 $ & $ \textbf{3.68} \pm 0.039 $\\
MIP2017 & $ 10.96 \pm 0.158 $ & $ 774.09 \pm 0.578 $ & $ \textbf{9.12} \pm 0.165 $ & $ 84.42 \pm 0.201 $ & $ 21.46 \pm 0.331 $ & $ 10.66 \pm 0.109 $ & $ 21.04 \pm 0.163 $\\
SAT2021    & $ 1292.76 \pm 1.641 $ & $ 2855.57 \pm 10.636 $ & $ \textbf{881.69} \pm 0.982 $ & $ 1058.97 \pm 3.283 $ & $ 1149.04 \pm 3.73 $ & $ 990.96 \pm 3.323 $ & $ 988.15 \pm 2.662 $\\
SAT2021-up & $ 1144.06 \pm 3.648 $ & $ 2393.85 \pm 4.799 $ & $ \textbf{780.02} \pm 6.009 $ & $ 903.93 \pm 2.517 $ & $ 1027.61 \pm 3.815 $ & $ 876.77 \pm 2.535 $ & $ 865.38 \pm 2.578 $\\
\hline
      \end{tabular}}\\
      \caption{Benchmark results on various sets of large, practical graphs (\textbf{not permuted}), timeout is $60s$. Running out of memory also counts as a timeout. The benchmarks compare solver configurations using the preprocessor (``\textsc{sy+}'') to state of the art \saucy{} and \Traces{}. Shown values are the sum over all instances in the set in seconds. The average and standard deviation of $3$ consecutive runs is used. Bold entries indicates the fastest running time for the given set.} \label{fig:benchmarks1_nopermute}
    \end{figure*}
    \begin{figure*}[p!]    
      \centering
 %   \vspace{0.3cm}\hspace{0.3cm}\\
 \scalebox{0.95}{
    \includegraphics{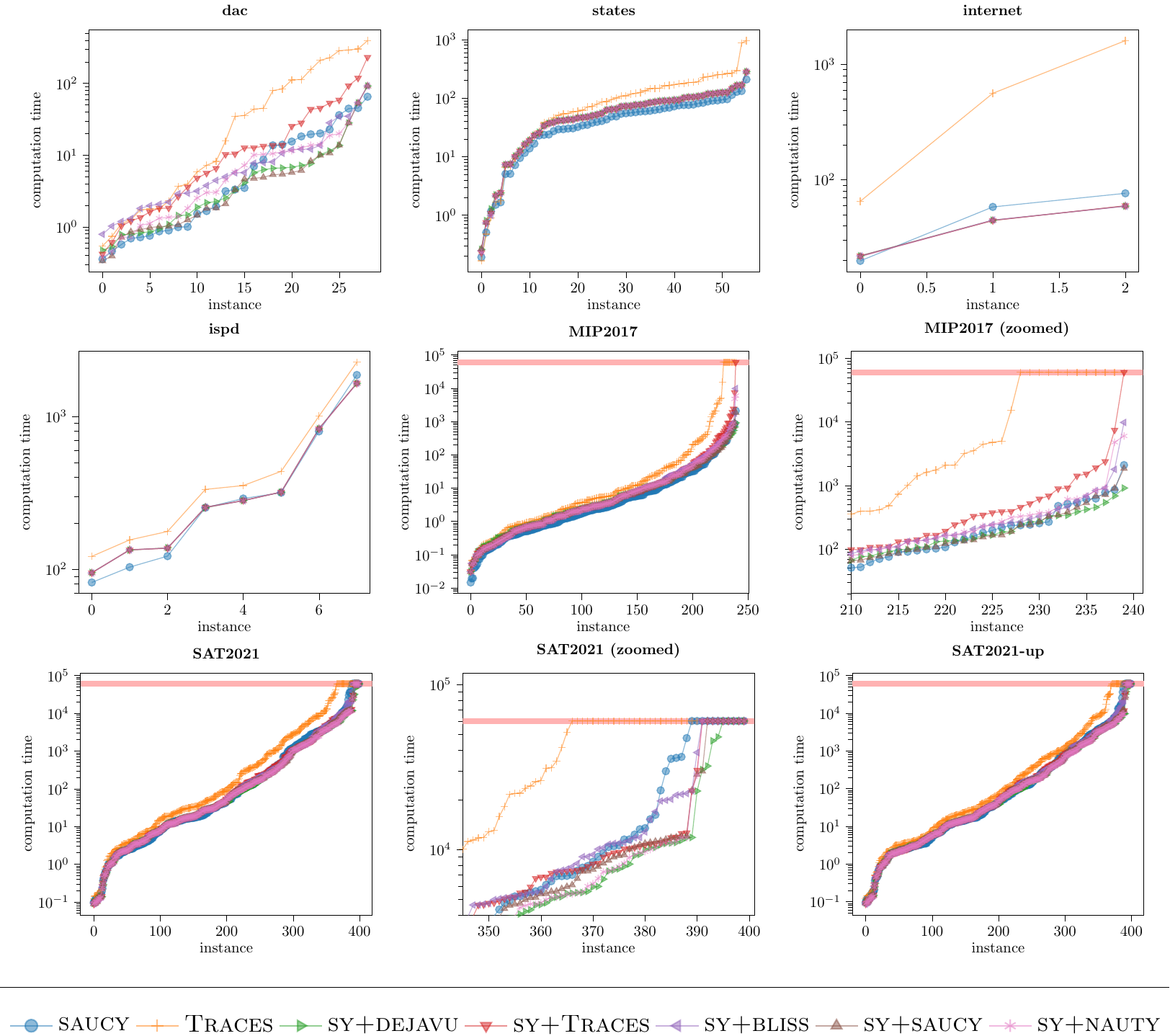}
 }
\caption{Detailed plots for the various sets of Figure~\ref{fig:benchmarks1_nopermute}. The red bar illustrates timeouts. Instances are sorted according to running time.} \label{fig:benchmarks2_nopermute}
\end{figure*}

\textbf{portfolio\_pract:} The goal of this set is to measure whether preprocessing is worthwhile for a given solver on graphs where there is a lot of exploitable structure. Thus, this set contains practical graphs. Note that we test practical graphs much more thoroughly in the next section. To make up \textbf{portfolio\_pract}, we picked the $5$ largest instances (if available) of all the \saucy{} benchmark sets, and for the sets arising from computational tasks (MIP and SAT) we picked $5$ instances uniformly at random (but of course the same instances were used for all the solvers).

The results are summarized in Figure~\ref{fig:respract_permute} and Figure~\ref{fig:respract_nopermute}.
We conclude for \bliss{}, \nauty{} and \dejavu{} that the preprocessor increases performance dramatically on most instances. 
For \Traces{}, performance also improves, in particular there are fewer timeouts. However, the improvement is not as dramatic.
For \saucy{}, there is no clear improvement.
Note that for most instances, standard deviation is too low to be visible in the figure.

\textbf{portfolio\_comb:} The goal of this set is to measure the overhead of applying the preprocessor on graphs where there is no or very little exploitable structure (i.e., where the preprocessor is expected to have no effect). For this purpose, we chose a large variety of graphs from combinatorics, on which solvers are routinely evaluated~\cite{DBLP:journals/jsc/McKayP14}. The subset we chose contains a graph from almost every graph class of the benchmark library from \cite{nautyTracesweb} (cfi, grid, grid-sw, had, had-sw, hypercubes, kef, latin, latin-sw, lattice, mz, paley, pp, ran10, ransq, sts, sts-sw, ranreg, tran, triang and shrunken multipedes). Whenever applicable, we chose a graph of around $1000$ vertices: note that here, we apply a size restriction, since combinatorial graphs are generally difficult for their size. We choose an even smaller graph or left out sets entirely whenever a solver had trouble finishing the instance quickly. Note that, since we want to measure the preprocessing overhead, only instances for which the solvers finish in a reasonable amount of time are of interest. If solvers take a long time solving an instance to begin with, the overhead of the preprocessor is always negligible.
Note that these restrictions \textbf{only apply to portfolio\_comb}: all the other sets tested in this paper have no restriction on the size of instances and instances were not chosen manually.

The results are summarized in Figure~\ref{fig:rescomb_permute} and Figure~\ref{fig:rescomb_nopermute}.
Note that again, for many instances, standard deviation is too low to be visible in the figure.
Overall, for the tested solvers, we conclude that the overhead of the preprocessor is negligible.
There are however two eye-catching instances: first, there is an instance with very high standard deviation for \dejavu{}. The instance in question is a Kronecker eye flip graph, which \dejavu{} is however known to struggle with \cite{DBLP:conf/alenex/AndersS21}.
Secondly, there is a particular expensive outlier for \Traces{}. 
We analyze and discuss the instance in detail in Appendix~\ref{sec:outlier}. There, we conclude that the increased running time is caused through an undesired interaction with a heuristic used by \Traces{}: in fact, we can produce a graph that is structurally equivalent to the preprocessed instance and that runs $6$ times faster than the original instance. 

\subsection{Comparison to state-of-the-art}
The state-of-the-art solver for solving large practical graphs is \saucy{}. Therefore, we compare all the solvers with the preprocessor to \saucy{} without the preprocessor. 
Since \Traces{} also contains low-degree techniques and a flavor of sparse automorphism detection, we also compare performance to \Traces{} without the preprocessor.
The timeout used is 60$s$ (also if a solver runs out of memory).

The benchmarks contain all sets from the \saucy{} distribution. We also test $3$ contemporary sets of practical graphs: the MIP2017 set contains graphs stemming from the mixed integer programming library (see \cite{miplib}). The SAT2021 library contains graphs stemming from SAT instances from the SAT competition 2021 \cite{satComp2021}. The set SAT2021-up is similar. However, instances were first preprocessed using the unit and pure literal rule (for a discussion on why this is relevant, see \cite{DBLP:conf/sat/Anders22}).
We want to remark that the SAT sets contain many large graphs with tens of millions of vertices -- in particular the largest graphs out of all the tested sets. 

The results are summarized in Figure~\ref{fig:benchmarks1_permute} and Figure~\ref{fig:benchmarks2_permute}, as well as Figure~\ref{fig:benchmarks1_nopermute} and Figure~\ref{fig:benchmarks2_nopermute}. First of all, we observe that the previous state of the art (\saucy{}) is outperformed on all but one set by several solvers using the preprocessor.
This demonstrates that the approach of using our universal preprocessor in conjunction with different solvers can outperform state-of-the-art.
Moreover, both \saucy{} and \Traces{} also visibly speed up by applying the preprocessor on most of the sets.

We want to mention that \dejavu{} and \Traces{} do run out of memory on some of the very large graphs in the SAT2021 sets. Hence, depending on the amount of RAM and how this is weighed into the evaluation, other solvers may be preferable. In all cases where \dejavu{} runs out of memory, all the other solvers time out. 
In any case, on all the sets where this is relevant, \textsc{sy+nauty} and \textsc{sy+saucy} also outperform \saucy{}. 

Going into more detail, we often observe that \textsc{saucy} performs best on instances that are solved quickly, falling behind on the larger or slower instances. 
We feel that this might indicate unused low-level optimization potential in the preprocessor implementation itself.

\section{Conclusion and Future Development}
We introduced the new \textsc{sassy} preprocessor for symmetry detection. 
We demonstrated that \textsc{sassy} indeed speeds up state-of-the-art solvers on large, practical graphs.

Future additions to the preprocessor could include more heuristics for degree $2$ removal, stronger invariants or even more efficient implementations and tuning of the existing heuristics.
Since we have observed a high sensitivity of state-of-the-art solvers to their choice of cell selectors, a more extensive study into the topic would be of interest.

\section*{Acknowledgements}
We thank Marc E.\ Pfetsch and Christopher Hojny for giving us further insights into the user-side of symmetry detection software, as well as providing us with the MIP2017 graphs.

\bibliography{main}
\bibliographystyle{plain}

\newpage
\appendix
{\Huge\noindent Appendix}
\section{The Outlier in Combinatorial Graphs} \label{sec:outlier}
There is one particular outlier in the evaluation of \Traces{} comparing preprocessed vs. unprocessed instances. 
The instance in question is a shrunken multipede on $408$ vertices. Without preprocessing, it is solved in $0.75s$, while with preprocessing it is solved in $6.3s$. This is at first glance confusing: first of all, the preprocessor finishes within less than $0.5$ms. The preprocessor also does not change the graph -- no vertices or edges are removed -- other than coloring it with its coarsest equitable coloring. 
This is however almost the same coloring \Traces{} would also compute for the graph.

The only difference is that \Traces{} might name the colors differently internally, e.g., color $3$ might be named color $6$ instead. 
While this does not structurally make the graph harder or easier, heuristics internally might always, for example, choose the ``first largest color'' (this is similar to, e.g., variable ordering in SAT solvers). Thus, renaming the colors might influence the decisions made by the solver.
Using the ``first'' color is however usually not a deliberate decision.
In fact, if we simply reverse the order of colors, the graph is indeed solved in $0.12s$.
We also want to mention that the $0.75s$ of \Traces{} on the graph is already slower than all the other solvers.

In \cite{DBLP:conf/esa/AndersS21}, it is also argued that cell selector choice has a significant impact on the set of shrunken multipedes.
We believe that the solution to this issue is to make structurally better choices, and has indeed little to do with the role of the preprocessor.

We also want to mention that in other testing, we also found graphs that become easier through preprocessing in the same manner for most of the solvers.

\section{Ablation study} \label{sec:ablation}
\begin{figure}
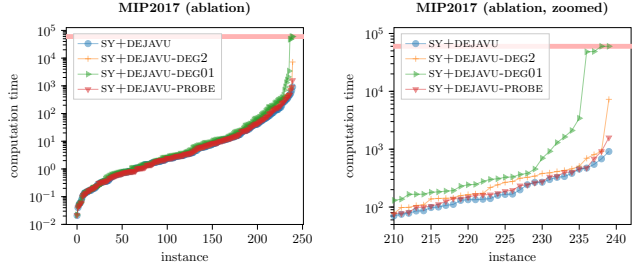

  \centering
  \scalebox{0.46}{
    \input{results/MIP_ablation_comp.tex}
      }
      \scalebox{0.46}{
        \input{results/MIP_ablation_comp_zoom.tex}
          }
  \caption{Ablation study for \textsc{sy+dejavu} on the MIP2017 graphs (times: $9.15$s, $17.87$s, $234.47$s, $10.65$s), timeout is $60$s.} \label{fig:ablation}
\end{figure}

In Figure~\ref{fig:ablation} we evaluate for \dejavu{} on the MIP2017 set  the effect of each of the preprocessing techniques separately. 
We do so by running the configuration \textsc{SY+dejavu}, but performing a separate run for each technique, deactivating the respective technique.
For example, \textsc{SY+dejavu-deg2} runs \textsc{SY+dejavu} without any of the degree $2$ removal techniques.
The data shows that each of the techniques has a beneficial impact on running time. 
By far the most impactful technique is the removal of degree $0$ and $1$, followed by the removal of vertices of degree $2$.
Even without the use of probing the set is solved faster than previous state-of-the-art (compare with Figure~\ref{fig:benchmarks1_nopermute}), however, the probing technique is still beneficial on top of the other techniques. 

\end{document}